\documentclass[aps,prx,twocolumn,superscriptaddress,nofootinbib]{revtex4-2}
\pdfoutput=1
\usepackage{graphicx,amssymb,amsmath,amsfonts,empheq}
\usepackage[dvipsnames]{xcolor}
\usepackage{color}
\usepackage{braket}
\usepackage{latexsym}
\usepackage{upgreek}
\usepackage{dsfont}
\usepackage{bm}
\usepackage{makecell}
\usepackage{multirow}
\usepackage{enumitem}
\usepackage[normalem]{ulem}
\usepackage{url}
\usepackage{hyperref}
\hypersetup{
colorlinks = true,
linkcolor = blue, 
citecolor = blue,
urlcolor = [rgb]{0.25, 0.41, 0.88}}
\usepackage{bbm}
\usepackage[utf8]{inputenc}
\usepackage[english]{babel}
\usepackage[T1]{fontenc}
\usepackage{lipsum}
\usepackage{physics}
\usepackage{comment}
\usepackage{wasysym} %
\usepackage{slashed}
\usepackage{booktabs}
\usepackage{xspace}

\usepackage{tikz}
\usepackage{tikz-cd}
\usepackage{tkz-base}
\usetikzlibrary{arrows}
\usetikzlibrary{intersections}
\usetikzlibrary{shapes.geometric}
\usetikzlibrary{decorations.pathmorphing, patterns,shapes}
\usetikzlibrary{decorations.markings}
\usetikzlibrary{calc}
\usepackage{diagbox}

\tikzset{
	partial ellipse/.style args={#1:#2:#3}{
		insert path={+ (#1:#3) arc (#1:#2:#3)}
	}
}

\tikzset{
	mid arrow/.style={postaction={decorate,decoration={
				markings,
				mark=at position .575 with {\arrow[#1]{stealth}}
	}}},
	near arrow/.style={postaction={decorate,decoration={
				markings,
				mark=at position .275 with {\arrow[#1]{stealth}}
	}}},
	far arrow/.style={postaction={decorate,decoration={
				markings,
				mark=at position .800 with {\arrow[#1]{stealth}}
	}}},
}

\pgfdeclarepatternformonly{south west lines}{\pgfqpoint{-0pt}{-0pt}}{\pgfqpoint{3pt}{3pt}}{\pgfqpoint{3pt}{3pt}}{
	\pgfsetlinewidth{0.4pt}
	\pgfpathmoveto{\pgfqpoint{0pt}{0pt}}
	\pgfpathlineto{\pgfqpoint{3pt}{3pt}}
	\pgfpathmoveto{\pgfqpoint{2.8pt}{-.2pt}}
	\pgfpathlineto{\pgfqpoint{3.2pt}{.2pt}}
	\pgfpathmoveto{\pgfqpoint{-.2pt}{2.8pt}}
	\pgfpathlineto{\pgfqpoint{.2pt}{3.2pt}}
	\pgfusepath{stroke}}

\usepackage[dvipsnames]{xcolor}

\renewcommand{\bar}{\overline}
\renewcommand{\leq}{\leqslant}
\renewcommand{\geq}{\geqslant}

\newcommand{\rd}{\mathrm{d}}
\newcommand{\ri}{\mathrm{i}}

\newcommand{\nn}{\nonumber}

\newcommand{\U}{\operatorname{U}}
\renewcommand{\O}{\operatorname{O}}

\newcommand{\SO}{\operatorname{SO}}

\newcommand{\T}{\operatorname{\textbf{T}}}

\newcommand{\bbZ}{\mathbb{Z}}

\newcommand{\calC}{\mathcal{C}}
\newcommand{\calD}{\mathcal{D}}
\newcommand{\calE}{\mathcal{E}}
\newcommand{\calF}{\mathcal{F}}
\newcommand{\calG}{\mathcal{G}}
\newcommand{\calH}{\mathcal{H}}

\newcommand{\calN}{\mathcal{N}}
\newcommand{\calO}{\mathcal{O}}
\newcommand{\calP}{\mathcal{P}}
\newcommand{\calQ}{\mathcal{Q}}
\newcommand{\calR}{\mathcal{R}}

\newcommand{\calZ}{\mathcal{Z}}

\newcommand{\sfa}{\mathsf{a}}
\newcommand{\sfb}{\mathsf{b}}

\newcommand{\sfv}{\mathsf{v}}

\newcommand{\bfr}{\mathbf{r}}

\newcommand{\bfZ}{\mathbf{Z}}
\newcommand{\bfx}{\mathbf{x}}
\newcommand{\bfy}{\mathbf{y}}
\newcommand{\bfe}{\mathbf{e}}

\newcommand{\NLSM}{NL$\sigma$M\xspace}

\newlength{\PanelPairWidth}
\newsavebox{\PanelPairArtwork}

\definecolor{red1}{RGB}{240,83,90}
\definecolor{blue1}{RGB}{91,98,165}
\definecolor{myturquoise}{RGB}{83,195,189}
\definecolor{lightturquoise}{RGB}{64, 224, 208}
\definecolor{violet}{RGB}{207, 159, 255}
\definecolor{darkturquoise}{RGB}{54, 194, 178}  %
\definecolor{darkviolet}{RGB}{177, 129, 225}    %
\definecolor{canaryyellow}{RGB}{255, 255, 143}
\definecolor{mygreen}{RGB}{134, 175, 142}
\definecolor{mint}{RGB}{152, 251, 152}
\definecolor{mint2}{RGB}{218, 247, 166}
\definecolor{myred}{RGB}{240,83,90}
\definecolor{myblue}{RGB}{73,103,189}
\definecolor{shockingpink}{RGB}{252, 15, 192}
\definecolor{neonpink}{RGB}{255, 16, 240}

\begin{document}

\title{Algebraic strong-to-weak symmetry breaking\\ and partial decodability in the surface code}

\author{Stephen W. Yan}
\affiliation{Department of Physics, University of California,
Santa Barbara, CA 93106, USA}
\author{Yimu Bao}
\affiliation{Kavli Institute for Theoretical Physics, University of California,
Santa Barbara, CA 93106, USA}
\affiliation{Department of Physics, Stanford University, Stanford, California 94305, USA}
\author{Sagar Vijay}
\affiliation{Department of Physics, University of California,
Santa Barbara, CA 93106, USA}

\begin{abstract}
Coherent noise can give rise to unconventional regimes of quantum information recovery and mixed-state order.
We study these phenomena in two settings related by Wegner duality: the triangular-lattice surface code subject to single-qubit Pauli-$Z$ rotations followed by syndrome measurements, and a $\mathbb{Z}_2$ paramagnet subject to strongly symmetric decoherence.
Increasing the rotation angle drives a transition from a decodable phase to a \emph{partially decodable} critical phase in the code, and from a strongly symmetric phase to a phase with \emph{algebraic} $\mathbb{Z}_2$ strong-to-weak spontaneous symmetry breaking (SWSSB) in the dual paramagnet. 
We develop a common description of these phenomena in terms of a weakly coupled $\mathrm{SO}(n)$ nonlinear sigma model in the replica limit $n\to1$.
Taking this replica limit analytically, we derive universal signatures of the critical phase: an optimal Pauli decoding fidelity that approaches a universal function of the system's aspect ratio in the thermodynamic limit, and a dual fidelity correlator that decays as a universal power law.
These observables relate to correlations of twist operators in the sigma model description. %
Large-scale numerical simulations confirm these analytical predictions.
We also argue that the algebraic SWSSB phase cannot be realized by any Pauli dephasing channel acting on the $\mathbb{Z}_{2}$ paramagnet. We probe this ``non-classical'' character through the output state's sensitivity to well-separated $\mathbb{Z}_{2}$ fluxes inserted in the initial state, which reveals persistent coherence between distinct $\mathbb{Z}_{2}$ charge trajectories at long distances.

\end{abstract}

\maketitle

\section{Introduction}

Local noise can drive phase transitions in both quantum information recovery and mixed-state many-body order. In quantum error-correction, increasing noise strength can induce a decoding transition beyond which logical information can no longer be reliably recovered using local ``syndrome'' measurements~\cite{dennis_surface_code}.
In quantum many-body systems, strongly symmetric coupling to an environment can induce strong-to-weak spontaneous symmetry breaking (SWSSB), a form of mixed-state order revealed by nonlinear correlation functions, but invisible to conventional symmetry-breaking diagnostics~\cite{lessa2025strong,lee2023quantum,sala2024spontaneous,temkin2025charge,hauser2025information,sala2026climbing}.

These seemingly distinct phenomena are directly connected in a simple setting: the surface code subject to Pauli-$Z$ dephasing.
Under Wegner duality~\cite{wegner1971duality}, the initial code state of the surface code maps to an Ising paramagnet with an on-site $\mathbb{Z}_2$ symmetry, while local Pauli errors become processes that incoherently create and transport symmetry charges.
The decoding transition induced by increasing the dephasing strength in the surface code then corresponds to an SWSSB transition in the dual mixed state~\cite{fan2024diagnostics,lee2023quantum,bao2023mixed,wang2025intrinsic}.
Both transitions admit a classical statistical-mechanical description.
Decoding is an inference problem closely related to the two-dimensional random-bond Ising model along the Nishimori line~\cite{fan2023diagnostics,tang2025phases,wang2003confinement,katzgraber2009error,bombin2010topological,bombin2012strong,kubica2018three,chubb2021statistical,song2022optimal,ferris2014tensor,bravyi2014efficient,Darmawan_2017,Darmawan_2018}, while the dual SWSSB transition is entirely encoded in the classical probability distribution over symmetry-charge configurations.

More generic noise introduces an ingredient absent from this classical picture: interference between distinct error configurations, including those that produce the same syndrome~\cite{bravyi_coherent,cheng2025emergent,venn2022coherent,behrends2024surface,behrends_general_x,phases_of_decodability_bao,lavasani2025stability}.
Correspondingly, the dynamics of the symmetry charges in the dual system can  retain contributions from interfering processes.
This raises a central question: can such interference survive at long distances and give rise to universal regimes of decodability and mixed-state order with no counterpart under Pauli dephasing?

We address this question through two dual descriptions of a locally decohered quantum system.
In the error correction setting, we consider a triangular-lattice surface code initialized in a logical state $\ket{\psi_0}$ and subject to coherent single-qubit $Z$ rotations followed by syndrome measurements.  The corrupted code state is 
\begin{align}
    \rho
    &=
    \mathcal{E}_{\theta}\left[
        \ketbra{\psi_0}
    \right]
    \,,
    \label{eq:surface_code_setup}
    \\
    \mathcal{E}_{\theta}\left[\cdot\right]
    &=
    \mathcal{D}_{\text{syn}}\!\left[
        e^{\ri\theta\sum_\ell Z_\ell}
        \left(\cdot\right)
        e^{-\ri\theta\sum_\ell Z_\ell}
    \right]
    \,.
\end{align}
where $\mathcal{D}_{\text{syn}}$ is complete dephasing in the syndrome basis, giving rise to an ensemble of post-measurement states. 
We study the optimal recovery fidelity achievable by transversal Pauli corrections conditioned on the measured syndrome.

\begin{figure}[t]
\centering
\includegraphics[width=\columnwidth]{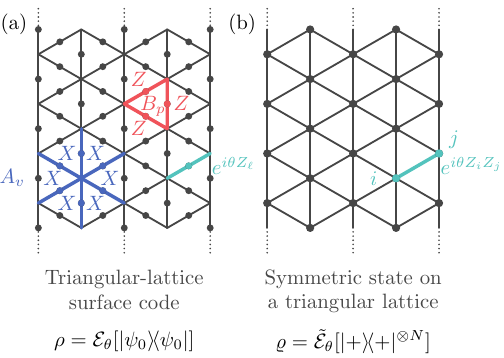}
\caption{Setup.
(a) Triangular-lattice surface codes with coherent errors. The code operates on physical qubits on edges and consists of vertex stabilizers $A_v$ (blue) and plaquette stabilizers $B_p$ (red).
The qubits are subject to coherent rotations $e^{\ri \theta \sum_\ell Z_\ell}$ followed by syndrome measurements, with their combined effect modeled by $\calE_\theta$.
(b) The Wegner dual problem involves qubits on the vertices initialized in the symmetric state $\ket{+}^{\otimes N}$.
This state is evolved under the strongly symmetric channel $\tilde{\mathcal{E}}_\theta$ consisting of two-qubit coherent rotation $e^{\ri\theta\sum_{\langle i, j \rangle} Z_i Z_j}$ followed by single-qubit Pauli-X measurements.
}
\label{fig:setup}
\end{figure}

The complementary mixed-state description follows from Wegner duality.
The initial code state maps to an Ising paramagnet $\ket{+}^{\otimes N}$, while the noise and measurement process maps to a local channel comprising symmetric coherent rotations followed by complete dephasing in the $X$ basis,
\begin{align}
    \varrho
    &=
    \tilde{\mathcal{E}}_{\theta}\left[
        \ketbra{+}^{\otimes N}
    \right]
    \,,
    \label{eq:sswb_setup}
    \\
    \tilde{\mathcal{E}}_{\theta}\left[\cdot\right]
    &=
    \mathcal{D}_{X}\!\left[
        e^{\ri\theta\sum_{\langle ij\rangle} Z_iZ_j}
        \left(\cdot\right)
        e^{-\ri\theta\sum_{\langle ij\rangle} Z_iZ_j}
    \right]
    \,.
\end{align}
While this channel is strongly symmetric under an global on-site $\mathbb{Z}_2$ symmetry generated by $\prod_j X_j$, 
its output nevertheless supports distinct patterns of symmetry-breaking, which we diagnose through fidelity correlators~\cite{lessa2025strong}.
The duality thus connects logical recoverability to a nonlinear diagnostic of mixed-state order.  These setups are summarized in Fig.~\ref{fig:setup}.

\begin{figure}[t]
\centering
\includegraphics[width=\columnwidth]{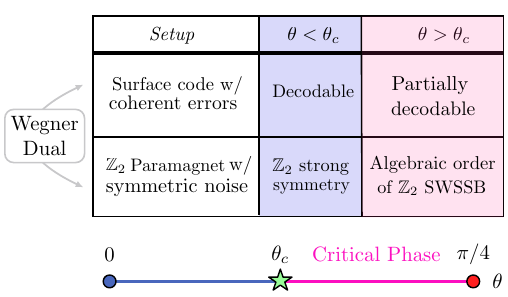}
\caption{Main results. 
In both setups, the coherent rotation tunes a phase transition at $\theta_c$.
The critical phase $\theta > \theta_c$ exhibits partial decodability and algebraic SWSSB, as summarized in the table.
}
\label{fig:results}
\end{figure}

Our principal finding is an extended critical phase at sufficiently large rotation angles, $\theta>\theta_c$, with two dual manifestations: \emph{partial decodability} in the surface code and \emph{algebraic} $\mathbb{Z}_2$ \emph{SWSSB} in the paramagnet, as summarized in Fig.~\ref{fig:results}.
For weak rotations, $\theta<\theta_c$, the surface code exhibits a conventional decodable phase in which optimal Pauli recovery achieves perfect fidelity in the thermodynamic limit.
For $\theta>\theta_c$, the optimal decoding fidelity approaches a universal function of the system's aspect ratio, which is independent of $\theta$ in the thermodynamic limit. 

In the dual paramagnet, the same critical regime exhibits algebraic $\mathbb{Z}_2$ SWSSB, characterized by a fidelity correlator with universal power-law decay throughout the phase.
This behavior distinguishes it both from the strongly symmetric phase, where fidelity correlations are short-ranged, and from a conventional $\mathbb{Z}_{2}$ SWSSB, where the fidelity correlator approaches a nonzero constant at long distances~\cite{lessa2025strong}.

We develop a unified long-distance description of these phenomena in terms of an $\mathrm{SO}(n)$ nonlinear sigma model (NL$\sigma$M) in the replica limit $n\to1$~\cite{yan_bao_coherent_error_nlsm,yang_coherent_hexagonal,fava2023nonlinear,jian2023measurement}.
The extended critical phase is governed by the weak-coupling fixed point of this theory.
Its description is closely related to that of the thermal metal phase of disordered superconductors in symmetry class DIII of the Altland--Zirnbauer classification, although the relevant replica limit here is $n\to1$, as opposed to $n\to0$ in the disordered-fermion setting~\cite{senthil2000quasiparticle,AZ_10fold,zirnbauer1996riemannian}.
Correlations of twists in the sigma model's order-parameter field govern both the decoding fidelity and the fidelity correlator of the dual mixed state.
The same field theory therefore accounts for the universal aspect-ratio dependence of optimal recovery and the algebraic decay of the SWSSB diagnostic.

We also introduce a probe of the non-classical character of the critical phase by testing whether interference between distinct error strings survives at long distances.
Our diagnostic compares the states obtained when widely separated $m$-anyon pairs are created incoherently before or removed after the decoherence channel.
Under Wegner duality, this probes the response of the algebraic SWSSB phase to background $\mathbb{Z}_2$ fluxes.
For stochastic Pauli noise, error strings proliferate incoherently, and their dynamics are unchanged under flux insertion.
In the critical phase, by contrast, coherent contributions from distinct error strings persist at long-distances, leaving the dynamics sensitive to background fluxes even at large separations.
We argue that this sensitivity rules out an effective stochastic Pauli description in the infrared.
In the associated random-bond Ising model, the flux response is encoded in spin correlations, complementary to the disorder-operator correlations that govern the decoding fidelity and the mixed-state fidelity correlator.

\tableofcontents

\section{Overview}
In this section, we provide a detailed overview of the key results in this work.  The subsequent sections are organized as follows.
Sec.~\ref{sec:setup} introduces the setups and quantities of interests studied in this work.
Sec.~\ref{sec:nlsm} develops the effective non-linear sigma model description of decoding and strong-to-weak symmetry breaking, and discusses the observables associated with our quantities of interests in various regimes.
Sec.~\ref{sec:decoding} discusses the sigma model predictions of decoding fidelity corroborated by large-scale numerical simulations.
Sec.~\ref{sec:mixed-state} investigates the mixed-state phases arising when an Ising paramagnet is subject to a strongly-symmetric decoherence channel, which provides a (Wegner) dual description of the decoding problem. 
We investigate the emergence of an algebraic SWSSB phase and explain the coherent nature of this phase that cannot be reproduced by stochastic Pauli channels.
Sec.~\ref{sec:decoding_others} studies the additional probes of the critical phase and presents numerical results that verify our effective theory.
We close with discussion and outlook in Sec.~\ref{sec:discussion}.

\subsection{Partial decodability in the surface code}
We first consider decoding in the triangular-lattice surface code subject to coherent rotations [see Fig.~\ref{fig:setup}(a)].
We investigate the optimal Pauli recovery fidelity based on random syndrome measurement results.
Our key finding is a decoding transition, tuned by the rotation angle $\theta$, as summarized in Fig.~\ref{fig:results}.
At small rotation angle $\theta < \theta_c$, we find a conventional \emph{decodable phase}, in which the decoding fidelity approaches unity in the thermodynamic limit. 
At larger angles ($\theta >\theta_c$), we instead find a critical phase that is \emph{partially decodable}; the fidelity approaches a universal function of the aspect ratio, independent of the angle $\theta$:
\begin{equation}
    \lim_{\substack{L, \, T\rightarrow\infty\\T/L=\kappa}}
    \mathcal{F}(\theta; \, L,\,T)
    =
    \mathcal{F}_{\ast}(\kappa),
    \qquad
    \kappa:=\frac{T}{L}.
    \label{eq:overview_universal_fidelity}
\end{equation}
The universal function $\mathcal{F}_{\ast}(\kappa)$ for the surface code on the cylinder (height $T$, circumference $L$) is given by Eq.~\eqref{eqn:order_one_AR_fid_integral_rep}, and interpolates between $1$ (at large $\kappa$) and $1/2$ (at small $\kappa$).
Thus, increasing the code size at fixed aspect ratio does not generally yield asymptotically perfect recovery.

\begin{table*}[t!]
\setlength{\tabcolsep}{2pt}
\renewcommand{\arraystretch}{1.5}
\begin{tabular*}{\textwidth}{wc{.19\textwidth}|wc{.19\textwidth}|wc{.31\textwidth}|wc{.31\textwidth}}
\hline\hline
 & $\theta < \theta_c$  & $\theta_c < \theta < \pi/4$ & $\theta=\pi/4$ \\ 
 \hline
Decoding Fidelity 
& \multirow{2}{*}{$1-e^{-\mu T}$}
& \multirow{2}{*}{\shortstack[c]{
$\theta$-independent function
$\mathcal{F}_{\ast}(\kappa)$ \\
as $L \to \infty$  [Eq.~\eqref{eqn:order_one_AR_fid_integral_rep}]}}

    & $\mathrm{Zig}$-$\mathrm{Zag}$ $\mathrm{Boundary}$: \; $1$ \\
 $\mathcal{F}$ &  &  

& $\mathrm{Armchair}$ $\mathrm{Boundary}$:\; $1/2$ 
    \\

 \hline
Fidelity Correlator &\multirow{2}{*}{$e^{-r/\xi_F}$}  & \multirow{2}{*}{$\sim {r^{-1 / 16}}$}  & \multirow{2}{*}{Oscillates with $r$}
 \\
$C_F(r)$ &  &  & \\
    
 \hline
Conductivity &\multirow{2}{*}{$e^{-L/\xi_G}$}    & \multirow{2}{*}{$\displaystyle\frac{1}{2\pi}\ln L + \calO(1)$}  & \multirow{2}{*}{$\propto L$}
 \\
$G$ &  &   & \\
\hline\hline
\end{tabular*}
\caption{Summary of our results for the decoding fidelity in the surface code, and the fidelity correlator in the Wegner-dual description of a $\mathbb{Z}_{2}$ paramagnet subject to symmetric noise are shown in the first two rows.  We also investigate the conductivity in a Chalker-Coddington network model associated with this problem, which is in Altland-Zirnbauer symmetry class DIII.  At the critical point $\theta=\theta_{c}$, the conductivity and decoding fidelity attain a universal dependence on aspect ratio, while the fidelity correlator is expected to decay as a power-law at long distances. 
}\label{tab:results_table}
\end{table*}

To arrive at this conclusion, we formulate the figure of merit, recovery fidelity averaged over the Born probability of syndrome measurements, as  observables in an Ising model with complex couplings and quenched bond disorder (similar descriptions arise in related problems, see Ref.~\cite{venn2022coherent,yan_bao_coherent_error_nlsm}).
Analytic study of this quantity is challenging as the recovery fidelity is a non-linear function of the quantum state $\rho$ in Eq.~\eqref{eq:surface_code_setup}.
We therefore invoke a replica trick to formulate the average fidelity in terms of a sequence of linear observables in the replicated description of the complex, random bond Ising model.
The fidelity of the physical decoding problem is recovered in the replica limit $n \to 1$~\cite{yan_bao_coherent_error_nlsm}.

The $n$-replica partition function exhibits a microscopic $\mathrm{SO}(n)\times\mathrm{SO}(n)$ symmetry.
Expanding in long-wavelength fluctuations about a saddle point that spontaneously breaks this symmetry yields a nonlinear sigma model with target space $\mathrm{SO}(n)$,
\begin{equation}
{S}[O]
=
\frac{1}{g_0}
\int \mathrm{d}^{2}r\,
\operatorname{tr}
\left(
    \nabla O^{T}\!\cdot\nabla O
\right),\hspace{.1in}
    O\in\mathrm{SO}(n).
    \label{eq:overview_sigma_model}
\end{equation}
with $1/g_0$ the bare stiffness.

The same nonlinear sigma model arises in the description of a disordered ``thermal metal" phase that arises in symmetry class DIII of the Altland--Zirnbauer classification, but with the distinct replica limit $n\rightarrow0$ appropriate to quenched disorder.
Crucially, the weak-coupling fixed point is stable in both replica limits.
In the $n\rightarrow1$ limit relevant to our optimal Pauli decoding problem, the renormalized stiffness $1/g_R(L)$ grows with scale, driving the theory toward the infrared fixed point at $g=0$.
The partially decodable phase therefore constitutes a stable thermodynamic phase, rather than a finite-size crossover governed by proximity to an unstable fixed point.

We connect this field-theoretic description directly to the recovery
of logical information.  The decoding fidelity relates to twist-defect correlations of the order-parameter field $O$.
An explicit evaluation of these correlations in the $n\to 1$ replica limit yields the universal aspect ratio dependence in Eq.~\eqref{eqn:order_one_AR_fid_integral_rep}.

This behavior contrasts sharply with optimal Pauli decoding in
the square-lattice surface code under the same noise channel.  The bipartite nature of the lattice yields a distinct SO$(2n)$ replica symmetry, yielding a coarse-grained nonlinear sigma model with target space
$\mathrm{SO}(2n)/\mathrm{U}(n)$, whose weak-coupling fixed point
is unstable in the physical replica limit
$n\rightarrow1$ as emphasized in Ref.~\cite{yan_bao_coherent_error_nlsm,yang_coherent_hexagonal}.  
The slow renormalization group (RG) flow nevertheless produces a large crossover scale below which the fidelity can exhibit pronounced behavior associated with the proximate weak-coupling fixed point.
Beyond this scale, the system leaves the weak-coupling regime, and exhibits an infrared flow toward conventional decodable or non-decodable behavior.

We test the $\SO(n)$ sigma-model predictions using large-scale simulations.
The main numerical challenge is to isolate the finite-size regime $L\gtrsim\xi,\lambda$ in which the sigma-model description applies.
The correlation length $\xi$ diverges at the transition between the decodable and partially decodable phases, $\theta=\theta_c$, while the crossover scale $\lambda$ diverges as $\theta\to\pi/4$.
In the disordered-fermion description, the latter limit corresponds to a diverging mean free path leading to a \emph{ballistic} metal.
Similar behavior has been observed in other coherent-noise models~\cite{venn2022coherent,bejan_projected_ensemble,yan_bao_coherent_error_nlsm,yang_coherent_hexagonal}.
To diagnose the crossover governed by $\lambda$, we compare surface codes with ``armchair'' and ``zig-zag'' boundaries, shown in Fig.~\ref{fig:surface-code-boundaries}.
Although the decoding fidelity is insensitive to this choice in the sigma-model regime, we show analytically that it takes the distinct values $1/2$ and $1$, respectively, at $\theta=\pi/4$.
Boundary dependence therefore signals the crossover regime $\lambda\gtrsim L$.
A complementary diagnostic identifies finite-size effects near $\theta_c$: for $L<\xi$, the fidelity decreases with increasing system size, in contrast to its behavior in the sigma-model regime.

\subsection{Algebraic $\mathbb{Z}_{2}$ strong-to-weak symmetry-breaking}
In Sec.~\ref{sec:mixed-state}, we consider the Wegner-dual of this decoding problem, an Ising paramagnet subject to a local quantum channel $\tilde{\calE}_\theta$ that respects the global strong $\mathbb{Z}_2$ symmetry.
In this dual picture, the rotation angle $\theta$ tunes a transition between the strongly-symmetric phase ($\theta<\theta_c$) and a critical phase ($\theta > \theta_c$) with an algebraic $\mathbb{Z}_{2}$ SWSSB.
In this critical phase, the $\mathbb{Z}_{2}$ charge dynamics are coherent at long-distances, so that this phase cannot be approximated by any symmetric Pauli dephasing channel acting on the paramagnet.

To probe the phase transition as $\theta$ is tuned, we consider the fidelity correlator~\cite{lessa2025strong},
\begin{align}\label{eq:fidelity_corr}
    C_F({\bf x, y}) :=  F(\varrho,\, Z_{\bf x}Z_{\bf y}\varrho Z_{\bf x}Z_{\bf y})
    \,,
\end{align}
where the fidelity $F(\rho_{1},\rho_{2}) := \mathrm{tr} \sqrt{\sqrt{\rho_{1}}\rho_{2}\sqrt{\rho_{1}}}$.
This quantity is used as a probe of the SWSSB transition driven by the symmetric dephasing channel acting on a paramagnetic state~\cite{lessa2025strong}.
The fidelity correlator is short-ranged in the symmetric phase, while it is long-ranged (non-decaying) in the SWSSB phase.

Sec.~\ref{sec:mixed_state_fid_corr} demonstrate that, unlike the usual SWSSB phase, the critical phase for $\theta > \theta_c$ exhibit algebraic order 
\begin{align}\label{eq:fid_corr_swssb}
C_F({\bf x,\, y}) \sim \frac{1}{|{\bf x} - {\bf y}|^{1/16}}
\,,
\end{align}
where the power-law exponent is universal within the phase.
Here, we have ignored multiplicative prefactors that are poly-logarithmic functions of the distance, arising from corrections-to-scaling as the sigma model flows to the infrared fixed-point. 

In the surface code decoding problem, this result implies a power-law decay of the fidelity correlator between the corrupted surface code, and the state where $e$-anyons are incoherently inserted at spatially-distant sites, 
\begin{align}\label{eq:fid_corr_surf_code}
    C_F({\bf x,\, y}) :=& F(\rho,\, Z^\calC\rho Z^\calC)
    \,,
\end{align}
where $\bf x$, $\bf y$ are the endpoints of the path $\calC$.
Since the creation of $e$-anyons by $Z^\calC(\cdot)Z^{\calC}$ commute with the channel $\calE_\theta$, this quantity also measures the distinguishability between two states with and without $e$-anyons after passing them through the error channel $\calE_\theta$.
This distinguishability transition is known to coincide with the decoding transition in the surface code with incoherent errors~\cite{fan2024diagnostics}.

We obtain these predictions based on the mapping to an effective NL$\sigma$M (\ref{eq:overview_sigma_model}). 
In the weak-coupling limit, the coarse-grained matrix field can be parametrized as $O({\bf x}) = O_0 e^{\ri \sum_\sfa \theta_\sfa({\bf x}) T^\sfa}$, where $T^\sfa$ form a basis of the $\mathfrak{so}(n)$ Lie algebra and $\theta_\sfa(\bf x)$ are non-interacting compact bosonic fields.
The fidelity correlator maps to correlations of $\mathbb{Z}_2$-orbifold twist fields, which implement the transformation $\theta_\sfa\to-\theta_\sfa$ on a subset of these bosons along a defect line, yielding the result (\ref{eq:fid_corr_swssb}) in the replica limit.
We extend this construction to more general twist operators, whose correlations correspond to natural generalizations of the fidelity correlator.
This yields analytic predictions for the scaling dimensions of a family of operators, which we verify with numerical simulations presented in Sec.~\ref{sec:mixed_state_fid_corr}.

Sec.~\ref{sec:mixed-state_non-classical} investigates the ``non-classical'' character of the algebraic SWSSB phase.
A standard realization of $\mathbb{Z}_2$ SWSSB applies symmetric Pauli dephasing channels,
$\calN_{ZZ,\,ij}(\rho) = (1-p)\rho + p Z_iZ_j \rho Z_iZ_j$,
to the paramagnetic state $\ket{+}^{\otimes N}$.
These channels generate classical stochastic dynamics of the $\mathbb{Z}_2$ charges: histories leading to the same final charge configuration contribute incoherently, with each history paired with itself in the bra and ket of the density matrix.
This ``gluing'' excludes interference between distinct charge histories, allowing both the charge dynamics and the resulting SWSSB transition to be described classically.

We ask whether the algebraic SWSSB phase likewise admits an effective description in terms of symmetric dephasing.
Our central result is that this phase cannot be reproduced by symmetric dephasing acting on the same paramagnetic initial state, even when this channel is not local. Physically, this obstruction is probed by investigating the sensitivity of the output state to the insertion of static $\mathbb{Z}_2$ fluxes in the initial state.
If the dynamics admit an effective dephasing description, the charge trajectories in the ket and the bra are glued at long distances, and the paired trajectories have trivial statistics with the fluxes.
Sensitivity to well-separated flux insertions thus signals a failure of this pairing, revealing interference between distinct charge trajectories at long wavelengths.

Using this intuition, we demonstrate our key result by comparing the output states of the channel $\tilde{\calE}_\theta$ when the initial state is an Ising paramagnet on a 2D disk with two distinct patterns of symmetry-breaking boundary conditions. The fidelity of the corresponding output states $F(\varrho_1,\,\varrho_2)$ effectively probes the $\mathbb{Z}_{2}$ charge dynamics with different flux configurations. In Sec.~\ref{sec:absence_dual_swssb}, we show that this fidelity is governed by boundary correlations of the same $\mathbb{Z}_2$ twist field in the NL$\sigma$M and vanishes as the distance between two flux insertions increases.
The two outputs therefore become asymptotically distinguishable.
This persistent sensitivity to flux insertion rules out an effective dephasing description of the channel generating the algebraic SWSSB phase, in contrast to the Pauli dephasing channels conventionally used to induce $\mathbb{Z}_2$ SWSSB.

This result implies that the decoherence channel $\calE_\theta$ also remains coherent at long-distances in the partially-decodable phase of the surface code.
Specifically, we formulate the fidelity $F(\rho_1 , \, \rho_2)$ between $\rho_1$, which is the coherently-corrupted surface code state, and $\rho_2$, which differs by the presence of a pair of $m$-anyons inserted at $\bfx, \, \bfy$, introduced prior to the application of $\mathcal{E}_\theta$ and removed afterwards.  This fidelity is unity for any Pauli dephasing dynamics, which necessarily commutes with the incoherent insertion of $m$ anyons.
The long-distance behavior of this fidelity in the partially-decodable phase is also dictated by the correlations of twist fields in the sigma model.
The fidelity exhibit a universal power-law scaling in the weak coupling regime, $F(\rho_1, \, \rho_2) \sim {|\bfx - \bfy|^{-1/16}}$ and vanish at infinite separation. 
Thus, the coherent nature of the channel in the partially decodable phase persists in the thermodynamic limit.

\section{Setup}\label{sec:setup}
In this section, we set up the problems studied in this work.
Section~\ref{sec:setup_model} introduces the triangular-lattice surface code under coherent errors and reviews the statistical mechanics model that governs the optimal decoding fidelity.
Section~\ref{sec:setup_fid_corr} concerns the Wegner dual of the decoding problem, i.e. a paramagnetic state with $\bbZ_2$ symmetry subject to strongly-symmetric coherent quantum channel.
We review the fidelity correlator that probes the symmetry breaking in the mixed states in this setup~\cite{lessa2025strong}.

\subsection{Decoding the triangular lattice surface code}\label{sec:setup_model}

\begin{figure}[t]
\centering
\includegraphics{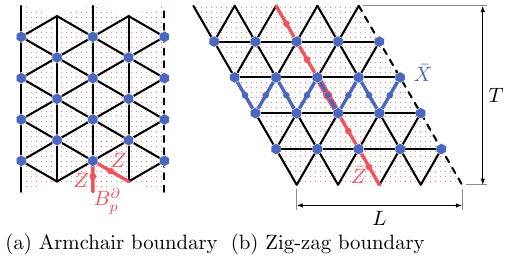}
\caption{Triangular-lattice surface code on a cylinder with (a) armchair and (b) zig-zag rough boundaries in the vertical direction.
Red dotted triangles denote plaquette stabilizers, and blue markers denote vertex stabilizers.
The highlighted incomplete plaquette $B_p^{\partial}$ acts on two edge qubits.
Panel (b) shows the logical operators $\bar X$ and $\bar Z$.
In our convention, the system sizes are $(L, T) = (2\sqrt{3}, 5)$ for (a) and $(L, T)=(4, 5\sqrt{3}/2)$ for (b).}
\label{fig:surface-code-boundaries}
\end{figure}

We consider the triangular-lattice surface code on the cylinder.
The surface code contains two types of stabilizers, $A_{v}$ and $B_{p}$, associated with the vertices and plaquettes, respectively.
In particular, $A_v$ is a six-body Pauli-X operator, while $B_p$ is a three-body Pauli-Z operator, as shown in Fig.~\ref{fig:setup}(a).
We consider the code with rough boundary conditions at the top and the bottom of the cylinder; we include the incomplete plaquette operators $B_p^{\partial}$ as a part of the stabilizer group.
The excitations in the surface code are labeled by the $A_v$ and $B_p$ stabilizers with $-1$ eigenvalues, which are commonly referred to as the $e$ and $m$ anyons, respectively.
The rough boundaries, from which we can create a single $e$-anyon, are known as $e$ condensates.
We note that on the triangular lattice, there are two different microscopic realizations of $e$-condensed (rough) boundary.
The two choices are illustrated in Fig.~\ref{fig:surface-code-boundaries} and referred to as the ``zig-zag'' and ``armchair'' boundaries in the rest of the work.
We consider both boundary conditions; they play a crucial role in numerically verifying the field-theoretic predictions for the decoding fidelity.
On this geometry, the surface code encodes one logical qubit with logical operators illustrated in Fig.~\ref{fig:surface-code-boundaries}(b).

We focus on the surface code subject to single-qubit unitary rotations that generate e-anyons coherently,
\begin{equation}\label{eqn:corrupted_state}
    \ket{\psi} = U_{Z, \, \theta}\,\ket{\psi_0} \,, \hspace{.1in}
    U_{Z,\,\theta} := \prod_{\ell} e^{\ri \theta Z_\ell} \,.
\end{equation}
Such unitary errors may result from a variety of experimental miscalibrations such as imperfect state preparation, magnetic field in the device, or stabilizer measurements in a tilted basis.
We are interested in active error correction by first measuring the stabilizers in the rotated surface code $\ket{\psi}$ and then apply Pauli operators based on the measurement results to recover the encoded state $\ket{\psi_0}$.
The decoding problem with such unitary errors was first studied in Ref.~\cite{bravyi_coherent} and later analyzed in the square-lattice surface code in~\cite{venn2022coherent,venn_entanglement_phases,behrends_general_x,yan_bao_coherent_error_nlsm}.

The potential decoding transition, when tuning $\theta$, is a phase transition in the ensemble of post-measurement wave functions
\begin{align}
    \left\{ \calQ_s, \quad \ket{\psi_s} = \frac{1}{\sqrt{\calQ_s}}\hat{P}_s \ket{\psi} \right\}.
\end{align}
The transition can be equivalently formulated as a phase transition in the density matrix~\cite{fan2024diagnostics},
\begin{align}
    \rho = \calE_\theta\left[ \ketbra{\psi_0} \right] = \sum_s \calQ_s \ketbra{\psi_s}
\end{align}
where the decoherence channel $\calE_\theta := \calD_{\text{syn}}[U_{Z,\theta}(\cdot)U_{Z,\theta}^\dagger]$ with $\calD_{\text{syn}}$ being the complete dephasing in the syndrome basis.
We uncover the universal features of this transition from this mixed-state perspective in Sec.~\ref{sec:mixed-state}.

\subsubsection{Pauli decoding fidelity}\label{sec:setup_decoding_fidelity}

We study the practically motivated \emph{optimal Pauli decoder} in this work.
We restrict the recovery operator to be a product of single-qubit Pauli operators, which is a transversal unitary and is easy to implement experimentally.
In this case, the entire decoding scheme, syndrome measurement and feedback, consists of only Clifford operations, which are easy to benchmark.

The optimal decoding based on Pauli recovery evaluates the coset probability $\calQ_{\alpha,\,s}$, i.e., the probability of the error string belonging to a homology class $\alpha$, given the observed syndromes $s$.
We apply the recovery operator belonging to the homology class $\alpha$ with a probability $\calQ_{\alpha|s} = \calQ_{\alpha,\,s}/\calQ_s$.
We note that the recovery operators belonging to the same homology class are equivalent as they recover the same state in the code space.
This decoding scheme, which we call \emph{Bayesian decoder}, has a decoding fidelity
\begin{align}\label{eqn:defintion_of_bayesian_fidelity}
    \calF = \sum_s \calQ_s \sum_\alpha \calQ_{\alpha|s}^2.
\end{align}
The Bayesian decoder is slightly different from the other optimal Pauli decoder, known as the maximum-likelihood decoder, which chooses the recovery operator from the most likely homology class.
However, they exhibit the same threshold~\cite{yan_bao_coherent_error_nlsm}.
We focus on the Bayesian decoder as its fidelity is easier to analyze.

We note that in the optimal decoding scheme, we assume the decoder has complete knowledge of the underlying error model.
Hence, the probability $\calQ_{\alpha|s}$ of the homology class estimated by the decoder is the same as that in the underlying quantum device.
In practice, one can also consider a suboptimal decoding scheme, in which the decoder only has an estimate of the rotation angle $\theta$, and apply the recovery string with probability $\calP_{\alpha|s}$ based on its imperfect knowledge.
Such a suboptimal decoding scheme is analyzed for the square lattice surface code in Ref.~\cite{yan_bao_coherent_error_nlsm}.
However, its distinction from the optimal decoder is less dramatic on the triangular lattice.
For simplicity, we mainly focus on the optimal decoder in this work and comment on the suboptimal decoder only in Sec.~\ref{sec:suboptimal}.

The fidelity of the Bayesian decoder can be analyzed based on a statistical mechanics mapping.
To begin, we can always express the rotated surface code after syndrome measurements as
\begin{align}
\ket{\psi_s} = \frac{1}{\sqrt{\calQ_s}}\left(\calZ_{0,\,s}\ket{\psi_{0,\,s}} + \calZ_{1,\,s}\ket{\psi_{1,\,s}}\right),
\end{align}
where we choose $\ket{\psi_{\alpha,\,s}}$ with $\alpha = 0,1$ to be the basis states labeling the two-dimensional Hilbert space in the surface code with fixed syndrome configuration $s$.
The basis state $\ket{\psi_{\alpha,\,s}} = Z^\calC\ket{\psi_0}$ is created from the encoded state $\psi_0$ by a string operator $Z^\calC$ with $\partial \calC = s$ and $[\calC] = \alpha$.
The coset probability $\calQ_{\alpha,\,s} = \abs{\calZ_{\alpha,\,s}}^2$ is then determined by the amplitudes of these two basis states, which takes the form
\begin{align}
    \calZ_{\alpha,\,s} = \sum_{\calC;\, \partial\calC = s, \,[\calC] = \alpha} (\cos\theta)^{N - \abs{\calC}} (\ri \sin\theta)^{\abs{\calC}}.
\end{align}
The amplitude can be formulated as the partition function of the random-bond Ising model on the honeycomb lattice with complex couplings~\cite{venn2022coherent},
\begin{align}
    \calZ_{\alpha,\,s} = \sum_\sigma e^{\sum_{\langle \bfr,\, \bfr'\rangle}\left(J - \frac{\ri\pi}{4} \right) \eta_{\bfr\bfr'} \sigma_\bfr \sigma_{\bfr'}} \label{eqn:rbim_single_replica_partition_function}
\end{align}
where $J = (1/2)\log(1/\tan\theta)$, $\sigma_\bfr$ is an Ising spin on the vertex of the honeycomb lattice (plaquette center of the original triangular lattice), and $\eta_{\bfr\bfr'} = \pm 1$ is a random bond variable that takes $-1$ value along a path $\calC$ with $\partial \calC = s$ and $[\calC] = \alpha$.
The choice of path $\calC$ is not unique; different choices are related by gauge transformation and give the same partition function.

The statistical mechanics mapping we obtain is distinct from that in the surface code with dephasing errors~\cite{dennis_surface_code}.
The coset probability in the case of dephasing errors is given by the RBIM partition function, rather than two copies of it.
Moreover, the RBIM there contains only real couplings and can be simulated using classical Monte-Carlo algorithms~\cite{tobias2026optimal}.
The ferromagnet transition in the RBIM manifests as the decoding transition in the surface code.

Nevertheless, the decodable phase and phase transition in the surface code with unitary rotations bear similarities with that in the case of dephasing errors.
The fidelity is governed by the partition functions $\calZ_{\alpha,\,s}$ with $\alpha = 0,1$. 
The two partition functions are related by flipping the random bond along a path spanning between two boundaries of the cylinder.
We expect in the decodable phase, the absolute value of one partition function is exponentially larger than the other, leading to an almost perfect decoding fidelity up to an exponentially small correction.
Tuning $\theta$ can lead to a potential decoding transition with a suppressed decoding fidelity.
However, performing the optimal decoding here is complicated due to the complex coupling in the RBIM.
A common strategy is to evaluate $\calZ_{\alpha,\,s}$ using transfer matrix contraction or similar $1+1$D simulation algorithms~\cite{bravyi_coherent,venn2022coherent,yan_bao_coherent_error_nlsm}.

\subsubsection{Replica formulation}
To gain analytic understanding of the decoding problem, we regard it as a problem in the disordered system.
In particular, we treat the syndrome $s$ as the disorder realization in the stat-mech model with its probability $\calQ_s$ given by the Born rule.
Here, for each realization $s$, the quantity of interest $\sum_\alpha \calQ_{\alpha|s}^2$ is a nonlinear function of the underlying post-measurement state $\ket{\psi_s}$.
Hence, an exact analysis of the disorder-averaged quantity $\calF$ becomes challenging.

We here formulate the fidelity $\calF$ as the $n \to 0$ limit of a replica sequence~\cite{yan_bao_coherent_error_nlsm},
\begin{equation}\label{eqn:replica_sequence}
    \mathcal{F} = \lim_{n \to 0} \mathcal{F}^{(n)} \,, \quad \calF^{(n)} = \frac{\sum_s \calQ_s^{n-1} \sum_\alpha \calQ_{\alpha,\,s}^2 }{\sum_{s} \calQ_{s}^{n+1}}.
\end{equation}
We identify the denominator with the partition function of replicated RBIM,
\begin{equation}\label{eqn:replicated_Z_RBIM}
    \bfZ_0 = \sum_{s, \, \alpha} 
    \mathcal{Q}_{\alpha, \, s}^{n+1}  = \sum_{s,\,\alpha} (\calZ_{\alpha,\,s}\calZ_{\alpha,\,s}^*)^{n+1}
\end{equation}
which formally describes $n+1$ copies of the RBIM (\ref{eqn:rbim_single_replica_partition_function}) along with $n+1$ copies of its complex conjugate, with fixed syndromes and homology classes across all replicas. 
In this way, we can formulate the replicated fidelity $\calF^{(n)}$ in terms of observables in the replicated partition function
\begin{equation}\label{eqn:rbim_replicated_fid_as_defects}
    \mathcal{F}^{(n)} = \frac{2\sum_{k=0}^{n-1}\binom{n-1}{k}\Phi_{k}}{\sum_{k=0}^{n+1}\binom{n+1}{k}\Phi_k}.
\end{equation}
where $\Phi_k := {\bfZ_k}/{\bfZ_0}$ and
\begin{equation}    
    \bfZ_{k} = \sum_{s, \, \alpha} 
    \mathcal{Q}_{\alpha, \, s}^{n+1-k} \mathcal{Q}_{1-\alpha, \, s}^{k}.
\end{equation}
Physically, $\Phi_k$ is the ratio between partition functions with and without defect insertions in $k$ copies of $\calQ_{\alpha,\,s}$, which can be interpreted as the correlation function of disorder operators on two boundaries.
The defect spans between two boundaries of the cylinder and flips the Ising couplings in $k$ copies of $\calZ_{\alpha,\,s}$ and $\calZ^*_{\alpha,\,s}$.

\subsection{Duality and strong-to-weak symmetry breaking}\label{sec:setup_fid_corr}

The second setup we study in this work is the paramagnet subject to symmetric coherent channel.
Specifically, we consider a 2D array of qubits on the vertices of triangular lattice [as shown in Fig.~\ref{fig:setup}(b)] and subject to two-qubit coherent rotations,
\begin{align}
    |\tilde{\psi}\rangle = U_{ZZ,\,\theta}\ket{+}^{\otimes N}, \quad U_{ZZ,\,\theta} = \prod_{\langle i,\,j\rangle}e^{\ri \theta Z_i Z_j}.
\end{align}
We consider the ensemble of wave functions obtained by sampling in the Pauli-X basis, $\{ \ket{m}, \calQ_m \}$, where $\calQ_m = |\langle m | \tilde{\psi}\rangle|^2$, and $m$ is a binary string labeling the charge configuration.
This ensemble can be equivalently formulated as a diagonal density matrix,
\begin{align}
    \varrho := \tilde{\mathcal{E}}_{\theta}
    \left[
    \ketbra{+}^{\otimes N}
    \right] = \sum_m \calQ_m \ketbra{m},\label{eq:mixed-state_dual_paramagnet}
\end{align}
where $\tilde{\calE}_\theta(\cdot) = \calD_X[U_{ZZ,\,\theta}(\cdot)U_{ZZ,\,\theta}^\dagger]$ with $D_X$ being complete dephasing in the Pauli-X basis.
The density matrix exhibits a strong $\bbZ_2$ symmetry generated by $\prod_i X_i$.

This setup is the Wegner dual of the surface code with coherent error in Sec.~\ref{sec:setup_decoding_fidelity}~\cite{wegner1971duality}.
The duality maps the surface code with no magnetic anyons to the paramagnet with global $\bbZ_2$ symmetry; it is achieved by the following identifications:
\begin{align}
    Z_\ell \mapsto Z_i Z_j, \quad A_v \mapsto X_i.
\end{align}

The potential phase transition in the strongly-symmetric density matrix has been discussed in the case of paramagnet subject to symmetric dephasing channel, $\calN_{ZZ,\,ij}(\cdot) = (1-p)(\cdot) + p Z_i Z_j (\cdot) Z_i Z_j$.
It has been shown that the channel, which delocalizes the $\bbZ_2$ symmetry charges, can induce a symmetry-breaking transition that spontaneously breaks the strong symmetry down to the weak~\cite{lessa2025strong}.
The channel $\tilde{\calE}_\theta$ studied in this work also delocalizes the charges, however, in a coherent manner, which allows the charges moving along different paths to coherently interfere.
We are interested in the potential strong-to-weak spontaneously symmetry breaking (SWSSB) transition induced by this channel.

\subsubsection{Fidelity correlator}
A standard probe of the SWSSB transition in the mixed-state is the fidelity correlator~\eqref{eq:fidelity_corr},
\begin{align}
    C_F({\bf x, \, y}) :=& F(\varrho, \,  Z_{\bf x} Z_{\bf y} \varrho  Z_{\bf x} Z_{\bf y}) \nn\\
    =&
    \tr
    \sqrt{
    \sqrt{\varrho}
    Z_{\bf x}Z_{\bf y}\varrho Z_{\bf x}Z_{\bf y}
    \sqrt{\varrho}
    }
    \,.
\end{align}
The correlator measures the distinguishability of the density matrices with and without charge insertions at two distant locations $\bf x$ and $\bf y$.
In the symmetric phase, the two states are distinguishable, leading to an fidelity correlator that decays exponentially with $|\bf x - \bf y|$.
In the conventional setup studied in Ref.~\cite{lessa2025strong}, the SWSSB phase features a nondecaying fidelity correlator $C_F = \calO(1)$ as $|\bf x - \bf y| \to \infty$.

In our setup, the density matrix $\varrho$ is diagonal, the fidelity correlator takes the form
\begin{align}
    C_F = \sum_m \sqrt{\calQ_m \calQ_{m+\mathbf{e}_{\bfx}+\mathbf{e}_{\bfy}}} = \sum_m |\calZ_m \calZ_{m+\mathbf{e}_{\bfx}+\mathbf{e}_{\bfy}}| \,,\label{eq:fid_corr_diagonal}
\end{align}
where $m+\mathbf{e}_{\bfx}+\mathbf{e}_{\bfy}$ flips the binary string $m$ at location $\bfx$ and $\bfy$, and $\calQ_m = \abs{\calZ_m}^2$.
Here, $\calZ_m$ is the partition function of random bond Ising model associated with the outcome $m$,
\begin{align}
    \calZ_m = \sum_{\sigma}
    e^{\left(J - \frac{\ri\pi}{4} \right)\left(\sum_{\langle \bfr,\, \bfr'\rangle} \eta_{\bfr\bfr'} \sigma_\bfr \sigma_{\bfr'} + \sum_{\bfr \in \partial D} \eta_\bfr \sigma_{\bfr}\right)}
    \,,
\end{align}
where the measurement outcomes $m$ labels the endpoints of the random bonds $\eta = -1$.
Specifically, at site $\bfr$, $(-1)^{m_{\bfr}} = \prod_{\langle \bfr,\, \bfr'\rangle} \eta_{\bfr\bfr'}$.
The two partition functions $\calZ_m$ and $\calZ_{m+\mathbf{e}_{\bfx}+\mathbf{e}_{\bfy}}$ are related by flipping the random bond $\eta$ along a path that connect $\bfx$ and $\bfy$.

The partition function of RBIM here only differs slightly from that in Eq.~\eqref{eqn:rbim_single_replica_partition_function}.
Since we consider the paramagnet with open boundary condition, the corresponding RBIM has a symmetry breaking boundary.
Moreover, different paths of error chains that produce the syndrome $m$ are always homologically equivalent.
We therefore neglect the subscript $\alpha$ that labels the homology class.

We also consider other probes of the SWSSB phase in this work to further probe the universal data in the critical phase.
In particular, we consider the $k$-th Rényi fidelity correlator
\begin{align}
C_{F,\, k}(\bfx,\, \bfy) := \sum_m \mathcal{Q}_m^{1-k} \calQ_{m+\mathbf{e}_{\bfx}+\mathbf{e}_{\bfy}}^k\,,
\end{align}
which is given by the Chernoff coefficient between two probability distributions $\calQ_m$ and $\calQ_{m+\mathbf{e}_{\bfx}+\mathbf{e}_{\bfy}}$~\cite{chernoff1952measure,fuchs1996distinguishability}.
The fidelity correlator $C_F$ is recovered in the limit $k \to 1/2$.
Similar to $C_F$, this quantity decays exponentially in the symmetric phase and is non-decaying in the SWSSB phase.
However, it exhibits a $k$-dependent scaling in the critical phase we identified in this work, which is discussed in Sec.~\ref{sec:mixed-state}.

\subsubsection{Replica formulation}
To develop analytic understanding of the fidelity corelator, we formulate it as the $2n \to 1$ limit of the replica sequence
\begin{gather}\label{eqn:fidelity_corr_replica_sequence}
    C_F^{(2n)}(\bfx, \, \bfy) 
    = \frac{\sum_m \left(\mathcal{Q}_m\mathcal{Q}_{m+\mathbf{e}_\bfx+\mathbf{e}_\bfy}\right)^n}{\sum_m \mathcal{Q}_m^{2n}}
    \,.
\end{gather}
The replicated fidelity correlator corresponds to the correlator function of disordered operator in the replicated partition function,
\begin{align}\label{eqn:fidelity_correlator_as_defects}
    C_F^{(2n)} = \langle \phi_{n}({\bfx}) \phi_{n}({\bfy})\rangle_{\bfZ_0^{(2n)}} \,.
\end{align}
Here, the disorder operator $\phi_n(\bfx)$ insert a symmetry defect in $n$ copies of $\calZ_m$ and $\calZ_{m}^*$ that terminates at $\bfx$.
We use $\bfZ_0^{(2n)}$ to denote the replicated partition function that involves $2n$ copies of $\calQ_m$.

We formulate a similar replica sequence for the $k$-th R\'enyi fidelity correlator 
\begin{equation}\label{eqn:k_EA_replica_sequence}
    C_{F,\, k}^{(n)}(\bfx,\, \bfy)
    :=
    \frac{
        \sum_m\mathcal{Q}_m^{1-k+n} \mathcal{Q}_{m+\bfe_{\bfx}+\bfe_{\bfy}}^k 
    }   
    {
    \sum_m \mathcal{Q}_m^{1+n}
    }
    \,,
\end{equation}
which recovers $C_{F,k}$ in the replica limit $n \to 0$.
This quantity also maps to the correlation function of symmetry defects in the replicated statistical mechanics model $\bfZ_0^{(n+1)}$,
\begin{equation}
    C_{F,\, k}^{(n)}(\bfx, \, \bfy) = \langle \phi_k(\bfx)\phi_k(\bfy)\rangle_{\bfZ_0^{(n+1)}}
    \,.
\end{equation}

\section{Effective SO$(n)$ non-linear sigma model}\label{sec:nlsm}

In this section, we present the non-linear sigma model with target space $\SO(n+1)$ as an effective theory that governs the decoding fidelity in Eq.~\eqref{eqn:replica_sequence} and the fidelity correlator in Eq.~\eqref{eqn:fidelity_corr_replica_sequence}.
The key result is
\begin{gather}
    \bfZ_0 = \sum_{s,\,\alpha} \calQ_{\alpha,\,s}^{n+1} = \int D O\; e^{-S[O]} \,,\label{eq:sigma_model_key_result} \\
    S[O] = \frac{1}{g_0}\int \rd^2 x \tr\left(\nabla O^T \cdot \nabla O\right) \,, \label{eqn:nlsm_action}
\end{gather}
where $O \in \SO(n+1)$, and $g_0$ is the bare coupling which depends on the rotation angle $\theta$.
In this effective sigma model description, our quantities of interest, the decoding fidelity and the fidelity correlator, are governed by the twist field correlations in the replica limit $n \to 0$. 

Before proceeding, we overview the derivation of the sigma model.
We start by identifying the symmetry of the replicated partition function $\bfZ_0 = \sum_{s,\,\alpha} (\calZ_{\alpha,\,s} \calZ_{\alpha,\,s}^*)^{n+1}$, which consists of $n+1$ copies of the partition sum $\calZ_{\alpha,\,s}$ and $\calZ^*_{\alpha,\,s}$.
To this end, we express each partition sum $\calZ_{\alpha,\,s}$ in 2D as a transition amplitude in 1+1D of the transfer matrices.
The transfer matrices are generated by Majorana bilinears after Jordan-Wigner transformation, and describe free fermion evolution.
In $\bfZ_0$, the transfer matrix is replicated $n + 1$ times, leading to a continuous $\O(n+1)$ rotational symmetry that acts as $\gamma^\sfa \mapsto \sum_{\sfb = 1}^{n+1} O^{\sfa\sfb} \gamma^\sfb$, where $\gamma^\sfa$ is a Majorana mode in the $\sfa$-th replica.
These Majorana modes are associated with $\calZ_{\alpha,\,s}$ characterizing the error strings acting on the density matrix $\rho$ from the left, i.e., in the ket Hilbert space.
Separately, we have a separate $\O(n+1)$ associated with rotations among Majorana modes in the bra Hilbert space associated with $n+1$ copies of $\calZ^*_{\alpha,\,s}$ together with a $\bbZ_2$ Hermiticity symmetry that exchanges the Majorana modes in the ket and bra Hilbert spaces.
In total, the replicated partition sum has a $(\O(n+1)\times \O(n+1)) \rtimes \bbZ_2$ global symmetry.

To derive the effective \NLSM, we follow Ref.~\cite{yan_bao_coherent_error_nlsm}.
In particular, we express $\bfZ_0$ as a fermionic coherent state path integral.
Although $\calZ_{\alpha,\, s}$ admits a free fermion representation, averaging over disorder realizations specified by $s$ and $\alpha$ generates inter-replica interactions in the path integral.
Thus, we introduce a Hubbard-Stratonovich matrix field to decouple the interaction and obtain the effective action after integrating out the fermions.
The \NLSM in Eq.~\eqref{eq:sigma_model_key_result} emerges as the effective description of the fluctuations around the saddle point of the matrix field.
The $\O(n+1) \times \O(n+1)$ symmetry acts on the sigma model matrix field $O$ as $O \mapsto V_L O V_R^T$ with $V_{L, \, R} \in \O(n+1)$ and leaves the action invariant.
We omit a detailed derivation of the \NLSM in this work.

The effective \NLSM enables perturbative analysis around the weak-coupling fixed point at $g_0 = 0$.
The coupling $g_0$ is set by $\theta - \pi/4$.
The perturbative analysis around $g_0 = 0$ allows us to investigate the stability of a putative non-decodable phase; this is distinct from the standard approach around the clean limit without error used to establish the finite decoding threshold of the surface code~\cite{dennis_surface_code}.

We note that the \NLSM is a valid description when $\theta$ is close to, but strictly less than $\pi/4$, and when the system size is above a characteristic length scale $\lambda$ set by the inter-replica interaction, called the \textit{mean-free path}.
The mean-free path diverges as $\theta \to \pi/4$.
At $\theta = \pi/4$, the inter-replica coupling vanishes and the \NLSM ceases to be a valid description at any scale~\cite{yan_bao_coherent_error_nlsm}.

For the readers familiar with the study of disordered fermion systems in 2D, the $\SO(n)$ sigma model also emerges as an effective description for the Anderson localization transition in Class DIII~\cite{evers2008anderson}.
We can map the decoding problem to the problem in disordered fermions by writing the coset probability $\calQ_{\alpha,\,s}$ (or similarly the Born probability $\calQ_m$) as the transition amplitude in the associated Chalker-Coddington network model description~\cite{chalker1988percolation,cho_fisher,merz_chalker}.
We detail this mapping in Appendix~\ref{app:conductance_explicit_forms} and demonstrate that the network model exhibits a time-reversal symmetry in addition to the particle-hole symmetry and therefore belongs to Class DIII.
The characteristic length scale $\lambda$ is the mean-free path in the network model, set by the scattering rate.
At $\theta = \pi/4$, the network model consists only of swap gates, describing a ballistic metal with $\lambda \to \infty$.
Non-linear sigma models are established as long-distance effective descriptions for the network model above the scale $\lambda$ for various symmetry classes~\cite{zirnbauer1996riemannian,ludwig2016topological,ryu2010topological,gruzberg2001random,jian2022criticality}.

One important distinction is that optimal decoding is associated with the $n \to 1$ limit of the $\SO(n)$ sigma model, while the Anderson localization problem is related to the $n \to 0$ limit~\cite{yan_bao_coherent_error_nlsm,yang_coherent_hexagonal}.
In principle, one can also consider the suboptimal decoding in the surface code, in which the decoder infers the homology class based on an inaccurate estimation of the rotation angle $\theta$.
As pointed out in Ref.~\cite{yan_bao_coherent_error_nlsm}, decoding is then governed by the sigma model in the $n \to 0$ limit, as in the study of Anderson localization.
On the other hand, the same $\SO(n)$ sigma model in the limit $n \to 1$ also emerges as an effective description of monitored free-fermion systems in 1+1D~\cite{fava2023nonlinear,jian2022criticality,jian2023measurement}.
There, the $n \to 0$ limit becomes relevant when considering measurements with postselection~\cite{jian2023measurement}.

We further note that the surface code on the triangular lattice studied in this work acquires a distinct sigma model description from that on the square lattice.
This is pointed out in Refs.~\cite{yan_bao_coherent_error_nlsm,yang_coherent_hexagonal}.
The distinction originates from the symmetry of the replicated partition function.
When the syndromes live on a bipartite lattice (e.g. the square lattice), the RBIM $\calZ_{\alpha, \, s}$ is defined on a lattice with even coordination number.
In this case, the replicated partition function exhibits an enlarged $\O(2n+2)$ symmetry rotating between the Majorana modes in the ket and the bra Hilbert spaces~\cite{yan_bao_coherent_error_nlsm}.
This results in the $\SO(2n+2)/\U(n+1)$ sigma model.

In what follows, we discuss the phase diagram of the $\SO(n)$ sigma model in Sec.~\ref{sec:nlsm_phases}.
We predict the existence of a critical phase near $\theta = \pi/4$ and a phase transition when tuning the coherent rotation angle $\theta$.
In Sec.~\ref{sec:nlsm_twists}, we identify the symmetry defects in the replicated statistical mechanical model with twist defects in the sigma model. 
We then calculate the twist correlations associated with the physical observables of interest.
Namely, these results are used to derive the scaling of decoding fidelity and fidelity correlator in the critical phase in Sec.~\ref{sec:decoding} and~\ref{sec:mixed_state_fid_corr}, respectively.

\subsection{Phases of the \texorpdfstring{NL$\sigma$M}{NLsM}}\label{sec:nlsm_phases}
Here, we discuss the possible phases described by the $\SO(n)$ \NLSM, with a gapless phase near the weak-coupling fixed point and two possible gapped phases at strong-coupling in the limit $n \to 1$.
This indicates a decoding transition (or a SWSSB transition) in the setup discussed in Sec.~\ref{sec:setup_decoding_fidelity} (Sec.~\ref{sec:setup_fid_corr}), when tuning the rotation angle from $\theta = 0$ in a gapped phase to $\theta \to \pi/4$ in the gapless phase.

The $\SO(n)$ sigma model~\eqref{eqn:nlsm_action} is parameterized by a single coupling constant $g_0$, where $g_0$ can be regarded as the effective ``curvature'' of the target space manifold.
The renormalization group (RG) flow of the coupling near the weak-coupling fixed point $g_0 = 0$ has been computed perturbatively~\cite{hikami_three_loop_beta,wegner_four_loop_beta},
\begin{equation}\label{eqn:RG_equation}
    \frac{\rd g_R}{\rd \ln L} = (n-2) g_R^2 + \mathcal{O}\left(g_R^3\right)
    \,.
\end{equation}
In the replica limit $n \to 1$ describing optimal decoding, the coupling is marginally irrelevant, and the fixed point at $g_R = 0$ is stable.
Thus, in our setups, when $\theta$ is close to $\pi/4$, the system belongs to a stable gapless phase, which we call the \emph{critical phase}.
Increasing the bare coupling in the sigma model (i.e., reducing the rotation angle $\theta$) can potentially induce a phase transition out of the critical phase at a finite coupling strength $g_c$.

This transition is driven by vortex configurations, which are allowed due to the non-trivial fundamental group $\pi_1(\SO(n)) \cong \bbZ_2$ for $n > 2$.
This leads to a $\bbZ_2$ classification of the resulting gapped phases of the $\SO(n)$ \NLSM. 
The two gapped phases are distinguished by the sign of the vortex fugacity~\cite{fava2023nonlinear,jian2023measurement}, with one of them characterizing the decodable phase near $\theta = 0$.
The existence of two gapped phases raises the following question regarding our setups in Sec.~\ref{sec:setup}: if increasing $\theta$ drives a phase transition into a distinct gapped phase before entering the critical phase at large $\theta$.
Our numerics in Sec.~\ref{sec:decoding} suggest that this is not the case.
This is in line with previous studies of related models governed by the same effective \NLSM ~\cite{fulga_class_diii_helical_superconductor,fava2023nonlinear}, in addition to the present model~\cite{yang_coherent_hexagonal}.
Instead, increasing $\theta$ induces a direct transition into the critical phase.

Finally, we note that the weak-coupling fixed point is also stable in the $n \to 0$ limit of the $\SO(n)$ NL$\sigma$M, giving rise to a critical phase.
The replica limit $n \to 0$ is associated with suboptimal decoding schemes in which the decoder has an imperfect characterization of the underlying noise model~\cite{yan_bao_coherent_error_nlsm}.
Thus, the phase diagram for the suboptimal decoder is similar, while the universal features of the phases and phase transitions may differ.
We comment on suboptimal decoding in Sec.~\ref{sec:suboptimal}.

\subsection{Twist fields in the \texorpdfstring{NL$\sigma$M}{NLsM}}\label{sec:nlsm_twists}

\begin{figure}[t]
\centering
\begin{tikzpicture}[baseline={(lab.base)}, scale=0.67]
  \def\h{3.5}    %
  \def\r{1.85}    %
  \draw (-\r,0) rectangle (\r,\h);
  \draw[red,decorate, decoration={snake, amplitude=0.5mm, segment length=2mm}]
        (-0.7,\h/2) -- (0.7,\h/2);
  \fill[red] (-0.7,\h/2) circle (0.07);
  \fill[red] (0.7,\h/2) circle (0.07);
  \node[red, right] at (-1.35,\h/2+.1) {$\phi$};   %
  \draw[-Stealth,red] (0,\h/2+0.5) node[above]{$O$} -- (0,\h/2-0.5) node[below]{$\Lambda O \Lambda$};
  \draw[-Stealth] (-2.4,0) -- (-2.4,2) node[left] {$T$};
  \draw[-Stealth] (-1.8,-0.6) -- (1.1,-0.6) node[right] {$L$};
  \node (lab) at (-3,3.5) {(a)};
\end{tikzpicture}
\quad
\begin{tikzpicture}[baseline={(lab.base)}, scale=0.67]
  \def\h{3}    %
  \def\r{1.5}  %
  \draw (0, \h) ellipse (1.5 and 0.5);
  \draw (0,0) [partial ellipse=180:360:1.5 and 0.5];
  \draw[dashed] (0,0) [partial ellipse=0:180:1.5 and 0.5];
  \draw (-\r,0) -- (-\r,\h);
  \draw (\r,0) -- (\r,\h);
  \draw[red,decorate, decoration={snake, amplitude=0.5mm, segment length=2mm}] (-0.474,-0.474) -- (-0.474,-0.474+\h);
  \draw[red,-Stealth] (0,\h/2) [partial ellipse=240:270:1.5 and 0.5];
  \node[red] at (-1.1,-0.474+\h/2+0.05) {$O$};
  \node[red] at (0.6,-0.474+\h/2-0.05) {$\Lambda O \Lambda$};
  \node[red] at (-0.95,2.15) {$\Phi$};   %
  \node (lab) at (-3,3.5) {(b)};
  \draw[-Stealth] (-1.8,0) -- (-1.8,2) node[left] {$T$};
  \draw[-Stealth] (0,0) [partial ellipse=210:330:1.9 and 0.7] node[right] {$L$};
\end{tikzpicture}
\caption{
The $\mathbb{Z}_2$-orbifold twist field of the NL$\sigma$M.
Across the defect indicated in red, the field satisfies the twisted boundary condition $O \mapsto \Lambda O \Lambda$.
(a) Twist $\phi$ associated with a bulk defect.
This twist is used to compute the fidelity correlator.
(b) A large twist defect $\Phi$ inserted along the longitudinal direction of the cylinder.
This twist is related to the fidelity of the surface code decoder.
}
\label{fig:twist_intro}
\end{figure}
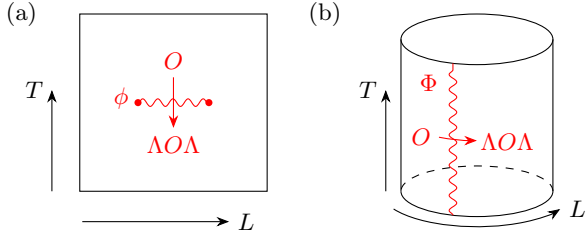

We turn to discuss correlations of the \NLSM twist fields that appear in 
the replicated fidelity correlator $C_F^{(n)}$ in Eq.~\eqref{eqn:fidelity_corr_replica_sequence}
and decoding fidelity $\calF^{(n)}$ in Eq.~\eqref{eqn:replica_sequence}.
These are associated with bulk twist correlations
and
twists along a large defect on the cylinder, respectively, illustrated in Fig.~\ref{fig:twist_intro}.
After outlining the general strategy, we present our results in the weak-coupling regime.
We use these results in Sec.~\ref{sec:decoding} and Sec.~\ref{sec:mixed-state} to obtain the scaling of $\mathcal{F}$ and $C_F$ after taking the replica limit.

The replicated decoding fidelity $\calF^{(n)}$ and fidelity correlator $C_F^{(n)}$ are expressed in terms of symmetry defects in the replicated RBIM as in Sec.~\ref{sec:setup}.
In the effective \NLSM, these symmetry defects are identified with defects across which the order parameter field satisfies a modified boundary condition.
The twist field is the operator which sources the defect at its endpoints.
This was discussed in Ref.~\cite{yan_bao_coherent_error_nlsm} for a related model governed by a similar \NLSM.

For the $\mathrm{SO}(n+1)$ \NLSM considered in this paper, the twist field is associated with the modified boundary condition $O \mapsto \Lambda_L O \Lambda_R$.
Here, $O \in \mathrm{SO}(n+1)$ and $\Lambda_{L, \, R} \in \mathrm{SO}(n+1)$ are diagonal matrices with entries $\pm 1$.
A $-1$ entry in the $\sfa$-th row of $\Lambda_L$ ($\Lambda_R$) corresponds to a symmetry defect insertion in the $\sfa$-th bra (ket) replica copy of the RBIM.
Next, note that the fidelity correlator [as in Eq.~\eqref{eqn:fidelity_corr_replica_sequence}] and decoding fidelity [as in Eq.~\eqref{eqn:replica_sequence}] are related to symmetry defect insertions in both $k$ bra and $k$ ket replica copies.
Thus, we may assume that $\Lambda_L = \Lambda_R := \mathrm{diag}(-\mathds{1}_k, \, \mathds{1}_{n+1-k})$.\footnote{
This is a consequence of the $S_{n+1} \times S_{n+1}$ permutation symmetry of the replicated RBIM and its microscopic boundary conditions.
We expect this symmetry to be preserved in the effective \NLSM, even though the full $\mathrm{SO}(n+1) \times \mathrm{SO}(n+1)$ symmetry is generically broken at the boundaries.}  We drop the subscript on $\Lambda$.

In the following, we obtain twist field correlations in the weak-coupling limit $g_R \to 0$.
This limit admits a useful parameterization of the $\mathrm{SO}(n+1)$ matrix field as $n(n+1)/2$ non-interacting compact bosons which transform simply under the twist.
As a review, note that the \NLSM field can be expanded around a translationally invariant reference $O_0 = \Lambda O_0 \Lambda$ which is also invariant under the twist
\begin{equation}\label{eqn:weak_coupling_expansion}
    O(x,\, t) = O_0 \exp \left( \ri \sum_{\sfa < \sfb=1}^{n+1} \theta_{\sfa \sfb}(x,\, t) \, T^{\sfa \sfb} \right) \,.
\end{equation}
Here, $\theta_{\sfa \sfb}$ are the $n(n+1)/2$ real compact boson fields while $T^{\sfa \sfb}$ is the standard basis for the vector representation of the $\mathfrak{so}(n+1)$ Lie Algebra as $(n+1)$-by-$(n+1)$ matrices with elements
\begin{equation}
    (T^{\sfa \sfb})_{ij} = \delta_{\sfa i} \delta_{\sfb j} - \delta_{\sfa j} \delta_{\sfb i} \,.
\end{equation}

Next, $\Lambda$ commmutes with $O_0$, while it acts on the $\mathrm{SO}(n+1)$ generators as $\Lambda T^{\sfa \sfb} \Lambda = \pm T^{\sfa \sfb}$. 
Equivalently, the potential sign acquired by $T^{\sfa \sfb}$ can be viewed as imposing a twisted boundary condition $\theta_{\sfa \sfb} \mapsto - \theta_{\sfa \sfb}$ for each associated boson $\theta_{\sfa \sfb}$.
The correlations of twist fields in the compact boson are known to exhibit universal scaling behavior independent of the stiffness~\cite{dixon_orbifold_cft}.
Correlations of the \NLSM twist field are obtained as the $k(n+1-k)$ power of this universal result, as a simple counting shows that $k(n+1-k)$ independent generators $T^{\sfa\sfb}$ acquire a sign.

We now present results for the \NLSM twist field correlation at weak-coupling.

\subsubsection{Bulk correlation of the twist fields}\label{sec:bulk_twists_for_fid_corr}
To begin, we discuss the bulk correlation of twist fields $\phi_k$ located at $\bfx, \, \bfy$ as pictured in Fig.~\ref{fig:twist_intro}(a).
These twists determine the scaling of the fidelity correlator.
As we compute the bulk correlations, we assume that $\bfx, \, \bfy$ are far from any boundaries and take the thermodynamic limit.
In that case, twist field correlations of a single real compact boson obey power law scaling with scaling dimension $1/8$~\cite{dixon_orbifold_cft}.
Thus, the scaling dimension of the \NLSM twist field is $k(n+1-k)/8$.
Following standard RG arguments [reviewed in Appendix~\ref{app:short_twist}], this scaling can be spoiled by logarithmic functions of the separation due to the marginally irrelevant flow of the coupling away from the fixed point.

To summarize, the bulk correlation takes the form
\begin{equation}
    \langle \phi_{k}(\bfx) \phi_{k}(\bfy) \rangle \sim  
    \frac{\left(
        \ln |\bfx- \bfy|
    \right)^{2  \Delta_1 }}{|\bfx-\bfy|^{k(n+1-k)/4}}
    \,.\label{eq:renyi_twist_corr}
\end{equation}
We assume the scaling dimension of $\phi_k(\bfx)$ is $\Delta_0 + \Delta_1 g$, where $\Delta_1$ is the coefficient of the first-order correction.
In general, it is a function of $n, k$ whose value cannot be fixed without an explicit perturbative calculation, which we do not do here.
Furthermore, our numerical studies are often in a regime where the renormalized coupling $g_R$ is dominated by its bare value over the entire range of accessible system sizes.
Thus, we often drop the corrections to scaling.

\subsubsection{Large twist defects on the cylinder}\label{sec:large_twists_for_decoding_fidelity}
We also discuss the expectation value of twists $\Phi_k$ along large defects in the longitudinal direction of the cylinder as in Fig.~\ref{fig:twist_intro}(b).
These twists determine the scaling of the decoding fidelity.
Following the above discussion in Sec.~\ref{sec:bulk_twists_for_fid_corr}, 
at weak-coupling the \NLSM twist is obtained as $k(n+1-k)$ copies of boundary correlations of the twist field in the compact boson.
On general grounds, the latter are expected to scale as $A\, \exp (- \pi \kappa / 4)$ where $A$ is a non-universal constant~\cite{cardy1984conformal}.
Strictly speaking, this scaling is valid for $\kappa \gg 1$, with corrections of order $e^{-\calO(\kappa)}$.
Additional corrections may also arise from the fusion of the twist with the boundary, which we do not discuss here.

To summarize, the twist expectation value takes the form
\begin{equation}\label{eqn:order_one_twist}
    \Phi_k 
    = \left[ A \, e^{- \pi \kappa /4} \right]^{k(n+1-k)}
    \,. 
\end{equation}
Along with the aforementioned corrections, $\Phi_k$ can also receive corrections to scaling from the marginal RG flow of the coupling.
Again, we neglect all these effects.

\section{Decoding fidelity in the coherently corrupted surface code}\label{sec:decoding}
In this section, we make predictions on the decoding fidelity based on the effective \NLSM. 
In particular, in Sec.~\ref{sec:decoding_fid} we investigate the scaling of the decoding fidelity in the partially decodable phase close to the weak-coupling fixed point.
We examine these theoretical predictions using large-scale numerical simulations in Sec.~\ref{sec:decoding_numerics}.

\subsection{Analytic results}\label{sec:decoding_fid}
In this section, we analytically take the replica limit of the replicated fidelity $\mathcal{F}^{(n)}$ based on the twist expectation value obtained in Sec.~\ref{sec:large_twists_for_decoding_fidelity}.
We briefly outline the technical steps for the analytic continuation and refer interested readers to Appendix~\ref{app:resummation}.
Then, we present the universal scaling of the fidelity at weak-coupling $\mathcal{F}_*$.

First, at weak-coupling we use the expectation value of the large twist defect in Eq.~\eqref{eqn:order_one_twist} and the expression for the replicated fidelity in Eq.~\eqref{eqn:rbim_replicated_fid_as_defects}
\begin{equation}\label{eqn:nlsm_rbim_fid}
    \mathcal{F}_* = \lim_{n\to0}
    \frac{
    2
    \sum_{k=0}^{n-1}
    \binom{n-1}{k}
    \left[ A \, e^{- \pi \kappa /4} \right]^{k(n+1-k)}
    }
    {
    \sum_{k=0}^{n+1}
    \binom{n+1}{k}
\left[ A \, e^{- \pi \kappa /4} \right]^{k(n+1-k)}
    }
    \,.
\end{equation}

In order to take the replica limit $n \to 0$, there are two technical steps:
\begin{itemize}
    \item First, we must remove the $n$-dependence of the limits of summation.
This is achieved by analytically continuing the binomial coefficient
\begin{equation}
    \binom{n}{k} = \frac{\Gamma(n+1)}{\Gamma(k+1)\Gamma(n-k+1)} \,,
\end{equation}
where $\Gamma(z)$ is Euler's Gamma function.
We also extend the upper limit of the sum to $k = \infty$.
This choice of analytic continuation agrees with the original sum for $n \in \mathbb{Z}_{\geq 1}$ due to the pole structure of $\Gamma(x)$.
    \item Second, the resulting infinite sum is often highly divergent.
    It is given an analytic continuation as a convergent integral through a generalized Borel resummation technique.
\end{itemize}

\begin{figure*}[t]
\centering
\includegraphics[width=.95\textwidth]{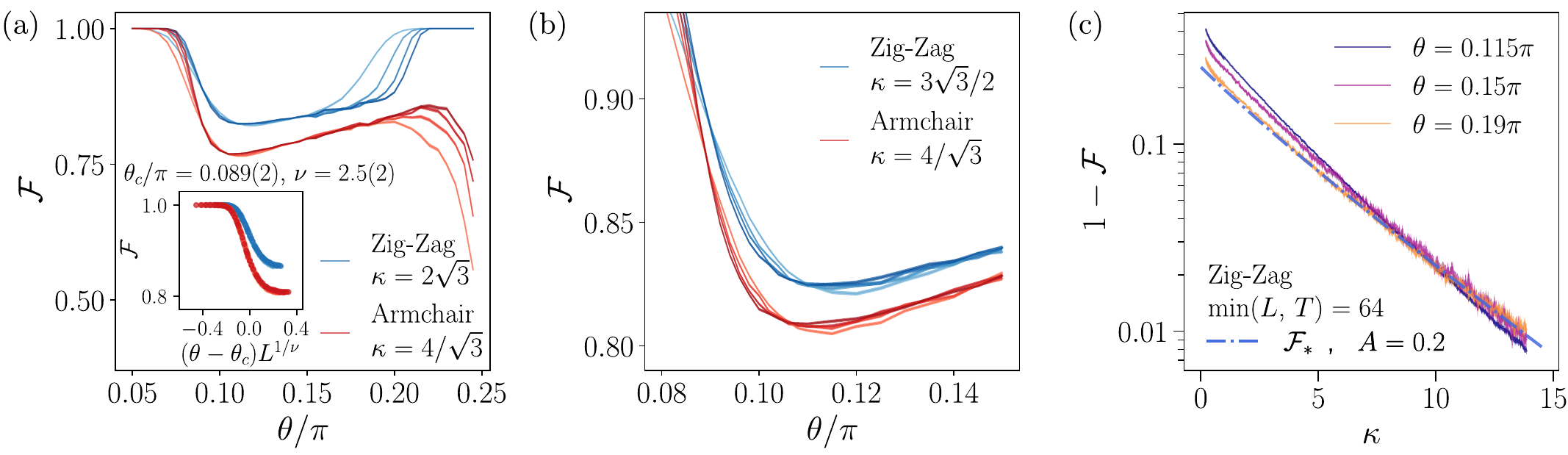}
    \caption{
    We plot the decoding fidelity.
    In (a) and (b), $\mathcal{F}$ is plotted against $\theta$ fixing $\kappa$, with lighter to darker color representing increasing system size $L = 8, 16, 24, 32$ ($L = 8\sqrt{3},
     16\sqrt{3}, 24\sqrt{3}, 32\sqrt{3}$) for the Zig-Zag (Armchair) boundary condition, respectively.
    In the inset of (a), the smallest system size is dropped and we perform a finite-size scaling collapse near $\theta_c$ while in (b) $\mathcal{F}$ is plotted for small $\theta$ demonstrating the crossover into the sigma model regime at $\theta \approx 0.11\pi$.
    In (c), $1- \mathcal{F}$ is plotted against $\kappa$ for fixed scale $\min(L, \, T) = 64$. The fidelity shows opposite trends as a function of coupling (tuned by $\theta$) for $\kappa \ll 1$ and $\kappa \gg 1$. 
    For the largest $\theta = 0.19\pi$, the fidelity is close to the universal scaling function at the weak-coupling fixed point $\mathcal{F}_*(\kappa)$, denoted by the blue dashed line.
        Data averaged over $1500$ to $200000$ samples, with error indicated by the line width.}
    \label{fig:fid_trends_combined}
\end{figure*}

Following the procedures detailed in Appendix~\ref{app:resummation}, we obtain the fidelity scaling in the weak coupling regime,
 \begin{align}\label{eqn:order_one_AR_fid_integral_rep}
 &\mathcal{F}_*(\kappa)
 =
 \frac{1}{\sqrt{\pi}}
    \int_{-\infty}^\infty \frac{
    e^{-u^2}
    \, \rd u
    }
    {
    1 + Ae^{-\pi \kappa/4} \, e^{2u \sqrt{\pi \kappa / 4 - \log A}}
    }
    \\
    &= 1 - 
   \frac{\sqrt{\pi} A^{1/4} \, e^{-\pi\kappa / 16}}{2 \sqrt{\pi \kappa /4 -\log(A)}}
   + \mathcal{O}
   \left(
   \frac{
    e^{- \pi \kappa /16}
   }
   {
   \left[
   \pi \kappa / 4 - \log (A)
   \right]^{3/2}
   }
   \right)
   \nonumber
   \,.
 \end{align}
The fidelity is a monotonically increasing function of $\kappa$ and is plotted as the dashed line in Fig.~\ref{fig:fid_trends_combined}(c).
In the second line, we have performed an asymptotic expansion in $\kappa \to \infty$ detailed in Appendix~\ref{app:asymptotic}.

Away from the weak-coupling fixed point, Eq.~\eqref{eqn:order_one_AR_fid_integral_rep} can receive corrections due to the marginally irrelevant flow of $g_R$.
As analyzed in Ref.~\cite{yan_bao_coherent_error_nlsm}, these corrections are expected to obey a ``trend-reversal'': for fixed $\kappa$, the behavior of the fidelity when tuning the overall scale $L$ is qualitatively different for small or large $\kappa$.
The precise form of the corrections to scaling will not play a role in the present work, however the qualitative features will be used in Sec.~\ref{sec:decoding_numerics} to identify the sigma model regime in numerics.

Finally, we note that the dependence of $\mathcal{F}$ on $g_R$ can be probed through the crossover into the quasi-1D regime.
This is achieved, for instance, by fixing $L$ and taking $\kappa \to \infty$.
After coarse-graining up to scale $L$, the effective description is the 1D \NLSM which is disordered and describes a decodable phase.
Here, the renormalized stiffness $1/g_R(L)$ becomes a correlation length while $\kappa$ becomes the renormalized system size while the dimensionless combination $\kappa g_R(L)$ serves as a renormalized code distance.
Thus, the fidelity is expected to scale as $\log(1 - \mathcal{F}) \propto \kappa g_R(L)$.
Similarly, we may fix $T$ and take $1/\kappa \to \infty$.
In this case, the disordered 1D \NLSM describes a non-decodable phase and $\log (\mathcal{F} - 1/2) \propto \kappa / g_R(T)$.
We provide a less heuristic argument for the fidelity in the quasi-1D regimes in Appendix~\ref{app:sec_quasi_1d_fid}.

\subsection{Numerical results}\label{sec:decoding_numerics}
In this section, we examine the prediction for the fidelity in the weak coupling regime with extensive numerical study.
Our numerical data are averaged over syndrome realizations.
Accessing the $n \to 1$ replica limit relevant for optimal decoding requires that the syndrome be sampled according to the Born rule, i.e., with probability given by $\mathcal{Q}_{\alpha, \, s}$.
To efficiently sample from $\mathcal{Q}_{\alpha, \, s}$, we use the algorithm introduced by Ref.~\cite{bravyi_coherent} and generalized by Ref.~\cite{venn_coherent_planar_graph,yan_bao_coherent_error_nlsm}.
A review of this algorithm can also be found in Appendix H of Ref.~\cite{yan_bao_coherent_error_nlsm}, which we modify for the triangular lattice.

Additionally, we systematically identify the sigma model regime by examining the crossovers in the vicinity of the decoding transition [see Sec.~\ref{sec:numerics_transition}] and near $\theta = \pi/4$ [see Sec.~\ref{sec:ballistic_metal_regime}].
We then present results for the partially decodable phase described by the \NLSM in Sec.~\ref{sec:numerics_thermal_metal}.

\subsubsection{Decoding transition}\label{sec:numerics_transition}
We begin by discussing the decoding transition between the decodable and partially decodable phase.
The universality of the transition is governed by a non-trivial strong-coupling fixed point of the $\mathrm{SO}(n)$ \NLSM, similar to that in the class DIII metal-to-insulator transition in disordered superconductors, but in the $n \to 1$ replica limit.

We now estimate the decoding threshold angle $\theta_c$ and the associated critical exponents.
First, the decoding fidelity is expected to be scale-free at $\theta_c$ and fixed $\kappa$.
Furthermore, in the vicinity of the critical point, $\mathcal{F}$ is expected to decrease with scale for $\theta > \theta_c$ and increase with scale for $\theta < \theta_c$.
Thus, we estimate the threshold rotation angle $\theta_c$ by locating the crossing point and extract the correlation length exponent $\nu$ through a scaling collapse of the form
\begin{equation}\label{eqn:scaling_form_fid}
    \mathcal{F} = \calH\left(L^{1/\nu}(\theta - \theta_c)\right) \,.
\end{equation}
The finite-size scaling collapse is performed for both zig-zag and armchair boundary conditions in Fig.~\ref{fig:fid_trends_combined}(a).
For both boundaries, we obtain $\nu = 2.5 \pm 0.2$.

The obtained exponent $\nu$ is consistent with the one extracted from the conductance, another observable of the associated network model, in Sec.~\ref{sec:conductance} and with previous results in related problems governed by the same effective theory~\cite{yang_coherent_hexagonal,wang_self_dual_cho_fisher,jian2023measurement}.
However, it differs slightly from~\cite{bejan_projected_ensemble} who used a different scaling form.

Finally, we discuss how the behavior of the scaling function $\mathcal{H}$ in Eq.~\eqref{eqn:scaling_form_fid} allows us to determine the correlation length $\xi$ associated with the transition.
First, we note that for fixed $\kappa$, $\mathcal{F}$ decreases with $L$ for $\theta > \theta_c$ and $L < \xi$ below the crossover scale.
This is in contrast with its behavior in the regime governed by the \NLSM and small $\kappa$ where it increases with $L$ [see Sec.~\ref{sec:decoding_fid}].
Although our argument there strictly applies to the quasi-1D regime $\kappa \ll 1$,  numerics reveal that $\mathcal{F}$ already increases with $L$ for $\kappa \lesssim 4$.
Thus, in Fig.~\ref{fig:fid_trends_combined}(b), we plot $\mathcal{F}$ for various system sizes and fixed $\kappa < 4$.
For a particular $\theta$, the corresponding crossover scale can then be estimated as the largest $L$ such that $\mathcal{F}$ begins to increase with scale.
The numerically accessible system sizes all appear to be below the crossover scale for $\theta \gtrsim 0.11\pi$.

\subsubsection{Crossover in the vicinity of \texorpdfstring{$\theta=\pi/4$}{theta=pi/4}}\label{sec:ballistic_metal_regime}
Next, we discuss the decoding fidelity in the vicinity of $\theta = \pi/4$.
Directly at $\theta = \pi /4$, the surface code is not described by the effective \NLSM.
Instead, the fidelity is independent of system size and aspect ratio but is sensitive to the choice of boundary conditions, with $\mathcal{F} = 1$ and $1/2$ for the zig-zag and armchair configurations, respectively.
The decoding properties of this point are derived in detail in Appendix~\ref{app:ballistic_metal}.

Features of the decoding fidelity at this point continue to play a role in finite-size systems in the vicinity of $\theta = \pi /4$ below a crossover scale set by the mean-free path $\lambda$.
This is presented in Fig.~\ref{fig:fid_trends_combined}(a) which shows $\mathcal{F}$ close to $1$ ($1/2$) for the zig-zag (armchair) boundary conditions in an extended region near $\theta = \pi/4$.
Furthermore, we leverage this observation to estimate $\lambda$ at a particular $\theta$ as the largest system size below which $\mathcal{F}$ begins to exhibit dependence on boundary condition.
For instance, in Fig.~\ref{fig:fid_trends_combined}(a), the behavior of the fidelity  for $L \sim 32$ appears to deviate at around $\theta = 0.2\pi$.
Thus, we may estimate $\lambda \sim 32$ at $\theta = 0.2\pi$.

\subsubsection{Partially decodable phase}\label{sec:numerics_thermal_metal}
In this section, we study the decoding fidelity in the partially decodable phase described by the weak-coupling regime of the NL$\sigma$M.

In order to test our analytic predictions, we must identify the regime which is described by the \NLSM.  This requires $L > \xi,\, \lambda$.
As discussed in the preceding Sec.~\ref{sec:numerics_transition} and Sec.~\ref{sec:ballistic_metal_regime}, for $L \sim 32$ this is roughly given by $0.11\pi \lesssim \theta \lesssim 0.20\pi$.
In the following, we consider $\min(L, \, T) = 64$ which is comfortably within the sigma model regime for $\theta \leq 0.19\pi$.

Our main result is the universal scaling of the fidelity $\mathcal{F}_*(\kappa)$ in Eq.~\eqref{eqn:order_one_AR_fid_integral_rep} which has the asymptotic scaling  $1 - \mathcal{F}_* \sim  e^{- \pi \kappa / 16}/\sqrt{\kappa}$.
We compare the analytic prediction to numerical simulations in Fig.~\ref{fig:fid_trends_combined}(c), where the non-universal constant $A$ is chosen to produce the closest fit to the numerical data for a particular value of $\theta$.
We find a particularly close agreement of $\mathcal{F}_*$ with the data for $\theta = 0.19\pi$.
The prediction also agrees closely with the data for $\theta = 0.17\pi$, which is omitted for visual clarity.

Away from the weak-coupling fixed point, the fidelity receives corrections at finite coupling $g_R$.
As discussed in Sec.~\ref{sec:decoding_fid}, these corrections are expected to exhibit a ``trend-reversal'' in their dependence on $g_R$ based on the value of $\kappa$, which is fixed.
This is most clearly illustrated when $g_R$ is tuned by its bare value set by $\theta$.
Indeed, in Fig.~\ref{fig:fid_trends_combined}(c), $1 - \mathcal{F}$ is a decreasing function of $\theta$ for $\kappa \lesssim 6$ while it appears to increase with $\theta$ for $\kappa \gtrsim 6$.
The trend-reversal in the dependence of $\mathcal{F}$ on $g_R$ can also be obtained in the crossover to the quasi-1D regime.
Numerical investigations of this regime are presented in Appendix~\ref{app:sec_quasi_1d_fid}.
There, we find that behavior of $\mathcal{F}$ is consistent with a $1/g_R$ which grows logarithmically with scale as predicted by the RG equation~\eqref{eqn:RG_equation}.

\section{Mixed-state order in the critical phase}\label{sec:mixed-state}
We investigate the critical phase from the perspective of mixed-state order.
In Sec.~\ref{sec:mixed_state_fid_corr}, we show that the density matrix $\varrho$ in Eq.~\eqref{eq:mixed-state_dual_paramagnet}, describing the paramagnet subject to a decoherence channel $\tilde{\calE}_\theta$, has a power-law decaying fidelity correlator.
This indicates that the critical phase exhibits an algebraic order of SWSSB.
In Sec.~\ref{sec:mixed-state_non-classical}, we further show that the mixed state in the critical phase generated by the channel $\tilde{\mathcal{E}}_\theta$ with $\theta > \theta_c$ depends sensitively on the boundary conditions.
This is a feature of the decoherence channel $\tilde\calE_\theta$ that cannot be reproduced by any symmetric dephasing channel $\calN_{ZZ}$, describing Glauber dynamics.
The decoding problem in the surface code subject to channel $\calE_\theta$ also exhibits a key distinction from that for the dephasing channel $\calN_Z$.
We show that in the critical phase, the probability of syndromes created by the channel $\calE_\theta$ depends sensitively on the far-separated $m$-anyons in the background.

\begin{figure*}[t]
\centering
\includegraphics[width=.95\textwidth]{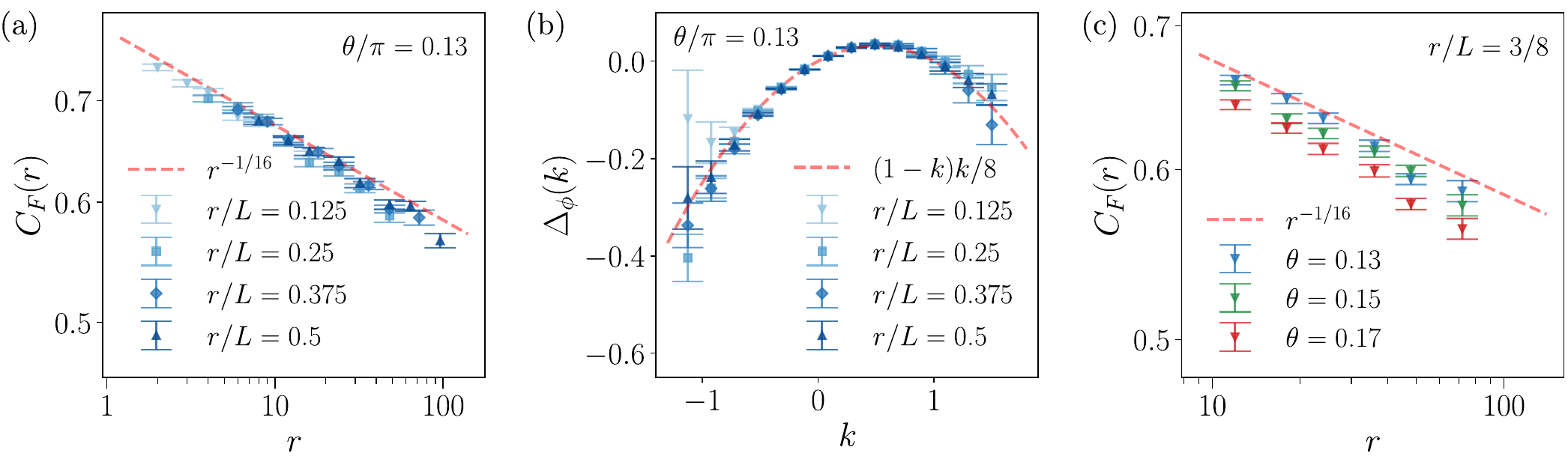}
    \caption{Fidelity correlator $C_F$ and the $k$-th Renyi fidelity correlator $C_{F, \, k}$ on the $L$-by-$L$ coherently-corrupted paramagnet state $\varrho$.
    In (a), the fidelity correlator is plotted as a function of separation $r := |{\bfx} - {\bfy}|$  
    Here, $\theta = 0.13 \pi$ is fixed, as is $r/L$, while $L$ is varied.
    In (b), the scaling dimension of the $k$-th Renyi fidelity correlator is extracted by performing a linear best fit on log-log scale, for fixed, but various values of $r/L$.
    The extracted scaling dimension is plotted as a function of $k$.
    In (c), the fidelity correlator is plotted against $r$ for fixed $r/L$ and various $\theta$.
    For all cases, the endpoints of the symmetry defect are chosen as ${\bfx} = (L/2-r/2, L/2)$ and ${\bf y} = (L/2+r/2, L/2)$.
    Note that in (a) and (c) both axes are on a logarithmic scale.
    Data averaged over $15000$ to $50000$ samples.}
    \label{fig:defect_correlator_combined}
\end{figure*}

\subsection{Fidelity correlator in the critical phase}\label{sec:mixed_state_fid_corr}
We begin with the fidelity correlator diagnostic of the mixed-state order.
This observable is given by the disorder field correlation functions in the RBIM description.

\subsubsection{Algebraic SWSSB}\label{sec:swssb}
We analyze the fidelity correlator in the coherently-corrupted paramagnet state $\varrho$ using the replica sequence in Eq.~\eqref{eqn:fidelity_correlator_as_defects}.
The replicated quantity $C_F^{(2n)}$ maps to twist field correlation functions in the effective NL$\sigma$M description, which exhibit a power-law decay at weak coupling as in Eq.~\eqref{eq:renyi_twist_corr}.
We then take the replica limit $2n \to 1$, predicting a power-law decaying fidelity correlator,
\begin{align}
    C_F(\bfx, \, \bfy) &=\lim_{2n \to 1}\langle \phi_n(\bfx) \phi_n(\bfy)\rangle_{\mathbf{Z}_0^{(2n)}} 
    \\
    &\sim \frac{1}{|{\bfx}-{\bfy}|^{1/16}} 
    \label{eqn:fidelity_correlator_universal_powerlaw}
    \,,
\end{align}
where we have ignored potential multiplicative corrections to scaling resulting from the RG flow of $g_R$.
We have also assumed that $\bfx, \, \bfy$ are deep within the bulk so that boundary conditions do not play a role.
The result indicates an algebraic order of the SWSSB in the critical phase.

We examine this prediction by numerically simulating the fidelity correlator $C_F$.
Specifically, we express the correlator in terms of the Born probability $\calQ_m$ as in Eq.~\eqref{eq:fid_corr_diagonal}.
We numerically estimate $C_F$ by sampling the measurement results $m$ from the Born probability $\mathcal{Q}_m$ and then compute the probabilities $\mathcal{Q}_m$ and $\mathcal{Q}_{m + \bfe_{\bfx}+\bfe_{\bfy}}$ associated with the configuration with two extra charges introduced at $\bfx, \, \bfy$.
We average the quantity $\sqrt{\mathcal{Q}_{m+\bfe_{\bfx}+\bfe_{\bfy}}/\mathcal{Q}_m}$ over measurement results $m$ to obtain an estimation of the fidelity correlator.
We sample from the Born distribution $\mathcal{Q}_m$ and compute the associated probabilities using the same algorithm as that used in Sec.~\ref{sec:decoding_numerics} since it is identical to the syndrome distribution $\mathcal{Q}_s$ in the surface code with appropriate boundary conditions, which we detail shortly.

To extract the power-law decay exponent of the fidelity correlator, we take $\varrho$ on the lattice of size $L$-by-$L$.
In general, the fidelity correlator takes the scaling form $C_F = L^{-1/16} \calF({\bfx}/L,\, {\bfy}/L,\, g_R)$.
We fix the location of the two endpoints relative to the boundary, i.e., ${\bfx}/L$ and ${\bfy}/L$. 
Furthermore, for the numerically accessible system sizes, the renormalized coupling $g_R$ remains approximately unchanged as shown in Fig.~\ref{fig:defect_correlator_combined}(c).
As a result, the potential correction to scaling is expected to be negligible, and the fidelity correlator closely follows a simple power-law scaling in the system size $L$.
Our numerics demonstrate agreement with this prediction shown in Fig.~\ref{fig:defect_correlator_combined}(a).
In these numerics, we choose $\theta = 0.13\pi$ which is deep within the sigma model regime following the analysis in Sec.~\ref{sec:decoding_numerics} for the smallest system sizes considered here ($L=8$).

To further examine our effective theory, we consider the $k$-th Rényi fidelity correlator, which is also expressed as a replica sequence of disorder operator correlators~\eqref{eqn:k_EA_replica_sequence}.
In the replica limit $n \to 0$, this quantity has the power-law scaling
\begin{align}\label{eqn:k_EA_powerlaw}
    C_{F, \, k}(\bfx, \, \bfy)&=\lim_{n \to 0} \langle \phi_{k}({\bfx}) \phi_{k}({\bfy}) \rangle_{\bfZ_0^{(n+1)}} \nn \\
    &\sim 
    \frac{1}{|{\bfx}-{\bfy}|^{(1-k)k/4}}
    \,.
\end{align}
We note that for $|2k-1| > 1$, the correlation function increases with distance.
Similar predictions for the correlation function of order and disorder parameters in the critical phase (also known as the thermal metal phase in the disordered fermion context) are described by a similar but distinct NL$\sigma$M in Ref.~\cite{read_absence_of_metallic}.

In Fig.~\ref{fig:defect_correlator_combined}(b), we extract the scaling dimension $\Delta_\phi(k)$ by fitting $C_{F, \, k} \sim 1/L^{2 \Delta_\phi(k)}$ while holding $\bfx/ L, \, \bfy / L$ fixed.
Our numerical simulations show close agreement with the theoretical prediction $\Delta_\phi(k) = (1-k)k/8$.
Note that for large $k$, Eq.~\eqref{eqn:k_EA_powerlaw} predicts that the variance becomes a rapidly increasing function of the separation.
Thus, we are limited to relatively small moments $|k| \lesssim 1$ in our simulations.

\subsubsection{Fidelity correlator in the surface code}
In the surface code, the mixed-state order of the critical phase manifests as algebraic decay of the fidelity correlator involving separated $e$-anyons~\eqref{eq:fid_corr_surf_code}.
To make an explicit connection to the dual problem, we consider the surface code on a square geometry with smooth boundaries on which $m$-anyons condense.
In this case, the surface code does not encode logical information, and the coherently corrupted state $\rho$ is completely diagonalized in the syndrome basis $\rho=\sum_s \mathcal{Q}_s \ketbra{s}$.
The fidelity correlator can then be expressed as 
\begin{align}
    C_{F}(\bfx, \, \bfy) &= \sum_s \sqrt{\mathcal{Q}_s \mathcal{Q}_{s + \partial \mathcal{C}}}\sim \frac{1}{|{\bfx}-{\bfy}|^{1/16}}
    \,.
\end{align}
Here, $\calQ_{s + \partial \mathcal{C}}$ maps to the RBIM with a defect inserted along path $\calC$. 
The endpoints of $\calC$ are $\bfx, \, \bfy$.
Such a fidelity correlator $C_e$ is the same as $C_F$ in the dual problem and therefore exhibits the same power-law decay.
We note that the distinguishability between two states with and without far-separated $e$-anyons has been used as a probe of the decoding transition for the surface code under dephasing errors in Ref.~\cite{fan2024diagnostics}.

\subsection{Coherent nature of the critical phase}\label{sec:mixed-state_non-classical}
\begin{figure}[t]
\centering
\includegraphics[width=.45\textwidth]{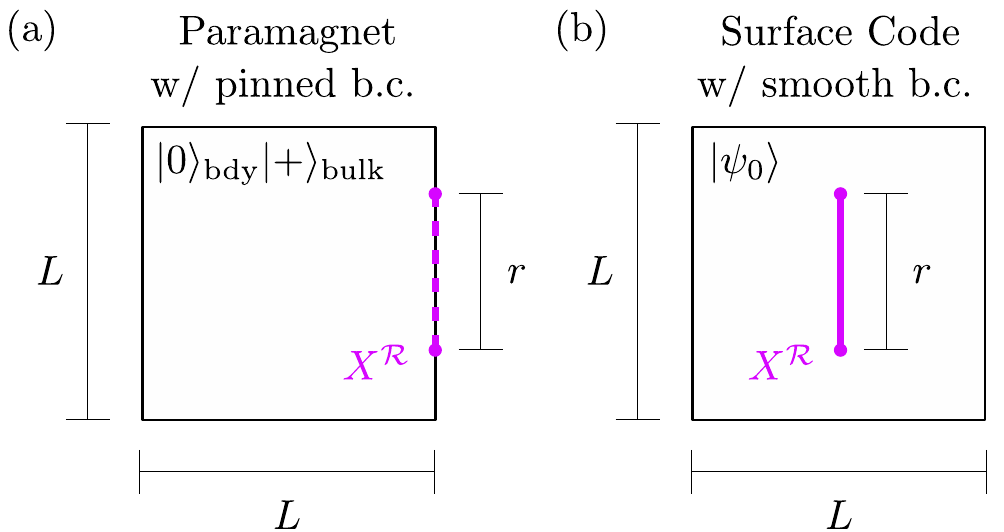}
\caption{To demonstrate the absence of an effective classical channel in the critical phase, we apply Pauli-$X$ to (a) qubits on boundary vertices in a contiguous region of size $r$ for the paramagnet 
(b) qubits on links crossed by a path $\mathcal{R}$ on the dual lattice of length $r$ for the surface code.
The initial state is prepared as $\ket{0}_\mathrm{bdy}\ket{+}_\mathrm{bulk}$ and $\ket{\psi_0}$ for the paramagnet and surface code, respectively.
We consider the $L \times L$ square geometry.
}
\label{fig:non_classical_setup}
\end{figure}

We have shown that the decoherence channel $\tilde{\calE}_\theta$ can drive a phase transition from a strongly-symmetric phase to a phase with algebraic SWSSB.
Conventionally, the phase transition between the strongly-symmetric and the $\mathbb{Z}_{2}$ SWSSB phase admits a classical description: applying a symmetric dephasing channel to a $\mathbb{Z}_2$ paramagnet produces a density matrix that is diagonal in the symmetry-charge basis. The SWSSB transition induced by tuning the dephasing strength is then entirely encoded in the classical probability distribution over $\mathbb{Z}_2$ charge configurations.

This raises the question of whether the transition into the algebraic $\mathbb{Z}_2$ SWSSB phase likewise admits an effective classical dynamical description. To make this distinction concrete, consider a strongly symmetric dephasing channel of the form
\begin{equation}\label{eqn:effective_glauber_channel}
    \mathcal{N}_{ZZ}[\cdot]
    =
    \sum_\calC
    p_{\calC} Z^{\partial \calC} [\cdot] Z^{\partial \calC} \,,
\end{equation}
where the Kraus operator creates charges at the endpoints $\partial \calC$ of paths $\calC$ on the triangular lattice with probabilities $p_{\calC}$.
The probability for each charge configuration $m$ is given by an incoherent sum over different paths, i.e.,
\begin{align}
    \calQ_m = \sum_{\calC;\, m = \partial \calC} p_\calC.
\end{align}
In this case, the paths $\calC$ appearing in the ket and the bra are identical, and the conventional SWSSB transition is associated with the onset of large-scale fluctuations of these glued paths for a fixed charge configuration. %

However, in our setup, the Born probability $\calQ_m$ is given by the squared magnitude of a coherent sum of amplitudes for distinct paths with the same endpoints
\begin{align}
    \calQ_m = \abs{\sum_{\calC, \, m = \partial \calC} (\cos\theta)^{N-\abs{\calC}} (\ri\sin\theta)^{\abs{\calC}}}^2. \label{eq:born_coherent}
\end{align}
The paths $\calC$ in the ket and the bra Hilbert space need not coincide.  The central question is whether these paths become effectively glued together at long distances in the algebraic SWSSB phase, recovering a classical stochastic description, or whether interference between distinct paths survives.

In this section, we show that the critical phase with algebraic $\mathbb{Z}_{2}$ SWSSB retains such a long-distance coherence which is absent in the conventional, dephasing-induced $\mathbb{Z}_{2}$ SWSSB phase.  We introduce a probe that diagnoses whether interference between distinct paths persists at large-scales and contrast the behavior of this probe in these two phases.  %
In particular, the resulting mixed state $\varrho$ for $\theta > \theta_c$ depends sensitively on the boundary condition, while this is not the case in the conventional SWSSB phase.
A related feature is present in the surface code under the channel $\calE_\theta$. 
The probability of $e$-anyon configurations created by $\calE_\theta$ depends sensitively on the presence of $m$-anyons in the code.

\subsubsection{Critical phase with algebraic SWSSB}\label{sec:absence_dual_swssb}

In the algebraic SWSSB phase $\theta > \theta_c$, the consequence of fluctuating error chains in the ket and the bra is the dependence of the resulting mixed state on the boundary condition of the paramagnet.

Specifically, we consider a two-dimensional array of qubits on the vertices of triangular lattice, initialized in a paramagnetic state.
We place the qubits in a finite geometry of $L$ by $L$ with symmetry-breaking boundary conditions that pin the qubits to the $\ket{0}$ or $\ket{1}$ state.
We now apply the decoherence channel $\tilde\calE_\theta$ to the state with and without boundary domain walls. We obtain two output states given by diagonal density matrices,
\begin{align}
    \varrho_1 &= \tilde{\mathcal{E}}_\theta
    \left[
    \ketbra{0}_\mathrm{bdy}
    \otimes
    \ketbra{+}_\mathrm{bulk}
    \right]
    = \sum_m \mathcal{Q}_m \ketbra{m}
    \,, \\
    \varrho_2 &= \tilde{\mathcal{E}}_\theta
    \left[X^\calR
    \ketbra{0}_\mathrm{bdy}
    \otimes
    \ketbra{+}_\mathrm{bulk}X^\calR
    \right] \nn \\
    &= \sum_m \tilde{\mathcal{Q}}_m \ketbra{m}
    \,,
\end{align}
where $X^\calR$ acts on a domain and creates two domain walls on the symmetry breaking boundary as shown in Fig.~\ref{fig:non_classical_setup}(a).

We consider the fidelity between two resulting density matrices $\varrho_{1,2}$, i.e. the Bhattacharyya coefficient associated with two probability distributions~\cite{bhattacharyya1946measure},
\begin{align}
    {F}(\varrho_1, \, \varrho_2) &= \sum_m \sqrt{\mathcal{Q}_m \tilde{\mathcal{Q}}_m}
    =
    \sum_m \mathcal{Q}_m |\langle \sigma_\bfx \sigma_\bfy\rangle_m|
    \,.
\end{align}
Here, we express $\calQ_m$ and $\tilde \calQ_m$ in terms of partition sums involving Ising spins,
\begin{align}
    \calQ_m &= \abs{\sum_\sigma e^{(J-\frac{\ri \pi}{4})\left(\sum_{\langle\bfr,\bfr'\rangle}\sigma_\bfr\sigma_{\bfr'}+ \sum_{\bfr\in\text{bdy}}\sigma_\bfr\right)}}^2, \\
    \tilde\calQ_m &= \abs{\sum_\sigma \sigma_{\bfx} \sigma_{\bfy}\, e^{(J-\frac{\ri \pi}{4})\left(\sum_{\langle\bfr,\,\bfr'\rangle}\sigma_\bfr\sigma_{\bfr'}+ \sum_{\bfr\in\text{bdy}}\sigma_\bfr\right)}}^2.
\end{align}

In the critical phase $\theta > \theta_c$, the spin-spin correlation function exhibits a power-law decay with the separation.
To see this, we introduce a replica sequence for the fidelity similar to that for the fidelity correlator in Eq.~\eqref{eqn:fidelity_corr_replica_sequence}.
The replicated fidelity is then expressed in terms of spin insertions in $n$ ket and $n$ bra copies.
The correlation functions of Ising spins in the RBIM correspond to the correlation functions of $\mathbb{Z}_2$ orbifold twist fields in the effective NL$\sigma$M, exactly the same as for the disorder operator correlations.\footnote{
However, in the present case, the locations of the spins $\bfx, \, \bfy$ are constrained to the boundary, leading to a modified boundary scaling.
Nevertheless, given that the boundary condition of the spins are free, on general grounds we expect $F(\varrho_1, \, \varrho_2) \sim 1/|\bfx - \bfy|^\delta$ with $\delta > 0$ since $F$ is bounded.}
This is because the critical phase is self-dual under the statistical Kramers-Wannier duality, as established by Ref.~\cite{read_absence_of_metallic} in related Ising models also described by similar replica NL$\sigma$Ms in the weak coupling regime.
Thus, the fidelity between two output states vanishes at large scale.

This is in sharp contrast with the fidelity between two output states when applying a symmetric dephasing channel followed by charge measurements, $\calD_X\circ\calN_{ZZ}$, to the two initial states.
In that case, the output state is unaffected by the change of boundary conditions; the two states have a perfect fidelity.

We note that the fidelity between two states $\varrho_{1,\,2}$ is non-decaying in the symmetric phase ($\theta < \theta_c$). 
The RBIM associated with the Born probability $\calQ_m$ is in the ordered phase and exhibits long-range correlations in the spins, indicating a $\calO(1)$ fidelity.
Intuitively, the error chains only fluctuate below a finite correlation length, and the fidelity saturates to an $\calO(1)$ value as the distance between two domain walls is above the correlation length.

\subsubsection{Partially decodable phase in the surface code}\label{sec:absence_decoding_setup}
In the surface code, the coherent nature of the channel $\calE_\theta$ lies in its actions in different flux sectors.
Namely, the presence of $m$-anyons affects the probability of syndromes created by $\calE_\theta$.
This is in sharp contrast with the general dephasing channel $\calN_Z$, which commutes with the incoherent process that creates $m$-anyons.

Specifically, we introduce two states $\rho_1$ and $\rho_2$.
The first state $\rho_1 = \mathcal{E}_\theta[\ketbra{\psi_0}]$ is the coherently-corrupted surface code state $\ket{\psi_0}$ defined on a square geometry with smooth boundaries.
On the other hand, $\rho_2$ is created by a three-step process: (1) A pair of $m$-anyons is created by applying $X^\mathcal{R}$ consisting Pauli-$X$ along a path $\mathcal{R}$ on the dual lattice as shown in Fig.~\ref{fig:non_classical_setup}(b); (2) Apply the coherent-corrupting channel $\mathcal{E}_\theta$; (3) Remove the $m$-anyon pair by applying $X^\mathcal{R}$.
This results in the corrupted density matrix
\begin{equation}
    \rho_2 = X^\mathcal{R}
    {\mathcal{E}}_\theta
    \left[
    X^\mathcal{R}
  \ketbra{\psi_0}
    X^\mathcal{R}
    \right]
    X^\mathcal{R}
    \,.
\end{equation}

We capture how the channel $\calE_\theta$ depends on the background $m$-anyons based on the fidelity between two states, i.e., the Bhattacharyya coefficient between two syndrome distributions,
\begin{align}
F(\rho_1, \, \rho_2)
&= \sum_s \sqrt{\mathcal{Q}_s \tilde{\mathcal{Q}}_{s}} = \sum_s \mathcal{Q}_s | \langle \sigma_\bfx \sigma_\bfy \rangle_s | \,,
\end{align}
where $\mathcal{Q}_s$ given by RBIM partition function with pinned boundary conditions and $\tilde{\mathcal{Q}}_s$ is given by the partition function with spin inserted at the endpoints $\bfx, \, \bfy$ of $\mathcal{R}$.
The average spin-spin correlation, in the replica formulation, maps to twist field correlations in the \NLSM, giving rise to the following scaling
\begin{equation}
    F(\rho_1, \, \rho_2) \sim \frac{1}{|\bfx-\bfy|^{1/16}} \,,
\end{equation}
in the thermodynamic limit $L \to \infty$.
For far separated $m$-anyons, $r\to \infty$ the fidelity approaches zero.
In contrast, when the decoherence channel is dephasing, the fidelity between two states is unity. 

We note that in the decodable phase $\theta  < \theta_c$, the RBIM is ordered and the fidelity approaches an $\mathcal{O}(1)$ constant asymptotically.

Finally, we comment that we can also consider the $k$-th Rényi fidelity as the Chernoff $k$-coefficient between $\mathcal{Q}_s$ and $\tilde{\mathcal{Q}}_s$.
We examine this quantity, in addition to the fidelity, numerically in Appendix~\ref{app:kw_dual_fid_corr}.
Our findings verify the power-law scaling predicted by the sigma model. 
The order parameter and disorder parameter in the RBIM exhibiting the same scaling exponent is evidence of the statistical Kramers-Wannier duality symmetry of the critical phase.

\section{Other probes of the critical phase}\label{sec:decoding_others}
In this section, we further examine the effective \NLSM description of the critical phase by considering two other quantities, defect free energy and conductivity, related to the decoding problem and its associated network model.
We corroborate our theoretical predictions using large-scale numerical simulations.

\begin{figure}[t]
\centering
\includegraphics[width=.44\textwidth]{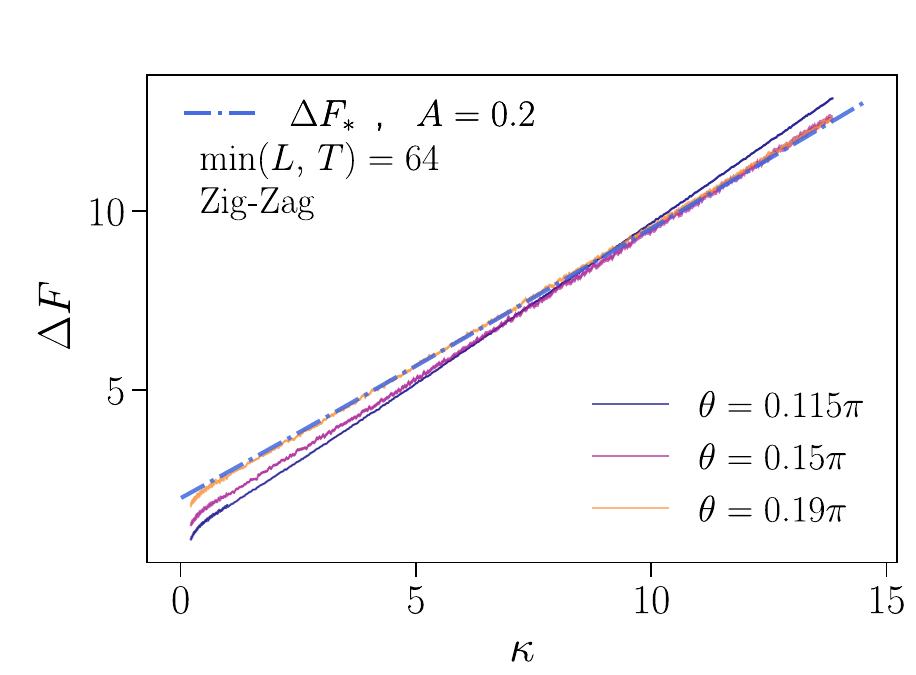}
    \caption{Defect free energy as a function of the aspect ratio while fixing the overall scale $\min(L, T) = 64$.
    The fidelity shows opposite trends as a function of coupling (tuned by $\theta$) for $\kappa \ll 1$ and $\kappa \gg 1$. 
     For $\theta = 0.19\pi$, the fidelity is close to the prediction at the weak-coupling fixed point $\Delta F_*$, as shown by the dashed blue line.
           Data averaged over $2000$ to $20000$ samples, with error indicated by the line width.}
    \label{fig:dfe_trends_combined}
\end{figure}
\subsection{Defect free energy}\label{sec:dfe}
In addition to the decoding fidelity, it is also useful to consider the defect free energy of inserting a symmetry defect in the statistical mechanical model
\begin{equation}\label{eqn:def_dfe}
    \Delta F = \sum_s \mathcal{Q}_s | \log \mathcal{Q}_{0, \, s} / \mathcal{Q}_{1, \, s}| \,.
\end{equation}
We introduce this quantity because it is expected to be self-averaging while the decoding fidelity, which is related to the quenched average of partition functions, is not.
Furthermore, it was shown in~\cite{yan_bao_coherent_error_nlsm} that the defect free energy exhibits the same qualitative behavior as the decoding fidelity in its dependence on the renormalized coupling $g_R$ and aspect ratio $\kappa$ and is therefore a useful proxy in numerical simulations.\footnote{We note that the two quantities can exhibit qualitatively distinct behaviors when considering suboptimal decoding in the surface code with coherent errors as shown in Ref.~\cite{yan_bao_coherent_error_nlsm}. However, we focus on optimal decoding in this work.}

The defect free energy can be expressed as the $n \to 1/2$, $m\to 0$ double replica sequence to handle the absolute value and logarithm in Eq.~\eqref{eqn:def_dfe}, respectively
\begin{equation}\label{eqn:dfe_rbim}
    \Delta F = 
    \lim_{\substack{m \to 0\\ n \to 1 /2}}
    \frac{2}{m^{2n}}
    \frac{
    \sum_{k=0}^{2n}
    \binom{2n}{k}
    (-1)^k
    \Phi_{mk+1}
    }
    {
    \sum_{k=0}^{2nm+1} \binom{2nm+1}{k} \Phi_k
    }
    \,.
\end{equation}
Here, $\Phi_k = {\bfZ_k} / {\bfZ_0}$ denotes the insertion of symmetry defects in the replicated statistical mechanical model derived from the RBIM picture with $2mn+1$ total replicas.

We take the replica limit for the defect free energy in Eq.~\eqref{eqn:dfe_rbim} following a similar strategy as outlined in Sec.~\ref{sec:decoding_fid} for the fidelity.
We use the expectation value of the large twist defect at weak-coupling in Eq.~\eqref{eqn:order_one_twist} and the procedure for taking the replica limit described in Appendix~\ref{app:dfe_int_rep}.
In this way, we find that $\Delta F$ takes a scale-invariant and $\theta$-independent value in the thermodynamic limit 
\begin{align}\label{eqn:dfe_prediction}
    &\Delta F_*(\kappa) := \lim_{L,\, T \to \infty} \Delta F(L, \, T = \kappa L)
    \nn\\
    &=
    \frac{
    \pi \kappa / 4 - \log A
    }
    {
    \sqrt{\pi}
}
\int_{-\infty}^\infty
\rd u \, e^{-u^2} 
\,
\Bigg|
1 + \frac{2u}{\sqrt{\pi \kappa / 4 - \log A}}
\Bigg|
    \,,
\end{align}
Here, $A$ is the non-universal constant which appears in the twist expectation value.

The defect free energy is approximately a linear function in the aspect ratio, up to exponential corrections.
To be precise, $\Delta F_*$ has the following asymptotic series as $\kappa \to \infty$
\begin{align}
    \Delta F_* &\sim \frac{\pi \kappa}{4} - \log A 
    \\
    &+ \frac{e^{- \pi \kappa / 16}}{\sqrt{\pi \kappa / 4 - \log A}}
    \left[
   \frac{4}{\sqrt{\pi}} +\mathcal{O}
    \left(\frac{1}{\pi \kappa /4 - \log A}
    \right)
    \right]
    \nonumber
    \,.
\end{align}

We compare the analytic prediction to numerical simulations in Fig.~\ref{fig:dfe_trends_combined}, with $A$ chosen to produce the closest fit to the numerical data for a particular value of $\theta$.
We find close agreement of $\Delta F_*$ with the data for $\theta = 0.19\pi$.
Our numerical data also demonstrates the trend-reversal in the dependence of $\Delta F$ on the coupling $g_R$ tuned by $\theta$.
Similar to our numerical study of the fidelity, we find that $\Delta F$ increases $\theta$ for $\kappa \lesssim 6$ while it decrease for $\kappa \gtrsim 6$.
The quality of the data is better than for the fidelity due to the self-averaging nature of $\Delta F$.
Finally, numerical study of the quasi-1D regime is presented in Appendix~\ref{app:sec_quasi_1d_dfe} where it again reveals behavior consistent with a logarithmically growing $1/g_R$.

\begin{figure*}[t]
\centering
\includegraphics[width=.95 \textwidth  ]{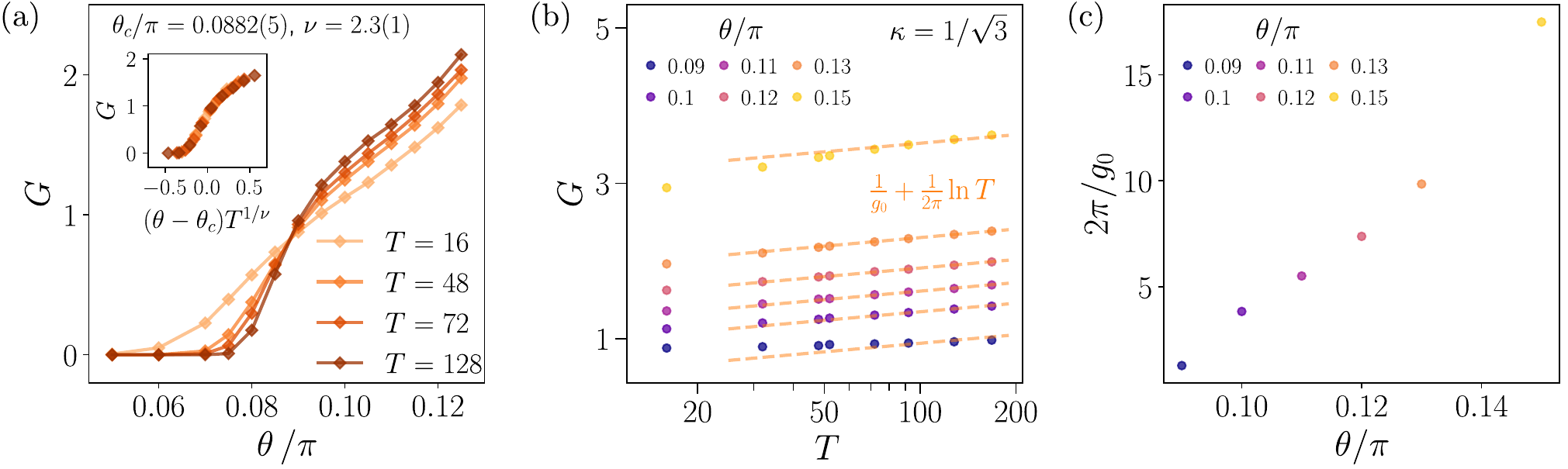}
    \caption{
    Conductivity data for various $\theta$.
    In (a), the conductivity is plotted against $\theta$ for different system sizes in increasing darkness and for fixed $\kappa = 1/\sqrt{3}$.  In the inset, we perform a finite-size scaling collapse near the transition.
    In (b), conductivity $G$ is plotted against system size $T$ in the armchair configuration with fixed aspect ratio $\kappa = 1/\sqrt{3}$.
    For $0.11\pi < \theta < 0.13\pi$, the available system sizes are above both the critical correlation length and mean-free path.
    We observe the $G \sim g_0^{-1} + \left(2 \pi \right)^{-1}\log T$ scaling predicted by the perturbative RG equation.
    Performing a linear fit on a linear-log scale allows us to estimate $g_0$, the bare coupling as a function of $\theta$, which is presented in (c).
    All errorbars are within the marker size, with data averaged over $3000$ to $9000$ samples.
    }
    \label{fig:bare_conductance}
\end{figure*}

\subsection{Conductivity}\label{sec:conductance}
Here, we discuss the conductivity of the Chalker-Coddington network model associated with the replicated RBIM partition function.
The conductivity is directly related to the renormalized coupling $g_R$ of the effective \NLSM.
This enables the numerical study of universal data that are not easily revealed by the decoding fidelity or fidelity correlator, such as the marginally irrelevant flow of $g_R$ away from the weak-coupling fixed point.

We briefly review the network model, referring readers to Appendix~\ref{app:conductance_explicit_forms} for details.
To begin, we write the coset probability $\mathcal{Q}_{\alpha, \, s}$ as 
a transition amplitude in 1+1D of the transfer matrix
\begin{equation}
    \mathcal{Q}_{\alpha, \, s } = |\mathcal{Z}_{\alpha, \, s}|^2 = \bra{\phi_0} 
\hat{\mathsf{T}}_{\alpha, \, s}
    \ket{\phi_0} \,.
\end{equation}
Here, the transfer matrix acts on the Majorana modes both the bra and ket Hilbert spaces while 
$\ket{\phi_0}$ are initial and final states associated with the boundary conditions of the Ising spins.
Since $\hat{\mathsf{T}}_{\alpha, \, s}$ describes free fermion dynamics, it is equivalently characterized by the 
single-particle transfer matrix $\mathsf{T}_{\alpha, \, s}$.

We define the network model by viewing $\mathsf{T}_{\alpha, \, s}$ as describing the scattering of quasi-particles of a disordered superconductor in 2+1D.
First, we label each mode as incoming or outgoing.
Then, $\mathsf{T}_{\alpha, \, s}$ is written as a scattering matrix from incoming to outgoing modes
\begin{equation}
    \mathsf{S}_{\alpha, \, s} = 
    \begin{pmatrix}
        \mathsf{t} & \mathsf{r} \\
        \mathsf{r}' & \mathsf{t}'
    \end{pmatrix}
    \,,
\end{equation}
where we have suppressed the $\alpha,\, s$ indices and introduced the 
matrices $\mathsf{t},\, \mathsf{t}'$ ($\mathsf{r},\, \mathsf{r}'$) describing the transmission (reflection) of energy carried by the quasi-particles.
The scattering matrix is unitary 
and indicates that the energy is conserved.

We note that both bra \emph{and} ket Hilbert spaces are required to formulate a unitary scattering problem, which is the generic case for monitored free fermion dynamics~\cite{jian2022criticality}.
The Hilbert spaces are exchanged by the
anti-unitary particle-hole and time-reversal symmetries, leading to the Class DIII symmetry classification. 
By contrast, when the syndromes reside on a bipartite lattice, the fermion transfer matrix in either the bra \emph{or} the ket Hilbert space by itself defines a unitary scattering problem.
The resulting network model does not exhibit the time-reversal symmetry and belongs to Class D~\cite{venn2022coherent,yan_bao_coherent_error_nlsm}.

We now introduce the longitudinal thermal conductivity, which is the figure of merit characterizing the thermal transport through the disordered network.
The thermal conductivity of the system is obtained from the transmission matrices by Landauer's formula~\footnote{Technically, the Landauer conductivity is defined for the electric conductivity of a network with $U(1)$ conservation.
However, the formula also holds for the thermal conductivity when $\mathsf{T}$ is the single-particle transfer matrix for a Majorana network without $U(1)$ conservation.
This may be verified by introducing a second decoupled copy of the original network as done in Refs.~\cite{cho_fisher,merz_chalker,yan_bao_coherent_error_nlsm}.}
\begin{equation}\label{eqn:conductance_definition}
    G = \kappa \tr
    \frac{1}{\mathsf{T}^\dagger \mathsf{T} + (\mathsf{T}^\dagger \mathsf{T})^{-1} + 2}
    = 
    \frac{\kappa}{2}
    \tr \mathsf{t}^\dagger \mathsf{t}
    \,,
\end{equation}
where the last expression holds with either transmission matrix, $\mathsf{t}'$ or $\mathsf{t}$.
Here, $\kappa$ relates the conductivity to the conductance while the factor of $2$ arises because both bra and ket modes define a $2$-channel network model~\cite{zirnbauer_continuum_CC}.

The conductivity of the network model exhibits distinct scalings in the phases predicted by the \NLSM.
First, in the two gapped phases described by the strong-coupling regime of the \NLSM, the associated network is insulating.
Thus, the conductivity is exponentially small $G \sim e^{- L / \xi_G}$ where $\xi_G$ is the localization length.

Next, in the critical phase the conductivity is proportional to the inverse coupling constant $G = 1/(2 \pi g_R)$~\cite{evers2008anderson,hikami_conductance_normalization}.
As a result, the scaling of the conductivity is given by the marginally irrelevant flow described by the perturbative RG
\begin{equation}\label{eqn:logarithmic_conductance}
    G = G_0 + \frac{1}{2\pi} \log L + \mathcal{O}(1)\,,
\end{equation}
where $G_0$ is the bare conductivity.
The prefactor in front of the logarithm is a universal constant following directly from the RG equation~\eqref{eqn:RG_equation} in the replica limit $n \to 1$.
We numerically examine the predicted conductivity scaling in the critical phase in Fig.~\ref{fig:bare_conductance}(b).
We observe agreement with the logarithmic scaling, including the universal prefactor $(2\pi)^{-1}$, over an extended range of $\theta$.

Finally, the conductivity is expected to be scale free at the decoding transition at an intermediate rotation angle $\theta_c$.
In the vicinity of this point, the conductivity follows a simple scaling
\begin{align}\label{eqn:conductivity_scaling_collapse}
    G = \calG\left(L^{1/\nu}(\theta - \theta_c)\right) 
    \,.
\end{align}
This scaling allows a numerical estimate of the threshold angle $\theta_c$ by identifying the scale-free crossing as shown in Fig.~\ref{fig:bare_conductance}(a).
We find $\theta_c / \pi \approx 0.0882 \pm .0005$, which is consistent with our estimate using the decoding fidelity.
Furthermore, we also estimate the correlation length exponent $\nu$ by performing a finite-size scaling collapse of the data according to~\eqref{eqn:conductivity_scaling_collapse}.
This is presented in the inset of Fig.~\ref{fig:bare_conductance}(a).
Our estimate $\nu = 2.3 \pm 0.1$ is also consistent with that extracted from the fidelity.

We note that the predicted scaling of the conductance in the critical phase enables us to estimate the bare coupling $g_0$ of the \NLSM at a particular rotation angle in numerics.
To do so, we fit the conductance data, holding $\theta$ fixed, to the predicted scaling in Eq.~\eqref{eqn:logarithmic_conductance} with $G_0 = 1 / (2 \pi g_0)$ as a free parameter.
The extracted bare coupling is shown in Fig.~\ref{fig:bare_conductance}(c).
We find that for $\theta > 0.13\pi$, $2 \pi /g_0$ is appreciable and the renormalized stiffness $1/g_R$ is dominated by the bare stiffness over accessible system sizes.
This data further justifies the dropping of corrections to scaling factors throughout the paper.

\section{Discussion}\label{sec:discussion}
In this work, we study certain mixed-state phases that arise from local unitary rotations followed by dephasing.
Viewed as a strongly $\mathbb{Z}_2$-symmetric channel acting on symmetric states, the channel induces a transition from a strongly symmetric phase to one with algebraic SWSSB.
Under Wegner duality, this transition acquires an operational interpretation as a surface-code Pauli decoding threshold separating a decodable and a partially decodable phases.
Finally, we provide a long-distance characterization of the coherence in these quantum channels: the interference between trajectories taken by the $\mathbb{Z}_{2}$ charges (equivalently, the error chains in the surface code) at long distances precludes a description of these channels in the critical phase as stochastic Pauli dephasing when acting on paramagnetic states.  %

We unify these perspectives through an effective $\mathrm{SO}(n)$ nonlinear sigma model (NL$\sigma$M) in the replica limit $n \to 1$.
Our central result is the emergence, at sufficiently large rotation angles, of a critical phase in which the sigma model flows to a weak-coupling fixed point.
The decoding fidelity and fidelity correlator---the primary observables for decoding and SWSSB, respectively---are related to correlations of $\mathbb{Z}_2$-orbifold twist operators in this theory, allowing us to derive their behavior analytically.
In the thermodynamic limit at fixed aspect ratio $\kappa$, the decoding fidelity approaches a universal function $\mathcal{F}_*(\kappa)$ that is independent of $\theta$ but depends nontrivially on $\kappa$.
The fidelity correlator instead decays as a power law, whose universal exponent we determine by analytically taking the replica limit.

\subsection{Suboptimal decoding}\label{sec:suboptimal}

The $\mathrm{SO}(n)$ NL$\sigma$M in the replica limit $n \to 1$ describes optimal Pauli decoding of the surface code.
Practical decoders are generally suboptimal, and their decoding fidelity can be governed by a different effective theory~\cite{yan_bao_coherent_error_nlsm}.

An instructive example is a Bayesian decoder that estimates coset probabilities using an assumed rotation angle $\theta'$ that differs from the true angle $\theta$.
Microscopically, the corresponding field theory is an $\mathrm{SO}(n+1)$ NL$\sigma$M in the replica limit $n \to 0$, perturbed by an external field that breaks the full
$\left(\mathrm{SO}(n+1) \times \mathrm{SO}(n+1)\right)\rtimes \mathbb{Z}_2$
symmetry down to
$\left(\mathrm{SO}(n) \times \mathrm{SO}(n)\right)\rtimes \mathbb{Z}_2$.
This symmetry-breaking perturbation is relevant and reduces the target space in the infrared to $\mathrm{SO}(n)$, yielding an $\mathrm{SO}(n)$ NL$\sigma$M in the replica limit $n \to 0$.

Optimal and suboptimal decoding can, in principle, have substantially different phase diagrams~\cite{yan_bao_coherent_error_nlsm}.
For the triangular-lattice problem, however, the perturbative RG equations predict a stable weak-coupling fixed point in both the $n \to 0$ and $n \to 1$ limits.
The weak-coupling critical phase is therefore supported in both replica limits, although its decoding signatures and the universal properties of the transition differ.

Several observables distinguish the two theories.
First, whereas the optimal decoding fidelity behaves as summarized in Table~\ref{tab:results_table}, the fidelity of the suboptimal decoder is controlled by the symmetry-breaking perturbation and exhibits the asymptotic scaling,
\begin{equation}
    \mathcal{F}_\mathrm{suboptimal} = \frac{1}{2}
    + 
    \exp \left[
        -L |\epsilon| \sqrt{\frac{g_0}{g_R(L)}}
    \right]
    \,,
\end{equation}
where
$\epsilon \sim (\theta' - \theta)/(\pi/4 - \theta')$
characterizes the strength of the perturbation.
Second, the conductivity of the corresponding network model has the logarithmic scaling $G = \beta_n \log L$ in both replica limits, but with distinct universal coefficients:
$\beta_{n \to 0} = 1/\pi$ and $\beta_{n \to 1} = 1/(2\pi)$.
These coefficients follow from the different RG equations in the two limits.
The conductivity associated with both decoders was studied in Ref.~\cite{bejan_projected_ensemble}.

The decoding transitions also have distinct universal data, including different correlation-length exponents $\nu$.
Such transition data have been investigated both in the statistical-mechanical description of surface-code decoding and in other systems governed by the same replica NL$\sigma$M~\cite{yang_coherent_hexagonal,jian2023measurement}.
Finally, the fidelity correlator of the Wegner-dual state probes universal properties intrinsic to the density matrix and is therefore described by the $n \to 1$ theory, independently of the decoder used.

\subsection{Outlook}

The coherent channels considered here generate an extended phase of critical mixed-state order.
Beyond the algebraic fidelity correlator, a natural expectation is that the conditional mutual information also decays as a power law, implying an infinite Markov length~\cite{sang_stability}.
Phases with critical conditional mutual information have also been studied in surface-code decoding with incoherent noise and higher $\mathbb{Z}_N$ symmetry~\cite{information_critical_phases}.
In the complex RBIM description, the conditional mutual information is related to free-energy differences between geometries with rough boundary conditions.
Identifying the corresponding boundary operator in the effective NL$\sigma$M and deriving its implications for the conditional mutual information is left for future work.

More broadly, our results motivate further exploration of mixed-state order and decoding phases induced by coherent noise.
The coherent channel with a uniform rotation angle on the triangular lattice realizes two of the three phases of the Class DIII NL$\sigma$M, while a generalized phase diagram containing all three was reported in Ref.~\cite{yang_coherent_hexagonal}.
It remains open whether other phases in the tenfold classification of non-interacting disordered fermions~\cite{AZ_10fold,evers2008anderson} are relevant for decoding and what their universal properties may imply for decodability.

\begin{acknowledgments}
We thank helpful discussions with Jan Behrends, Matthew P.A. Fisher, Jeongwan Haah, Chaoming Jian, Andreas W.W. Ludwig, Rushikesh A. Patil, Shengqi Sang, and we acknowledge GPT-6 for assistance with proofreading the manuscript.
S.W.Y. and S.V. acknowledge the support of the National Science Foundation under Grant No. DMR–2441671. 
Y.B. is supported in part by grant NSF PHY-2309135 and the Gordon and Betty Moore Foundation Grant No. GBMF7392 to the Kavli Institute for Theoretical Physics (KITP).
This work was performed in part at the Aspen Center for Physics, which is supported by National Science Foundation grant PHY-2210452.
This research was done using services provided by the OSG Consortium~\cite{osg_1,osg_2,osg_3,osg_4}, which is supported by the National Science Foundation awards No. 2030508 and No. 2323298.
\end{acknowledgments}

\appendix

\section{Decoding fidelity at \texorpdfstring{$\theta = \pi/4$}{theta=pi/4}}\label{app:ballistic_metal}

In this appendix, we derive the decoding fidelity for the triangular-lattice surface code on the cylinder at $\theta = \pi/4$.
We consider two types of rough boundary conditions ($e$-condensed boundaries) shown in Fig.~\ref{fig:surface-code-boundaries} and show that the codes with armchair and zig-zag boundary conditions have decoding fidelity $1/2$ and $1$ at $\theta = \pi/4$, respectively.
We note that unlike the square-lattice surface code on the cylinder studied in Ref.~\cite{yan_bao_coherent_error_nlsm}, the fidelity does not depend on $L$ and $T$.

\begin{figure}[t]
    \centering
    \includegraphics[width=\columnwidth]
        {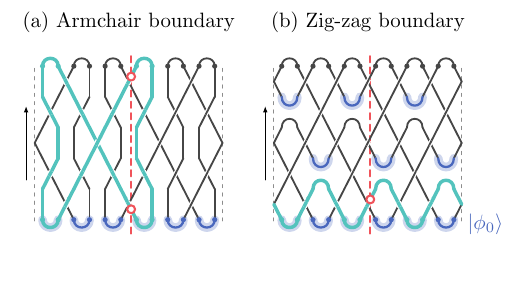}
    \caption{Majorana loop representations for $\calZ_{\alpha,\,s}$ at $\theta=\pi/4$ for (a) armchair and (b) zig-zag boundaries.
        A representative Majorana loop (in teal) intersects the defect (red dashed line) twice in (a) and once in (b).
        Blue highlights mark the Majorana pairs initialized in the definite parity state, described by $|\phi_0\rangle$.
    }
    \label{fig:majorana-loops}
\end{figure}

We start with the partition function $\calZ_{\alpha,\,s}$ of the RBIM on the honeycomb lattice in Eq.~\eqref{eqn:rbim_single_replica_partition_function}, which can be formulated as the transition amplitude generated by free-fermion transfer matrices
\begin{align}
    \calZ_{\alpha,\,s} = \bra{\phi_0} \hat{\mathsf{T}}_{\alpha,\,s}\ket{\phi_0}
\end{align}
where $\ket{\phi_0}$ is the initial and final state upon which the transfer matrix acts.
For the armchair boundaries, $\ket{\phi_0}$ describes the boundary qubits initialized in $\ket{+}$.
For the zig-zag boundaries, the transfer matrix acts not only on the boundary spins, but also on bulk spins initialized in the $\ket{+}$ state.
We formulate the transition amplitude in the fermion picture, in which $\hat{\mathsf{T}}_{\alpha,\,s}$ becomes a Gaussian transfer matrix, and $\ketbra{\phi_0} = \prod_{j} (1+\ri \gamma_{2j-1} \gamma_{2j})/2$.
We illustrate the transition amplitude $\calZ_{\alpha,\,s}$ for both boundary conditions in Fig.~\ref{fig:majorana-loops}.

At $\theta = \pi/4$, the transfer matrix $\hat{\mathsf{T}}_{\alpha,\,s}$ consists of unitary evolution by the swap gates $e^{\pm \pi/4 \gamma_j \gamma_{j+1}}$.
As shown in Fig.~\ref{fig:majorana-loops}, the transition amplitude is given by the product of contributions from Majorana loops, i.e.,
\begin{align}
    \calZ_{\alpha,\,s} = \prod_{\{C_i\}} \bra{\phi_0(C_i)} \mathsf{U}_{\alpha,\,s}(C_i) \ket{\phi_0(C_i)} \,,
\end{align}
where $\{C_i\}$ is a partition of the collection of $2N$ Majorana modes; $\mathsf{U}_{\alpha,\,s}(C_i)$ describes the unitary evolution that the swap gates implement on the Majorana subset.
The contribution of each subset $C_i$ to $\calZ_{\alpha,\, s}$ can be represented graphically as a ``Majorana loop.''
Different homology classes $\alpha = 0,1$ are related by inserting a large defect in the longitudinal direction, which flips the sign of swap gates along the path.

The decoding fidelity at $\theta = \pi/4$ hinges on whether the defect insertion affects the contribution from each Majorana loop.
In the case of armchair boundary conditions, the Majorana loops are contractible on the cylinder, which always intersect the large defect an even number of times.
For this reason, the transition amplitude $\calZ_{\alpha,\,s}$ is the same for $\alpha = 0,1$, giving rise to the decoding fidelity $\calF = 1/2$.
In the case of zig-zag boundary condition, there exist non-contractible Majorana loops on the cylinder which intersect the defect an odd number of times.
Such loops contribute non-vanishingly to the transition amplitude for only one of the two classes $\alpha = 0,1$.
Thus, $\calQ_{\alpha,\,s} = \abs{\calZ_{\alpha,\,s}}^2$ is non-vanishing only for one class, leading to a decoding fidelity $\calF = 1$.

\section{Bulk twist field correlator}\label{app:short_twist}
In this appendix, we calculate the bulk correlations of the \NLSM twist fields.
This analysis was carried out in Ref.~\cite{read_absence_of_metallic}, which we review here.
We assume that the surface code is defined on the infinite plane where there are no homologically non-trivial defects, but our results hold more generally when $\bfx, \bfy$ are far away from the boundaries.

First, at the weak-coupling fixed point, the \NLSM matrix field can be written as $n(n+1)/2$ non-interacting compact bosons, with the twist affecting $k(n+1-k)$ of them.
Thus, the twist correlator factors into $k(n+1-k)$ copies of a $\mathbb{Z}_2$-orbifold twist field correlator for compact bosons which is known in the literature~\cite{dixon_orbifold_cft}, leading to
\begin{equation}\label{app:eqn_twist_field_correlator}
    \langle \phi_{k}(\bfx) \phi_{k}(\bfy) \rangle \sim \frac{1}{|\bfx-\bfy|^{k(n+1-k) / 4}}
    \,.
\end{equation}

On the other hand, close to the fixed point $g = 0$, the correlator can pick up multiplicative logarithmic corrections due to the marginally irrelevant flow of $g_R$.
This is dictated by the Callan-Symanzik equation~\cite{peskin_schroeder,brezin_callan_symanzik}
\begin{equation}\label{app:eqn_callan_symanzik}
    \left[
        \mu \partial_\mu + 2 \Delta(g) + \beta(g) \partial_g
    \right] \langle \phi_{k}(\bfx) \phi_{k}(\bfy) \rangle = 0
    \,,
\end{equation}
where $\Delta(g)$ is the twist field scaling dimension expanded perturbatively in the coupling 
\begin{equation}
    \Delta(g) = \sum_{k=0}^\infty \Delta_k g^k \,,
\end{equation}
with $\Delta_0 = k(n+1-k) / 8$.
In the simplest setting, we assume that $\Delta_1$ is non-vanishing.
Additionally, in writing down Eq.~\eqref{app:eqn_callan_symanzik}, we have assumed that $g$ is the least irrelevant coupling to which the twist couples.
In that case, we may integrate Eq.~\eqref{app:eqn_callan_symanzik} up to scale $\mu = |\bfx- \bfy|$ and obtain, asymptotically at large separation $|\bfx-\bfy| \to \infty$ the twist field correlator
\begin{equation}
    \langle \phi_{k}(\bfx) \phi_{k}(\bfy) \rangle \sim  
    \frac{\left(
        \ln |\bfx- \bfy|
    \right)^{2 \Delta_1 }}{|\bfx-\bfy|^{k(n+1-k)/4}}
    \,.\label{app:eqn_renyi_twist_corr}
\end{equation}
Note that $\Delta_1$ is, in general, a function of $n, k$ which cannot be fixed without explicitly computing the
twist-field scaling dimension perturbatively at finite coupling.

\section{Nachbin resummation and replica limit}\label{app:resummation}
In this appendix, we detail the procedures to obtain the scaling of both the fidelity and defect free energy in the replica limit.
We use the expectation value of the twist associated with a large defect in the weak coupling regime in Eq.~\eqref{eqn:order_one_twist}.

Our analysis makes use of the fact that the twist takes the form
\begin{equation}
    \Phi_k  = f^{k(n+1-k)} \,,
\end{equation}
where $f$ does not explicitly depend on $n, k$.
In our case, $f = A \exp(-\pi \kappa / 4)$.
In the rest of this appendix, we use $f$ to denote the expectation value of elementary twist to simplify the equations.

\subsection{Integral representation of the fidelity}\label{app:sec_int_rep_fid}
We begin by detailing the replica trick for the fidelity.
The fidelity takes the form
\begin{equation}\label{eqn:app_replicated_fidelity_def}
    \mathcal{F} = \lim_{n\to0}
    \frac{
    2
    \sum_{k=0}^{n-1}
    \binom{n-1}{k}
    f^{k(n+1-k)}
    \Phi_{k}
    }
    {
    \sum_{k=0}^{n+1}
    \binom{n+1}{k}
    f^{k(n+1-k)}
    }
    \,.
\end{equation}
In order to take the replica limit, we must remove the $n$ dependence in the limits of summation.
Thus, we use the following analytic continuation for the binomial coefficient
\begin{equation}\label{eqn:binomial_continuation}
    \binom{n}{k} = \frac{\Gamma(n+1)}{\Gamma(k+1) \Gamma(n-k+1)}
    \,,
\end{equation}
and extend the upper limits of summation in both the numerator and denominator of~\eqref{eqn:app_replicated_fidelity_def} to $k = \infty$.
This analytic continuation reproduces the expressions for finite replica number $n \in \mathbb{Z}_{\geq 1}$ since $\Gamma(x)$ has simple poles at the non-positive integers which terminates the sum at finite $k$.

To motivate the next step, observe that if we take the $n \to 0$ limit, the denominator of~\eqref{eqn:app_replicated_fidelity_def} is identically $2$ and the fidelity takes the simpler form
\begin{equation}\label{eqn:app_rbim_divergent_series}
    \mathcal{F} = 
    \sum_{k=0}^{\infty}
    (-1)^k
   f^{k(1-k)}
    \,.
\end{equation}
The resulting sum bears a formal resemblance to the ``partial theta functions'' appearing in the study of mock modular forms~\cite{andrews_partial_1,andrews_partial_9,andrews_lostnotebook_book_1}
\begin{equation}
    \Theta_p(\omega,\, q) = \sum_{k=0}^\infty \omega^k q^{k^2} \,,
\end{equation}
defined for nome $|q| < 1$.
However, in our case $f < 1$ such that the fidelity~\eqref{eqn:app_rbim_divergent_series} is formally a highly divergent series.
Worse, $\Theta_p$ has a lacunary boundary at $|q|=1$ so there is no analytic continuation in the usual sense~\cite{krantz_handbook}.
Nevertheless, we prescribe a finite value to~\eqref{eqn:app_rbim_divergent_series} using Nachbin resummation, which generalizes the standard Borel resummation to divergent series which grow faster than $k!$ by using the Mellin transform instead of the Laplace transform.

To this end, we are interested in resumming divergent series of the form
\begin{equation}
    \Sigma_N(\omega,\, q) := \sum_{k=0}^\infty
    \binom{N}{k}  \, \omega^k q^{k^2} \,,
\end{equation}
where $\binom{N}{k}$ denotes the analytic continuation~\eqref{eqn:binomial_continuation} to $N \in \mathbb{C}$.
It will be important that $q, \omega \in \mathbb{R}_{>0}$ and $q > 1$.

In our case, the problematic term, which diverges as $q^{k^2}$, is Gaussian for which the Mellin transform coincides with the ordinary Gaussian integral
\begin{equation}
    e^{\log(q) k^2} = \frac{1}{\sqrt{4 \pi \log(q)}} \int_{-\infty}^\infty
    \rd u \, e^{uk} e^{
    - u^2/4 \log(q)
    } \,.
\end{equation}
We obtain the following integral representation of the resummed series
\begin{align}
    \Sigma_N(\omega,\, q) &= \frac{1}{\sqrt{4 \pi \log(q)}}\int_{-\infty}^\infty \rd u \,
    \sum_{k=0}^\infty
    \binom{N}{k} \left(\omega e^{u}\right)^k e^{-\frac{u^2}{4 \log(q)}}
    \\
    &=
     \frac{1}{\sqrt{4 \pi \log( q)}}
     \int_{-\infty}^\infty \rd u \, 
     \left(1 + \omega e^{u}\right)^N
     e^{-\frac{u^2}{4 \log(q)}}
     \,.
\end{align}
In the second line, we analytically continue the binomial series outside its formal radius of convergence.
This is reasonable as $\omega e^{u} \neq -1$.

Therefore, we obtain the replicated fidelity
\begin{align}
    \mathcal{F}^{(n)} &= 2 \frac{
    \Sigma_{n-1}\left(f^{n+1}, \, f^{-1}\right)
    }
    {
    \Sigma_{n+1}\left(f^{n+1}, \, f^{-1}\right)
    }
    \\
    &=
    2
    \frac{
    \int_{-\infty}^\infty
    \rd u \, \left(1 + f^{n+1} e^u\right)^{n-1} e^{-u^2/4 \log (1/f)}
    }
    {
    \int_{-\infty}^\infty
    \rd u \, \left(1 + f^{n+1} e^u\right)^{n+1} e^{-u^2/4 \log (1/f)}
    }
    \,.
\end{align}
Taking the replica $n \to 0$ limit, we obtain the fidelity
\begin{equation}\label{eqn:app_rbim_RL_fid}
    \mathcal{F} = \frac{1}{\sqrt{\pi}}
    \int_{-\infty}^\infty \rd u \, \frac{
    e^{-u^2}
    }
    {
    1 + f \,  e^{2 u \sqrt{\log (1/f)}}
    }
    \,.
\end{equation}
Substituting $A \exp (-\pi \kappa/4)$ for $f$ reproduces Eq.~\eqref{eqn:order_one_AR_fid_integral_rep}.

We comment that Eq.~\eqref{eqn:app_rbim_RL_fid} is consistent with Eq.~\eqref{eqn:defintion_of_bayesian_fidelity} if one assumes that $\log (\mathcal{Q}_{0, \, s}/\mathcal{Q}_{1, \, s})$ is normally distributed with variance $\pi \kappa / 2 - 2\log (A)$ and mean $\pi \kappa/4 - \log(A)$.
This is similar to the distribution of disorder operator correlations in Ref.~\cite{read2000absence}, but the non-vanishing mean here is a consequence of the $n \to 1$ limit.

\subsection{Integral representation of the defect free energy}\label{app:dfe_int_rep}
Next, we discuss taking the replica limit for the defect free energy.
Using Eq.~\eqref{eqn:dfe_rbim} and the weak-coupling expectation value of the large twist in Eq.~\eqref{eqn:order_one_twist}, the defect free energy takes the form
\begin{equation}
    \Delta F = 
    \lim_{\substack{m \to 0\\ n \to 1 /2}}
    \frac{2}{m^{2n}}
    \frac{
    \sum_{k=0}^{2n}
    \binom{2n}{k}
    (-1)^k
    f^{(mk+1)(2mn-mk)}
    }
    {
    \sum_{k=0}^{2nm+1} \binom{2nm+1}{k} f^{k(2mn+1-k)}
    }
    \,,
\end{equation}
where we have used that the twists are evaluated in the $\mathrm{SO}(2mn+1)$ \NLSM.
Next, we follow a similar analytic continuation as done in Appendix~\ref{app:sec_int_rep_fid}.
The defect free energy then takes the form
\begin{widetext}
\begin{align}
    \Delta F 
    &= 
        \lim_{\substack{m \to 0\\ n \to 1 /2}}
    \frac{2 f^{2mn}}{m^{2n}}
    \frac{
    \Sigma_{2n}(-f^{2m^2n-m}, \, f^{-m^2})
    }
    {
    \Sigma_{2nm+1}(f^{n+1} ,\,  f^{-1})
    }
    \\
    &= 
        \lim_{\substack{m \to 0\\ n \to 1 /2}}
    \frac{2 f^{2mn}}{m^{2n}}
    \frac{
    \int_{-\infty}^\infty \rd u \,
    (1 - f^{2m^2n - m}e^u)^{2n} e^{-u^2/4m^2 \log(1/f)}
    }
    {
    \int_{-\infty}^\infty \rd u \,
    (1 + f^{n+1} e^{u})^{2mn+1} e^{-u^2 / 4 \log (1/f)}
    }
    \,.
\end{align}

First, note that in the limit $m \to 0, n\to 1/2$, the denominator yields $2 \sqrt{4 \pi \log (1/f)}$.  Consequently, we have
\begin{align}
    \Delta F &=  
    \lim_{\substack{m \to 0\\ n \to 1 /2}}
    \frac{f^{2mn}}{m^{2n} \sqrt{\pi}}
    \int_{-\infty}^\infty
    \rd u \,
    \left[
    1 - f^{2m^2n-m} e^{2 u m \sqrt{\log (1/f)}}
    \right]^{2n}
    e^{-u^2}
    \\
    &=
    \lim_{\substack{m \to 0\\ n \to 1 /2}}
    \frac{1}{m^{2n} \sqrt{\pi}}
    \int_{-\infty}^\infty
    \rd u \,
    \left[
    m \left( \log(1/f) + 2 u \sqrt{\log(1/f)}\right) + \mathcal{O}(m^2)
    \right]^{2n}
    e^{-u^2}
    \\
    &= \lim_{n \to 1/2}
    \frac{1}{\sqrt{\pi}}
        \int_{-\infty}^\infty
    \rd u \,
\left( \log(1/f) + 2 u \sqrt{\log(1/f)}\right)^{2n}
e^{-u^2}
\\
&= \frac{1}{\sqrt{\pi}}
        \int_{-\infty}^\infty
    \rd u \,
\left| \log(1/f) + 2 u \sqrt{\log(1/f)}\right|
e^{-u^2}
\label{eqn:app_final_dfe_int_rep}
\,.
\end{align}
\end{widetext}
In the second line, we expanded in powers of $m$ allowing us to take $m \to 0$ in the third line.  In the last line, we used $\lim_{n \to 1/2} (\cdot)^{2n} = |\cdot|$~\cite{kurchan_norm_replica,calabrese_norm_replica}.
Finally, we obtain Eq.~\eqref{eqn:dfe_prediction} after replacing $f$ with $A \exp(- \pi \kappa /4)$.

Again, we remark that this result is consistent with Eq.~\eqref{eqn:def_dfe} if one assumes that $\log(\mathcal{Q}_{0, \, s}/\mathcal{Q}_{1, \, s})$ is normally distributed with a mean and variance which both grow linearly in $\kappa$.

\section{Asymptotic expansions for the integral formulas}\label{app:asymptotic}
In this appendix, we derive asymptotic expansions starting from the integral representations of the fidelity and defect free energy from App.~\ref{app:resummation}.
\subsection{Asymptotic expansion for the fidelity}\label{app:fid_asymptotic}
We derive an asymptotic expansion for the decoding fidelity in Eq.~\eqref{eqn:app_rbim_RL_fid} as $f \to 0$ corresponding to $\kappa \to \infty$.
First, observe that the integrand takes the form of a Gaussian integrated against a function which approaches the Heaviside step function $H(-u+\sqrt{\log (1/f)}/2)$ in the limit $f\to 0$.

As a result, we may expand the integral in terms of the step function
\begin{widetext}
\begin{align}
    \mathcal{F} 
    &=
    \frac{1}{\sqrt{\pi}}
     \int_{-\infty}^\infty
    \rd u \,
    \frac{
    e^{-\left(u + \sqrt{\log(1/f)}/2\right)^2}
    }
    {
    1 + e^{2 u \sqrt{\log(1/f)}}
    }
    \\
    &= 
    \frac{1}{\sqrt{\pi}}
    \int_{-\infty}^\infty
    \rd u \,
    H(-u) \, e^{-\left(u + \sqrt{\log(1/f)}/2\right)^2}
    +
        \frac{1}{\sqrt{\pi}}
    \int_{-\infty}^\infty
    \rd u \,
       \left[
       \frac{1}{1 + e^{2u \sqrt{\log(1/f)}}}
       -H(-u)
       \right] e^{-\left(u + \sqrt{\log(1/f)}/2\right)^2}
       \\
       &= 1 - \frac{1}{2} \mathrm{Erfc}\left(
       \frac{\sqrt{\log(1/f)}}{2}
       \right)
       + \frac{1}{\sqrt{\pi}}
        \int_{0}^\infty
    \rd x \,
    \frac{
    e^{-\left(u + \sqrt{\log(1/f)}/2\right)^2}
    -
        e^{-\left(u - \sqrt{\log(1/f)}/2\right)^2}
    }{1 + e^{2u \sqrt{\log(1/f)}}}
    \\
    &=
       1 - \frac{1}{2} \mathrm{Erfc}\left(
       \frac{\sqrt{\log(1/f)}}{2}
       \right)
       + \frac{f^{1/4}}{\sqrt{\pi\log(1/f)}}
        \int_{0}^\infty
    \rd z \,
        e^{-z^2/\log(1/f)} e^{-z}
        \tanh (z)
        \label{eqn:app_final_fidelity_expression_asymptotic}
        \,,
\end{align}
\end{widetext}
where $\mathrm{Erfc}(z)$ is the complementary error function
\begin{equation}
    \mathrm{Erfc}(z) = \frac{2}{\sqrt{\pi}} \int_z^\infty \rd x \, e^{-x^2}
    \,.
\end{equation}
The complementary error function has the following asymptotic expansion for $z \to \infty$~\cite{erfc_asymptotics}
\begin{equation}\label{eqn:app_erfc_asymptotic}
    \mathrm{Erfc}(z) \sim \frac{e^{-z^2}}{z \sqrt{\pi}}
    \sum_{k=0}^\infty \frac{(-1)^k}{k!} \frac{(2k)!}{(4z^2)^k} \,.
\end{equation}
Using the integral representation of $(2k)!$ 
\begin{equation}
    (2k)! = \int_0^\infty \rd x \, x^{2k} e^{-x} \,,
\end{equation}
Taylor expanding the Gaussian in the integrand of Eq.~\eqref{eqn:app_final_fidelity_expression_asymptotic}, and combining like terms we obtain
\begin{align}
    \mathcal{F} &\sim 
    1 - \frac{2f^{1/4}}{\sqrt{\pi \log(1/f)}}
    \\
    &
    \quad\quad
    \times \sum_{k=0}^\infty
    \frac{(-1)^k }{k! \left(\log(1/f)\right)^k}
    \int_0^\infty \rd z \, z^{2k} \frac{e^{-z}}{1 + e^{-2z}} 
    \nonumber
    \\
    &=
    1 - \frac{2f^{1/4}}{\sqrt{\pi \log(1/f)}}
    \sum_{k=0}^\infty
    \frac{(-1)^k (2k)! \beta(2k+1)}{k! \left(\log(1/f)\right)^k}
    \\
    &= 1 - \frac{\sqrt{\pi} f^{1/4}}{2 \sqrt{\log(1/f)}} + \mathcal{O}\left(
    \frac{f^{1/4}}{\left(\log(1/f)\right)^{3/2}}
    \right)
    \,.
\end{align}
In the second line, the integral is expressed in terms of the Dirichlet beta function $\beta(z)$ while in the third line, we used $\beta(1) = \pi/4$.

\subsection{Asymptotic expansion for the defect free energy}\label{app:dfe_asymptotic}
We now derive an asymptotic expansion for the defect free energy in Eq.~\eqref{eqn:app_final_dfe_int_rep}.
The defect free energy takes the form
\begin{widetext}
\begin{align}
    \Delta F &=
    2 \sqrt{\frac{\log (1/f)}{\pi}}
    \int_{-\infty}^\infty \rd u \, \Bigg|\sqrt{\log (1/f)}/2 + u\Bigg|\, e^{-u^2} 
    \\
    &=2 \sqrt{\frac{\log (1/f)}{\pi}}
    \int_{-\infty}^\infty \rd u \, |u| \, e^{-\left(u - \sqrt{\log(1/f)}/2\right)^2} 
    \\
    &=
    4 \sqrt{\frac{\log (1/f)}{\pi}} \int_{\sqrt{\log(1/f)}/2}^\infty
    \rd x \, e^{-x^2} + \frac{\log(1/f)}{\sqrt{\pi}}
    \int_{-\sqrt{\log(1/f)}/2}^{\sqrt{\log(1/f)}/2} \rd x \, e^{-x^2}
    \\
    &=
    2 \sqrt{\frac{\log(1/f)}{\pi}} e^{- \log(1/f)/4}
    + \log(1/f) - \log(1/f) \mathrm{Erfc}\left(\sqrt{\log(1/f)}/2
    \right)
    \\
    &= \log (1/f) - 2 f^{1/4} \sqrt{\frac{\log(1/f)}{\pi}}
    \sum_{k=1}^\infty \frac{(-1)^k}{k!} \frac{(2k)!}{\left(\log(1/f)\right)^k}
    \\
    &=\log (1/f) +  \frac{4 f^{1/4}}{\sqrt{\pi\log(1/f)}}
    +
    \mathcal{O}\left(
    \frac{f^{1/4}}{\left(\log(1/f)\right)^{3/2}}
    \right)
    \,,
\end{align}
\end{widetext}
where in the second to last line we used the asymptotic expansion for $\mathrm{Erfc}(z)$ in Eq.~\eqref{eqn:app_erfc_asymptotic}.

\section{Quasi-1D Regime}\label{app:sec_quasi_1d}
In this appendix we discuss the decoding fidelity in the quasi-1D regimes $\kappa \gg 1$ and $\kappa \ll 1$.
We estimate the expectation value of the large twist for the $\kappa \gg 1$ case shortly. 
For the $\kappa \ll 1$ regime, it is convenient to do the calculation using the \NLSM derived from the statistical mechanical model which is Kramers-Wannier dual to the RBIM.
This is done in Appendix~\ref{app:sec_kw_dual}.
Then, we obtain qualitative predictions for the decoding fidelity and defect free energy in Appendices~\ref{app:sec_quasi_1d_fid} and~\ref{app:sec_quasi_1d_dfe}, respectively.

For the quasi-1D regime $\kappa \gg 1$, we may coarse-grain up to scale $L$ at which point the \NLSM becomes effectively 1D and the RG is qualitatively very different.
Specifically, the 1D \NLSM is disordered, with $1/g_R(L)$ serving as a correlation length.
We are interested in the case that we are deep in the disordered regime, i.e., $\kappa \gg 1/g_R$.
Then, the partition function can be approximated by summing over decoupled block spins on the scale of the correlation length $1/g_R$.

Furthermore, the twist acts as a local potential which is relevant under RG and pins the spins to be invariant under the twist action.
Thus, the twist expectation value is exponentially decaying in the number of block spins, schematically
\begin{equation}\label{eqn:heuristic_twist_exp_volume_facotr}
    \Phi_{k} \sim \left(\frac{2\mathrm{Vol} \left(\SO(n-k) \times \SO(k)\right)}{\mathrm{Vol} \left(\SO(n)\right)}\right)^{\kappa g_R}
    \,.
\end{equation}
This is non-perturbatively small in $g_R \kappa$ so obtaining the correct behavior in the replica limit is subtle~\cite{verbaarschot_critique,zirnbauer_another_critique,lerner_non_perturbative}.\footnote{
One can attempt to perform the analytic continuation.
The volume factor appearing in~\eqref{eqn:heuristic_twist_exp_volume_facotr} has been computed in~\cite{macdonald_volume, abe_yokota_symmetric_space_volume} and an analytic continuation based on the Barnes $G$-function was given in Appendix F of Ref.~\cite{yan_bao_coherent_error_nlsm}.
However, one finds that the resulting volume factor is proportional to $G(1-k/2)G(1/2-k/2)$ in the limit $n \to 0$ and thus vanishes for $k \geq 1$.
As a result, $\mathcal{F} = 1$ ($1/2$) for $\kappa \gg 1$ ($1/\kappa \gg 1$) \textit{with no dependence on} $\kappa$ or $g_R$. This clearly fails to capture the correct non-perturbative corrections.}
To this end, we do not take the replica limit and instead focus on the qualitative behavior obtained for a finite number of replicas $n \geq 1$.
In that case, the twists are expected to scale as 
\begin{equation}\label{app:rbim_quasi_1d_twist}
    \Phi_k \sim e^{- \alpha(n+1,\, k) \kappa g_R} \,, 
\end{equation}
where $\alpha(n+1,\, k)$ is some function satisfying $\alpha(n+1,\, k) = \alpha(n+1,\, n+1-k)$, $\alpha(n+1,\, 0)=0$ and $\alpha(n+1,\, k) > 0$ for $0 \leq k \leq n+1$.

\subsection{Kramers-Wannier Dual Description of Decoding}\label{app:sec_kw_dual}
For the quasi-1D regime $\kappa \ll 1$, it is convenient to formulate the fidelity and defect free energy as a replica limit in the Kramers-Wannier dual of the replicated statistical mechanical model as done in Ref.~\cite{yan_bao_coherent_error_nlsm}.

The Kramers-Wannier dual replica theory can be written as
\begin{equation}
    \tilde{\bfZ}_0 = \sum_{s} | \mathcal{Z}_{+,\, s}|^{2n+2} \,,
\end{equation}
with the complex partition sum for the dual spins
\begin{equation}
    \mathcal{Z}_{\pm,\, s} = \mathcal{Z}_{0,\, s} \pm \mathcal{Z}_{1,\, s} \,.
\end{equation}
Importantly, the Kramers-Wannier duality requires a sum over both homology classes for each replica independently corresponding to twisted boundary conditions~\cite{aasen_mong_fendley_ising,yan_bao_coherent_error_nlsm}.

We also introduce the dual partition function in the presence of symmetry defect insertions. 
In the dual picture, the symmetry defect flips the sign of the coupling constant along a defect in the compact direction of the cylinder 
\begin{equation}
    \tilde{\bfZ}_{k, \, l} = \sum_s 
    \mathcal{Z}_{+, \, s}^{n+1-k}
    \mathcal{Z}_{-, \,  s}^k
    {\mathcal{Z}^{*}_{+, \,  s}}^{\! \!n+1-l}
    {\mathcal{Z}^{*}_{-, \,  s}}^{\!\! l}
    \,,
\end{equation}
where $k$ defects are inserted in the ``bra'' copies and $l$ defects in ``ket'' copies, with $k \neq l$ in general ().

\begin{figure}[t]

\begin{tikzpicture}[baseline={(lab.base)}, scale=0.67]
  \def\h{3}    %
  \def\r{2}  %
  \draw (0, \h) ellipse (2 and 0.5);
  \draw (0,0) [partial ellipse=180:360:2 and 0.5];
  \draw[dashed] (0,0) [partial ellipse=0:180:2 and 0.5];
  \draw (-\r,0) -- (-\r,\h);
  \draw (\r,0) -- (\r,\h);
  \draw[red,decorate, decoration={snake, amplitude=0.5mm, segment length=2mm}] (0,2) [partial ellipse=180:360:2 and 0.5];
  \draw[dashed,red,decorate, decoration={snake, amplitude=0.5mm, segment length=2mm}] (0,2) [partial ellipse=0:180:2 and 0.5];
  \draw[-Stealth,red] (0,0.2+\h/2) node[above]{$O$} -- (0,-0.4+\h/2) node[below]{$\Lambda_k O \Lambda_l$};
  \node[red] at (-1.45,1.15) {$\tilde{\Phi}_{k, \, l}$};   %
  \draw[-Stealth] (-2.3,0) -- (-2.3,2) node[left] {$T$};
  \draw[-Stealth] (0,0) [partial ellipse=210:330:2.3 and 0.7] node[right] {$L$};
\end{tikzpicture}
\caption{The $\mathbb{Z}_2$-orbifold twist field of the NL$\sigma$M computed in the Kramers-Wannier dual picture.
The twist expectation value $\tilde{\Phi}_{k, \, l}$ is associated with a large defect in the compact direction of the cylinder across which the field satisfies the twisted boundary condition $O \mapsto \Lambda_k O \Lambda_l$.}\label{app:fig_dual_defect}
\end{figure}
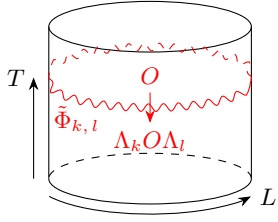

In this way, the fidelity can be expressed as the $n \to 0$ limit of a replica sequence $\tilde{\mathcal{F}}^{(n)}$ where
\begin{equation}\label{app:dual_replicated_fid_as_defects}
    \tilde{\mathcal{F}}^{(n)} = \frac{1}{2} + \frac{1}{2}
    \frac{
    \sum_{k=0}^{n-1} 
    \binom{n-1}{k}
    \tilde{\Phi}_{k+2,\, k}+
    2\tilde{\Phi}_{k+1,\, k+1}
    +
    \tilde{\Phi}_{k,\, k+2}
    }
    {
    \sum_{k=0}^{n+1}\binom{n+1}{k}
    \tilde{\Phi}_{k,\, k}
    }
    \,,
\end{equation}
where we have introduced the ratio of partition functions with defect insertion
\begin{equation}\label{eqn:dual_defect_definition}
    \tilde{\Phi}_{k, \, l} := \tilde{\bfZ}_{k,\, l} / \tilde{\bfZ}_0 \,.
\end{equation}
These objects can correspondingly be viewed as expectation values of operators which insert defects in the transverse direction.
The two indices $k$ and $l$ label the number of defects inserted in the $n+1$ copies of the partition function $\calZ_{+,\,s}$ and $\calZ^*_{+,\,s}$, respectively (see Fig.~\ref{app:fig_dual_defect}).

Similarly, the defect free energy can be expressed as the $m\to 0,\, n \to 1/2$ limit of the double replica sequence 
\begin{equation}\label{app:dfe_dual}
   \Delta F =  
   \lim_{\substack{m \to 0\\ n \to 1 /2}}
        \frac{
        \sum_{l,\, l' = 0}^{2nm+1}
        C(l,\, l';\, m,\, n) \, \tilde{\Phi}_{l,\, l'}
        }{
                m^{2n}
        2^{2nm}
        \sum_{l=0}^{2nm+1}
        \binom{2nm+1}{l}
        \tilde{\Phi}_{l,\, l}
        }
        \,,
\end{equation}
where the coefficients $C(l,\, l';\, m,\, n)$ are defined by
\begin{widetext}
\begin{equation}\label{app:dfe_dual_coefficients}
        C(l,\, l';\, m,\, n) = 
        \begin{cases}
        \sum_{q,\,q'=0}^{l,\, l'}
        \sum_{k=0}^{2n}
        (-1)^{q+q'+k}
        \binom{2n}{k}
        \binom{mk+1}{l-q}
        \binom{mk+1}{l'-q'}
        \binom{2nm-km}{q}
        \binom{2nm-km}{q'} &
        l + l' \equiv 0\mod 2
        \\
        0 & l+l' \equiv 1 \mod 2
        \end{cases}
    \,.
    \end{equation}
\end{widetext}

Next, following Ref.~\cite{yan_bao_coherent_error_nlsm}, the Kramers-Wannier dual replica theory is also governed by an effective $\mathrm{SO}(n+1)$ \NLSM, while the symmetry defect insertions $\tilde{\Phi}_{k, \, l}$ correspond to twist defects inserted in the transverse direction of the cylinder as depicted in Fig.~\ref{app:fig_dual_defect}.
The twist acts on the \NLSM order parameter field by imposing the modified boundary condition $O \mapsto \Lambda_k O \Lambda_l$ across the defect, where $\Lambda_k$ is the diagonal matrix with $k$ entries $-1$ and $n+1-k$ entries $+1$.

To evaluate the twist expectation values in the quasi-1D regime $\kappa \ll 1$, we coarse grain up to scale $T$, where the \NLSM is again disordered with correlation length $1/g_R(T)$.  
In the geometry where $\kappa \ll 1$, the twist in the transverse direction again acts as a local potential.
Similar arguments as above lead to the following estimate of the twists with $k = l$
\begin{equation}\label{app:dual_quasi_1d_twist}
        \tilde{\Phi}_{k,\, k} \sim e^{- \alpha(n+1,\, k) g_R / \kappa} \,.
\end{equation}

On the other hand, there are no elements of the target space that are invariant under the twist in the case that $k \neq l$.
For this reason $\tilde{\Phi}_{k, \, l}$ is suppressed, and we drop its contribution in the following.

\begin{figure*}[t]
\centering
\includegraphics[width=\textwidth]{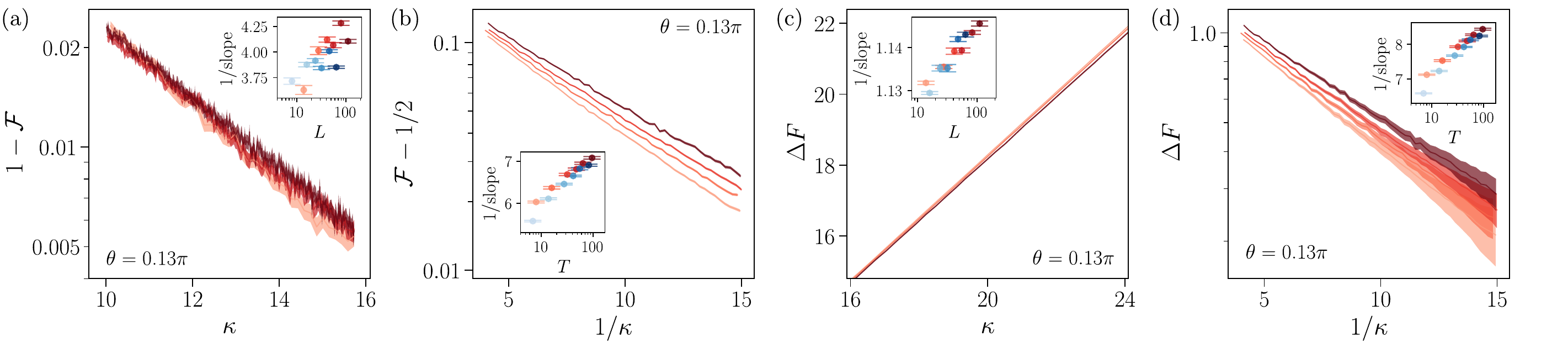}
    \caption{Fidelity and defect free energy in the quasi-1D regimes for fixed $\theta = 0.13\pi$.
    In (a) and (b), fidelity against $\kappa$ for $\kappa \gg 1$ and $\kappa \ll 1$, respectively.
    In (b) and (c), defect free energy against $\kappa$ for $\kappa \gg 1$ and $\kappa \ll 1$, respectively.
    The data is is linear on a log-linear scale for (a), (b), (d) and linear on a linear scale for (c).
    The inverse slope of the linear best fit, which is proportional to the stiffness, is presented in the inset against the system size.
    For clarity, only the data for the armchair boundary condition is shown in the main figures, while data in the inset is for both boundary conditions.
    Data averaged over $10000$ to $280000$ samples, with error indicated by the line width.}
    \label{app:fig_quasi_1d}
\end{figure*}
\subsection{Decoding Fidelity}\label{app:sec_quasi_1d_fid}
Here, we discuss the qualitative scaling of the decoding fidelity at finite replica number.

First, in the quasi-1D regime $\kappa \gg 1$, we use Eq.~\eqref{eqn:rbim_replicated_fid_as_defects} in the replica $n \to 1$ limit.
Using the estimated twist expectation value in Eq.~\eqref{app:rbim_quasi_1d_twist}, we have
\begin{equation}\label{eqn:large_kappa_fid_scaling}
     \mathcal{F} = 1 - e^{- \alpha \kappa g_R}  + \mathcal{O}(e^{- 2 \alpha \kappa g_R}) \,,
\end{equation}
where we suppress the arguments of $\alpha := \alpha(2, 1)$ for simplicity.

Similarly, in the regime $\kappa \ll 1$, an analogous calculation is done in the $n \to 1$ limit of Eq.~\eqref{app:dual_replicated_fid_as_defects} using Eq.~\eqref{app:dual_quasi_1d_twist}
\begin{equation}\label{eqn:small_kappa_fid_scaling}
    \mathcal{F} = \frac{1}{2} + \frac{1}{2} e^{- \alpha g_R / \kappa} + \mathcal{O}(e^{- 2 \alpha g_R / \kappa}) \,.
\end{equation}

We examine the qualitative features of the predicted scaling in numerical simulations.
First, in Fig.~\ref{app:fig_quasi_1d}(a), $1-\mathcal{F}$ is plotted against $\kappa$ on a log-linear scale for $\kappa \gg 1$.
The data is expected to be linear on this scale, with the slope proportional to $g_R(L)$.
Thus, we perform a best fit of the data on this scale, with the inverse slope plotted in the inset as a function of overall scale $L$.
The inverse slope is proportional to $1/g_R \sim \log L$, in agreement with the observed data.
Similarly, in Fig.~\ref{app:fig_quasi_1d}(b), $\mathcal{F} - 1/2$ is plotted against $\kappa$ for $\kappa \ll 1$.
The data is again expected to be linear on a log-linear scale, with the inverse of the slope proportional to $1/g_R(T) \sim \log T$.
This is plotted in the inset to Fig.~\ref{app:fig_quasi_1d}(b), which is consistent with the predicted scaling.

\subsection{Defect Free Energy}\label{app:sec_quasi_1d_dfe}
Next, we discuss the qualitative scaling of the defect free energy.

First, in the regime $\kappa \gg 1$, the twists are expected to scale as $\Phi_k \sim e^{- \alpha(R,\, k) \kappa g_R}$ where $R = 2mn+1$ is the total number of replicas.
We use the expression for $\Delta F$ in terms of twists in Eq.~\eqref{eqn:dfe_rbim}.
Observe that in the relevant replica limits, the denominator evaluates to $2$.
On the other hand, the numerator is $\Delta^{2n}_r \Phi_r |_{r=mn+1}$, where $\Delta^{2n}_r$ is the discretized $2n$-th derivative with respect to $r$ and step size $m$.
In the limit $m \to 0$, $\Delta^{2n}_r \to \partial^{2n}_r$, which becomes the first derivative as $n \to 1 / 2$.
Thus, we obtain the estimate
\begin{align}
    \Delta F 
    &\sim 
    \kappa g_R \, \partial_r \alpha(1,\, r)\Big|_{r=1} 
    \,,
    \label{eqn:dfe_large_AR_estimate}
\end{align}
where we have assumed that $\alpha(1,\, 1) = 0$ since $\alpha(R,\, R) = 0$ for $R \in \mathbb{Z}_{> 1}$ because twisting $R$ components of the $\mathrm{SO}(R)$ field is equivalent to the trivial twist.

Next, we establish a qualitative scaling in the regime $\kappa \ll 1$.
We work with a finite number of replicas $n=m=1$.
Using Eq.~\eqref{app:dfe_dual} and Eq.~\eqref{app:dual_quasi_1d_twist}, we have
\begin{align}
    \Delta F 
    &\underset{\substack{n,m \to 1}}{=}
    \frac{
    16 e^{- \alpha g_R / \kappa}
    }
    {
    4(2 + 6 e^{- \alpha g_R / \kappa})
    }
    \\ 
    &\sim e^{-\alpha g_R / \kappa} +\mathcal{O}(e^{-2\alpha g_R / \kappa}) \,, \label{eqn:dfe_small_AR_estimate}
\end{align}
where we have suppressed the argument of $\alpha := \alpha(3, 1)=\alpha(3, 2) > 0$ for simplicity.

We now examine these predictions using numerical simulations.
First, in Fig.~\ref{app:fig_quasi_1d}(c), $\Delta F$ is plotted against $\kappa$ for $\kappa \gg 1$.
In this case, the data is expected to be linear in $\kappa$, with the slope  proportional to $g_R(L)$.
Thus, we plot the inverse slope of the linear best fit of the data in the inset as a function of $L$.
The inverse slope is proportional to $1/g_R \sim \log L$, in agreement with the observed data.
Next, in Fig.~\ref{app:fig_quasi_1d}(d), $\Delta F$ is plotted against $\kappa$ for $\kappa \ll 1$ on a log-linear scale.
The data is expected to be linear on this scale, with the inverse of the slope proportional to $1/g_R(T) \sim \log T$.
This is plotted in the inset to Fig.~\ref{app:fig_quasi_1d}(d), which is consistent with the predicted scaling.

\section{Fidelity between corrupted surface code states in different flux sectors}\label{app:kw_dual_fid_corr}
\begin{figure*}[t]
\centering
\includegraphics[width=.95\textwidth]{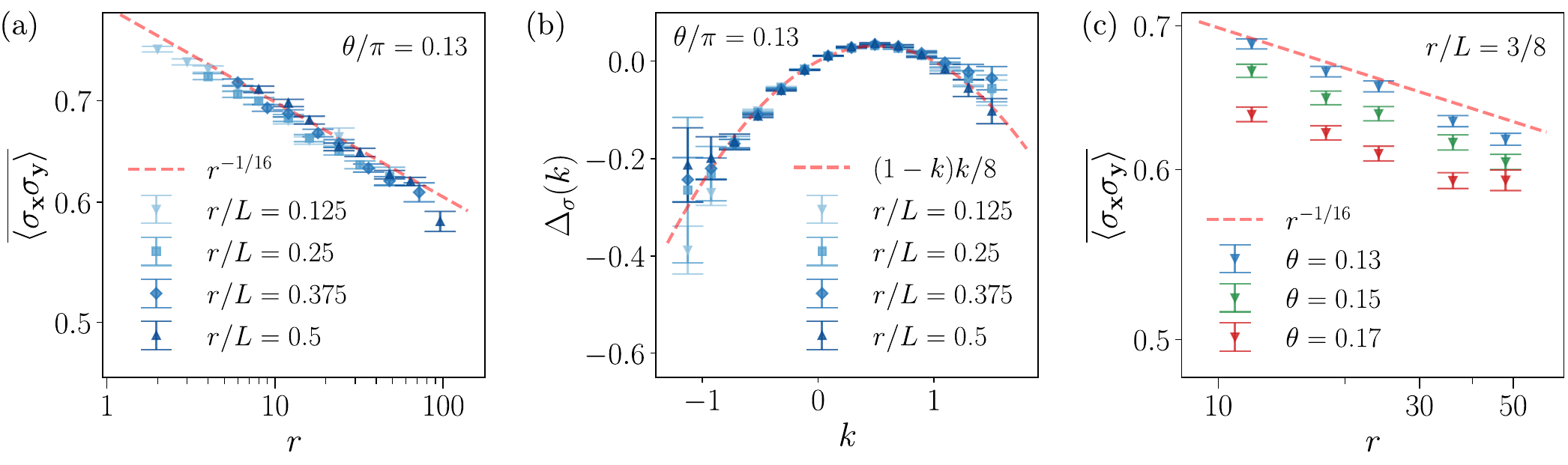}
    \caption{Fidelity $F(\rho_1, \, \rho_2) =\overline{|\langle \sigma_\bfx \sigma_\bfy \rangle|}$ and the $k$-th R\'enyi fidelity $\overline{|\langle \sigma_\bfx \sigma_\bfy \rangle|^{2k}}$ between two $L\times L$ coherently corrupted surface code states $\rho_1, \, \rho_2$.
    The state $\rho_2$ is prepared in the presence of an $m$-anyon pair created at $\bfx, \, \bfy$.
    In (a), the fidelity is plotted as a function of separation $r := |{\bfx} - {\bfy}|$  
    Here, $\theta = 0.13 \pi$ is fixed, as is $r/L$, while $L$ is varied.
    In (b), the scaling dimension of the $k$-th R\'enyi fidelity is extracted by performing a linear best fit on log-log scale, for fixed, but various values of $r/L$.
    The extracted scaling dimension is plotted as a function of $k$.
    In (c), the fidelity is plotted against $r$ for fixed $r/L$ and various $\theta$.
    For all cases, the endpoints of the symmetry defect are chosen as ${\bfx} = (L/2-r/2, L/2)$ and ${\bf y} = (L/2+r/2, L/2)$.
    Note that in (a) and (c) both axes are on a logarithmic scale.
    Data averaged over $8000$ to $40000$ samples.}
    \label{app:fig_spin_correlator_combined}
\end{figure*}
In this appendix, we provide additional numerics for the fidelity $F(\rho_1, \, \rho_2)$ between two corrupted surface code states studied in Sec.~\ref{sec:absence_decoding_setup}.
Additionally, we numerically determine the associated $k$-th Rényi fidelity 
\begin{equation}
    \sum_s \calQ_s^{1-k}\tilde{\calQ}_s^k = \sum_s \mathcal{Q}_s | \langle \sigma_\bfx \sigma_\bfy \rangle_s|^{2k}
    \,,
\end{equation}
as the Chernoff $k$-coefficient between the two distributions $\mathcal{Q}_s$, $\tilde{\mathcal{Q}}_s$.
This quantity is estimated in a similar Monte-Carlo procedure as outlined in Sec.~\ref{sec:swssb}.
Namely, we sample from the distribution $\mathcal{Q}_s$ utilizing the algorithm of Ref.~\cite{bravyi_coherent} (also, see Appendix H of Ref.~\cite{yan_bao_coherent_error_nlsm} and Ref.~\cite{venn_coherent_planar_graph}).
The output of this algorithm also computes $\mathcal{Q}_s$.
To compute $\tilde{\mathcal{Q}}_s$, related to the partition function with spin insertions, we run the algorithm again post-selected on the same outcomes $s$ but flipping $\theta \to - \theta$ on bonds which cross $\mathcal{R}$.
Here, $\mathcal{R}$ is a path on the dual lattice with endpoints at $\bfx, \, \bfy$.
Then, $|\langle \sigma_\bfx \sigma_\bfy \rangle_s|^2 = \tilde{\mathcal{Q}}_s / \mathcal{Q}_s$.
Averaging the appropriate moments of this quantity over many samples $s$ yields the $k$-th R\'enyi fidelity.
We indicate the average over syndromes $s$ as $\overline{|\langle \sigma_\bfx \sigma_\bfy \rangle|^{2k}}$.

Next, the correlator is expected to have the scaling form $\overline{|\langle \sigma_\bfx \sigma_\bfy \rangle|^{2k}} = L^{-2 \Delta_\sigma(k)} \mathcal{F}(\bfx / L , \,  \bfy / L, \, g_R)$ on the square $L$-by-$L$ geometry.
For the numerically accessible system sizes, $g_R$ is approximately unchanged [see Fig.~\ref{app:fig_spin_correlator_combined}(c)].
Fixing $\bfx / L$ and $\bfy / L$, $C_{F, \, k}$ is expected to follow a power law scaling in $L$.
Our numerical results are consistent with a power-law scaling with universal exponent $\Delta_\sigma(1/2) = 1/32$ for the $k=1/2$ fidelity $F(\rho_1, \, \rho_2)$ in Fig.~\ref{app:fig_spin_correlator_combined}(a).
By performing power law fits of $\overline{|\langle \sigma_\bfx \sigma_\bfy \rangle|^{2k}}$ against system size $L$ while holding $\bfx / L, \, \bfy / L$ fixed, we extract the scaling dimension $\Delta_\sigma(k)$ in Fig.~\ref{app:fig_spin_correlator_combined}(b).
Our results are consistent with the theoretical prediction for $|k| \lesssim 1$.

\section{Transfer matrix and network model}\label{app:conductance_explicit_forms}
In this appendix, we present a more detailed review of the Chalker-Coddington network model associated with the coset probability $\mathcal{Q}_{\alpha, s}$ and present explicit form of the node transfer matrices used in the conductivity simulations.

We begin by writing the coset probability as a transition amplitude in a 1+1D free fermion dynamics
\begin{equation}\label{app:coset_prob_as_transfer_matrices}
    \mathcal{Q}_{\alpha,\, s} = |\mathcal{Z}_{\alpha,\, s}|^2 = \bra{\phi_0} 
    \hat{\mathsf{H}}_{T}
    \hat{\mathsf{T}}_{T-1}
    \cdots 
    \hat{\mathsf{T}}_{1}
    \ket{\phi_0} \,.
\end{equation}
Here, $\hat{\mathsf{T}}_t = \hat{\mathsf{V}}_t \hat{\mathsf{H}}_t$ with the horizontal $\hat{\mathsf{H}}_t$ and vertical $\hat{\mathsf{V}}_t$ transfer matrices derived from the transfer matrices acting on spins using the Jordan-Wigner transformation.
We present their forms shortly.
These matrices depend on the disorder realization $\alpha,\, s$, whose indices we have chosen to suppress.
They act on $4W$ Majorana modes $\gamma_i^\sfa$ where $i = 1,\, \dots,\, 2W$ is the site index while $\sfa = 1,\, 2$ indicates that the mode lives on the ``bra'' or ``ket,'' respectively.
Here, $W = 2L/\sqrt{3},\, 2L$ for the armchair and zig-zag boundary condition, respectively while the boundary state 
is stabilized by $\ri \gamma_{2i}^\sfa\gamma_{2i+1}^\sfa$ for all $i = 1,\, \dots ,\, W$ and $\sfa = 1,\, 2$.

For numerical stability, we focus on the armchair configuration for which the transfer matrix at $\theta = \pi/4$ is unitary.

In that case, the row transfer matrices take the form
\begin{align}\label{eqn:app_row_matrices_general_form}
    \hat{\mathsf{H}}_t &=  
    \begin{cases}
    \bigotimes\limits_{\substack{i=1\\i \in 2\mathbb{Z} +1}}^W \hat{\mathsf{h}}_{\bfr,\,\bfr+\hat{e}_x} &  t \equiv 1 \mod 2 \\
    \\
    \bigotimes\limits_{\substack{i=1\\i \in 2\mathbb{Z}}}^W \hat{\mathsf{h}}_{\bfr, \, \bfr+\hat{e}_x}
        & t \equiv 0 \mod 2
    \end{cases}
    \\
    \hat{\mathsf{V}}_t &=  \bigotimes_{i=1}^W \hat{\mathsf{v}}_{\bfr, \, \bfr+\hat{e}_t} 
    \,.
\end{align}
The node transfer matrices corresponding to horizontal and vertical rows take the form
\begin{align}\label{eqn:app_node_transfer_matrices_explicit_form}
    \hat{\mathsf{h}}_{\bfr,\, \bfr+\hat{e}_x} &= 
    \exp\left[
    \sum_{\sfa=1}^2
    \left(J - \ri (-1)^\sfa \frac{\pi}{4}\right)
    \eta_{\bfr,\,\bfr+\hat{e}_x}
    \ri \gamma_{2i}^\sfa\gamma_{2i+1}^\sfa
    \right]
    \\
    \hat{\sfv}_{\bfr,\, \bfr+\hat{e}_t} &= 
    \exp\left[
    \sum_{\sfa=1}^2
    \ri(-1)^\sfa
    \left(
    \theta + \frac{\pi}{4}(1 - \eta_{\bfr,\, \bfr+\hat{e}_t})
    \right)
       \ri \gamma_{2i-1}^\sfa\gamma_{2i}^\sfa
    \right]
    \,,
\end{align}
The horizontal transfer matrices in Eq.~\eqref{eqn:app_row_matrices_general_form} act on even and odd horizontal bonds at even and odd time, leading to the dynamics pictured in Figures $2$ and $3$ of Ref.~\cite{yang_coherent_hexagonal}.
Here, $\hat{\mathsf{h}}$ and $\hat{\mathsf{v}}$ corresponds to the $M$ and $U$ gates, respectively, in their notation.
We refer to the full evolution after time $T$ as $\hat{\mathsf{T}}=\hat{\mathsf{H}}_{T}
    \hat{\mathsf{T}}_{T-1}
    \cdots 
    \hat{\mathsf{T}}_{1}$.

Because the dynamics are non-interacting, they are completely characterized by their action in the single-particle sector as the linear map $\hat{\mathsf{T}} \gamma_\alpha \hat{\mathsf{T}}^{-1} = \mathsf{T}_{\alpha\beta}\,\gamma_\beta$ where $\alpha = (i, \sfa)$ is a composite index including both
the site index labeled by $i$ and bra and ket modes labeled by $\sfa$.
The network model is then derived by associating the single-particle transfer matrix $\mathsf{T}$ with the scattering matrix $\mathsf{S}$ of a disordered superconductor in 2+1D.
Specifically, the connectivity of the network model defines a $\mathbb{Z}_2$-grading of the vector space on which the transfer matrices act corresponding to incoming and outgoing scattering modes.
In the canonical basis, this is represented by the matrix $\mathsf{Z} = \mathrm{diag}(
    \mathds{1}_{2W}, -\mathds{1}_{2W})
$ with $+1$ ($-1$) representing incoming (outgoing) modes.
We define the canonical basis below.
In this basis, the single-particle transfer matrix $\mathsf{T}$ satisfies the pseudo-unitarity condition $\mathsf{T}^\dagger \mathsf{Z} \mathsf{T} = \mathsf{Z}$.
The pseudo-unitary of $\mathsf{T}$ is equivalent to the unitary of $\mathsf{S}$.

The single-particle transfer matrices are defined by ${\mathsf{T}}={\mathsf{H}}_{T}
    {\mathsf{T}}_{T-1}
    \cdots 
    {\mathsf{T}}_{1}$ with ${\mathsf{T}}_t = {\mathsf{V}}_t {\mathsf{H}}_t$.
    Similar to Eq.~\eqref{eqn:app_row_matrices_general_form}, we have
    \begin{align}
    {\mathsf{H}}_t &=  
    \begin{cases}
    \bigoplus\limits_{\substack{i=1\\i \in 2\mathbb{Z} +1}}^W \left({\mathsf{h}}_{\bfr,\,\bfr+\hat{e}_x} \oplus \mathds{1}_4\right) &  t \equiv 1 \mod 2 \\
    \\
    \bigoplus\limits_{\substack{i=1\\i \in 2\mathbb{Z}}}^W \left(\mathds{1}_4 \oplus {\mathsf{h}}_{\bfr, \, \bfr+\hat{e}_x}\right)
        & t \equiv 0 \mod 2
    \end{cases}
    \\
    {\mathsf{V}}_t &=  \bigoplus_{i=1}^W {\mathsf{v}}_{\bfr, \, \bfr+\hat{e}_t} 
    \,.
\end{align}
The single-particle node transfer matrices have the form
\begin{equation}\label{eqn:app_pre_gauge_network_transfer_matrices_explicit_form}
\begin{aligned}
    \mathsf{h}_{\bfr,\,\bfr+\hat{e}_x} &= \frac{1}{\sin2\theta} \begin{pmatrix}
        -\ri\cos2\theta & \eta_{\bfr,\,\bfr+\hat{e}_x} &&\\
        -\eta_{\bfr,\,\bfr+\hat{e}_x} & -\ri\cos2\theta &&\\
        && \ri\cos2\theta & \eta_{\bfr,\,\bfr+\hat{e}_x} \\
        && -\eta_{\bfr,\,\bfr+\hat{e}_x} & \ri\cos2\theta
    \end{pmatrix}\,,\\
    \mathsf{v}_{\bfr,\,\bfr+\hat{e}_t} &= \eta_{\bfr,\,\bfr+\hat{e}_t}\begin{pmatrix}
        \cos2\theta & -\sin2\theta &&\\
        \sin2\theta & \cos2\theta && \\
                &&\cos2\theta & -\sin2\theta \\
        &&\sin2\theta & \cos2\theta 
    \end{pmatrix}\,.
\end{aligned}
\end{equation}
The transfer matrices satisfy~\cite{jian2022criticality} the conditions $\mathsf{h}^\dagger (\sigma^x \otimes \mathds{1}_2) \mathsf{h} = (\sigma^x \otimes \mathds{1}_2)$ and $\mathsf{v}^\dagger (\sigma^x \otimes \mathds{1}_2) \mathsf{v} = (\sigma^x \otimes \mathds{1}_2)$ and thus define a unitary scattering problem which conserves the probability amplitude/current (equivalently, the ``doubled'' network~\cite{shanker_rbim,cho_fisher} conserves the $U(1)$ particle number).
Here 
\begin{equation}
\sigma^x \otimes \mathds{1}_2 = \begin{pmatrix}
    && 1 & 0 \\
    && 0 & 1 \\
    1 & 0 && \\
    0 & 1 &&
\end{pmatrix}
\,.
\end{equation}
The canonical basis is the defined as the one in where the current operator takes the diagonal form $\sigma^z \otimes \mathds{1}_2$.
We work in this basis in numerics.
Additionally, it is numerically convenient to work with transfer matrices that are purely real, which is possible due to the particle-hole symmetry that squares to $+1$.
This may be obtained by performing an additional phase rotation.
Conjugating Eq.~\eqref{eqn:app_pre_gauge_network_transfer_matrices_explicit_form} by
\begin{equation}
    U = \begin{pmatrix}
        1 & 0 & \ri & 0 \\
        0 & 1 & 0 & \ri \\
        1 & 0 & -\ri & 0 \\
        0 & 1 & 0 & -\ri
    \end{pmatrix}
    \,,
\end{equation}
the transfer matrices then take the form
\begin{equation}
\begin{aligned}
    \mathsf{h}_{\bfr,\,\bfr+\hat{e}_x} &= \frac{1}{\sin2\theta} \begin{pmatrix}
    0 & \eta_{\bfr,\,\bfr+\hat{e}_x} & \cos 2 \theta & 0 \\
    -\eta_{\bfr,\,\bfr+\hat{e}_x} & 0 & 0 & \cos 2\theta \\
    -\cos 2 \theta & 0 & 0 & \eta_{\bfr,\,\bfr+\hat{e}_x} \\
    0 & - \cos 2 \theta & - \eta_{\bfr,\,\bfr+\hat{e}_x} & 0
    \end{pmatrix}\,,\\
    \mathsf{v}_{\bfr,\,\bfr+\hat{e}_t} &= \eta_{\bfr,\,\bfr+\hat{e}_t}\begin{pmatrix}
        \cos2\theta & -\sin2\theta &&\\
        \sin2\theta & \cos2\theta && \\
                &&\cos2\theta & -\sin2\theta \\
        &&\sin2\theta & \cos2\theta 
    \end{pmatrix}\,.
\end{aligned}
\end{equation}
Both node transfer matrices $\mathsf{h}, \mathsf{v}$ are purely real, i.e.,  they satisfy the anti-unitary particle-hole symmetry $\mathsf{h}^* = \mathsf{h}$ and $\mathsf{v}^* = \mathsf{v}$.  
In addition, the transfer matrix also satisfies an additional anti-unitary time-reversal symmetry $(\sigma^y \otimes \mathds{1}_2) \mathsf{h}^* (\sigma^y \otimes \mathds{1}_2) = \mathsf{h}$ and $(\sigma^y \otimes \mathds{1}_2) \mathsf{v}^* (\sigma^y \otimes \mathds{1}_2) = \mathsf{v}$.
As a result, the network model possesses both a particle-hole symmetry squaring to $+1$ and a time-reversal symmetry squaring to $-1$ and therefore belongs to symmetry class DIII.

\bibliographystyle{ieeetr}
\bibliography{refs.bib}

@article{read2000absence,
   title={Absence of a metallic phase in random-bond Ising models in two dimensions: Applications to disordered superconductors and paired quantum Hall states},
   volume={63},
   ISSN={1095-3795},
   url={http://dx.doi.org/10.1103/PhysRevB.63.024404},
   DOI={10.1103/physrevb.63.024404},
   number={2},
   journal={Physical Review B},
   publisher={American Physical Society (APS)},
   author={Read, N. and Ludwig, Andreas W. W.},
   year={2000},
   month=dec 
}

@article{fan2023diagnostics,
title={Diagnostics of Mixed-State Topological Order and Breakdown of Quantum Memory},
   volume={5},
   ISSN={2691-3399},
   url={http://dx.doi.org/10.1103/PRXQuantum.5.020343},
   DOI={10.1103/prxquantum.5.020343},
   number={2},
   journal={PRX Quantum},
   publisher={American Physical Society (APS)},
   author={Fan, Ruihua and Bao, Yimu and Altman, Ehud and Vishwanath, Ashvin},
   year={2024},
   month=may 
}

@article{cho_fisher,
  title = {Criticality in the two-dimensional random-bond Ising model},
  author = {Cho, Sora and Fisher, Matthew P. A.},
  journal = {Phys. Rev. B},
  volume = {55},
  issue = {2},
  pages = {1025--1031},
  numpages = {0},
  year = {1997},
  month = {Jan},
  publisher = {American Physical Society},
  doi = {10.1103/PhysRevB.55.1025},
  url = {https://link.aps.org/doi/10.1103/PhysRevB.55.1025}
}

@inproceedings{osg_1,
  title  = {The open science grid},
  author = {
    Pordes, Ruth
    and Petravick, Don
    and Kramer, Bill
    and Olson, Doug
    and Livny, Miron
    and Roy, Alain
    and Avery, Paul
    and Blackburn, Kent
    and Wenaus, Torre
    and W{\"u}rthwein, Frank
    and Foster, Ian
    and Gardner, Rob
    and Wilde, Mike
    and Blatecky, Alan
    and McGee, John
    and Quick, Rob
  },
  doi       = {10.1088/1742-6596/78/1/012057},
  booktitle = {J. Phys. Conf. Ser.},
  volume    = {78},
  series    = {78},
  pages     = {012057},
  year      = {2007},
}

@inproceedings{osg_2,
  title        = {The pilot way to grid resources using glideinWMS},
  author       = {
    Sfiligoi, Igor
    and Bradley, Daniel C
    and Holzman, Burt
    and Mhashilkar, Parag
    and Padhi, Sanjay
    and Wurthwein, Frank
  },
  doi          = {10.1109/CSIE.2009.950},
  booktitle    = {2009 WRI World Congress on Computer Science and Information Engineering},
  volume       = {2},
  series       = {2},
  pages        = {428--432},
  year         = {2009},
}

@misc{osg_3,
  doi = {10.21231/906P-4D78},
  url = {https://osg-htc.org/services/open_science_pool.html},
  author = {{OSG}},
  title = {OSPool},
  publisher = {OSG},
  year = {2006}
}

@misc{osg_4,
  doi = {10.21231/0KVZ-VE57},
  url = {https://osdf.osg-htc.org/},
  author = {{OSG}},
  title = {Open Science Data Federation},
  publisher = {OSG},
  year = {2015}
}

@article{venn_entanglement_phases,
  title = {Surface codes, quantum circuits, and entanglement phases},
  author = {Behrends, Jan and Venn, Florian and B\'eri, Benjamin},
  journal = {Phys. Rev. Res.},
  volume = {6},
  issue = {1},
  pages = {013137},
  numpages = {15},
  year = {2024},
  month = {Feb},
  publisher = {American Physical Society},
  doi = {10.1103/PhysRevResearch.6.013137},
  url = {https://link.aps.org/doi/10.1103/PhysRevResearch.6.013137}
}

@article{cheng2025emergent,
  title={Emergent unitary designs for encoded qubits from coherent errors and syndrome measurements},
  author={Cheng, Zihan and Huang, Eric and Khemani, Vedika and Gullans, Michael J and Ippoliti, Matteo},
  journal={PRX Quantum},
  volume={6},
  number={3},
  pages={030333},
  year={2025},
  publisher={APS}
}

@article{bao2023mixed,
  title={Mixed-state topological order and the errorfield double formulation of decoherence-induced transitions},
  author={Bao, Yimu and Fan, Ruihua and Vishwanath, Ashvin and Altman, Ehud},
  journal={Physical Review Letters},
  volume={136},
  number={22},
  pages={220402},
  year={2026},
  publisher={APS}
}

@article{wang2025intrinsic,
  title={Intrinsic mixed-state topological order},
  author={Wang, Zijian and Wu, Zhengzhi and Wang, Zhong},
  journal={PRX Quantum},
  volume={6},
  number={1},
  pages={010314},
  year={2025},
  publisher={APS}
}

@article{bravyi_coherent,
   title={Correcting coherent errors with surface codes},
   volume={4},
   ISSN={2056-6387},
   url={http://dx.doi.org/10.1038/s41534-018-0106-y},
   DOI={10.1038/s41534-018-0106-y},
   number={1},
   journal={npj Quantum Information},
   publisher={Springer Science and Business Media LLC},
   author={Bravyi, Sergey and Englbrecht, Matthias and König, Robert and Peard, Nolan},
   year={2018},
   month=oct 
}

@article{venn_coherent_planar_graph,
   title={Error-correction and noise-decoherence thresholds for coherent errors in planar-graph surface codes},
   volume={2},
   ISSN={2643-1564},
   url={http://dx.doi.org/10.1103/PhysRevResearch.2.043412},
   DOI={10.1103/physrevresearch.2.043412},
   number={4},
   journal={Physical Review Research},
   publisher={American Physical Society (APS)},
   author={Venn, F. and Béri, B.},
   year={2020},
   month=dec 
}

@article{evers2008anderson,
  title={Anderson transitions},
  author={Evers, Ferdinand and Mirlin, Alexander D},
  journal={Reviews of Modern Physics},
  volume={80},
  number={4},
  pages={1355--1417},
  year={2008},
  publisher={APS}
}

@article{chalker1988percolation,
  title={Percolation, quantum tunnelling and the integer Hall effect},
  author={Chalker, JT and Coddington, PD},
  journal={Journal of Physics C: Solid State Physics},
  volume={21},
  number={14},
  pages={2665},
  year={1988},
  publisher={IOP Publishing}
}

@article{gruzberg2001random,
  title={Random-bond Ising model in two dimensions: The Nishimori line and supersymmetry},
  author={Gruzberg, Ilya A and Read, N and Ludwig, Andreas WW},
  journal={Physical Review B},
  volume={63},
  number={10},
  pages={104422},
  year={2001},
  publisher={APS}
}

@article{read_absence_of_metallic,
   title={Absence of a metallic phase in random-bond Ising models in two dimensions: Applications to disordered superconductors and paired quantum Hall states},
   volume={63},
   ISSN={1095-3795},
   url={http://dx.doi.org/10.1103/PhysRevB.63.024404},
   DOI={10.1103/physrevb.63.024404},
   number={2},
   journal={Physical Review B},
   publisher={American Physical Society (APS)},
   author={Read, N. and Ludwig, Andreas W. W.},
   year={2000},
   month=dec }

@article{behrends2024surface,
  title={Surface codes, quantum circuits, and entanglement phases},
  author={Behrends, Jan and Venn, Florian and B{\'e}ri, Benjamin},
  journal={Physical Review Research},
  volume={6},
  number={1},
  pages={013137},
  year={2024},
  publisher={APS}
}

@article{hauser2025information,
  title={Information dynamics and symmetry breaking in generic monitored $\mathbb{Z}_{2}$-symmetric open quantum systems},
  author={Hauser, Jacob and Lavasani, Ali and Vijay, Sagar and Fisher, Matthew},
  journal={arXiv preprint arXiv:2512.03031},
  year={2025}
}

@article{lavasani2025stability,
  title={Stability of gapped quantum matter and error-correction with adiabatic noise},
  author={Lavasani, Ali and Vijay, Sagar},
  journal={Physical Review Research},
  volume={7},
  number={2},
  pages={023166},
  year={2025},
  publisher={APS}
}

@article{AZ_10fold,
   title={Nonstandard symmetry classes in mesoscopic normal-superconducting hybrid structures},
   volume={55},
   ISSN={1095-3795},
   url={http://dx.doi.org/10.1103/PhysRevB.55.1142},
   DOI={10.1103/physrevb.55.1142},
   number={2},
   journal={Physical Review B},
   publisher={American Physical Society (APS)},
   author={Altland, Alexander and Zirnbauer, Martin R.},
   year={1997},
   month=jan, pages={1142–1161} }

@misc{phases_of_decodability_bao,
      title={Phases of decodability in the surface code with unitary errors}, 
      author={Yimu Bao and Sajant Anand},
      year={2024},
      eprint={2411.05785},
      archivePrefix={arXiv},
      primaryClass={quant-ph},
      url={https://arxiv.org/abs/2411.05785}, 
}

@misc{wang_self_dual_cho_fisher,
      title={Decoherence-induced self-dual criticality in topological states of matter}, 
      author={Qingyuan Wang and Romain Vasseur and Simon Trebst and Andreas W. W. Ludwig and Guo-Yi Zhu},
      year={2025},
      eprint={2502.14034},
      archivePrefix={arXiv},
      primaryClass={quant-ph},
      url={https://arxiv.org/abs/2502.14034}, 
}

@article{abe_yokota_symmetric_space_volume,
author = {Kōjun Abe and Ichiro Yokota},
title = {{Volumes of Compact Symmetric Spaces}},
volume = {20},
journal = {Tokyo Journal of Mathematics},
number = {1},
publisher = {Publication Committee for the Tokyo Journal of Mathematics},
pages = {87 -- 105},
year = {1997},
doi = {10.3836/tjm/1270042402},
URL = {https://doi.org/10.3836/tjm/1270042402}
}

@article{aasen_mong_fendley_ising,
   title={Topological defects on the lattice: I. The Ising model},
   volume={49},
   ISSN={1751-8121},
   url={http://dx.doi.org/10.1088/1751-8113/49/35/354001},
   DOI={10.1088/1751-8113/49/35/354001},
   number={35},
   journal={Journal of Physics A: Mathematical and Theoretical},
   publisher={IOP Publishing},
   author={Aasen, David and Mong, Roger S K and Fendley, Paul},
   year={2016},
   month=aug, pages={354001} }

@article{macdonald_volume,
author = {Macdonald, I.G.},
journal = {Inventiones mathematicae},
pages = {93-96},
title = {The Volume of a Compact Lie Group.},
url = {http://eudml.org/doc/142687},
volume = {56},
year = {1980},
}

@article{senthil2000quasiparticle,
  title={Quasiparticle localization in superconductors with spin-orbit scattering},
  author={Senthil, T and Fisher, Matthew PA},
  journal={Physical Review B},
  volume={61},
  number={14},
  pages={9690},
  year={2000},
  publisher={APS}
}

@article{jian2022criticality,
  title={Criticality and entanglement in nonunitary quantum circuits and tensor networks of noninteracting fermions},
  author={Jian, Chao-Ming and Bauer, Bela and Keselman, Anna and Ludwig, Andreas WW},
  journal={Physical Review B},
  volume={106},
  number={13},
  pages={134206},
  year={2022},
  publisher={APS}
}

@article{jian2023measurement,
  title={Measurement-induced entanglement transitions in quantum circuits of non-interacting fermions: Born-rule versus forced measurements},
  author={Jian, Chao-Ming and Shapourian, Hassan and Bauer, Bela and Ludwig, Andreas WW},
  journal={arXiv preprint arXiv:2302.09094},
  year={2023}
}

@article{cardy1984conformal,
  title={Conformal invariance and universality in finite-size scaling},
  author={Cardy, John L},
  journal={Journal of Physics A: Mathematical and General},
  volume={17},
  number={7},
  pages={L385},
  year={1984},
  publisher={IOP Publishing}
}

@article{ferris2014tensor,
  title={Tensor networks and quantum error correction},
  author={Ferris, Andrew J and Poulin, David},
  journal={Physical review letters},
  volume={113},
  number={3},
  pages={030501},
  year={2014},
  publisher={APS}
}

@article{bravyi2014efficient,
  title={Efficient algorithms for maximum likelihood decoding in the surface code},
  author={Bravyi, Sergey and Suchara, Martin and Vargo, Alexander},
  journal={Physical Review A},
  volume={90},
  number={3},
  pages={032326},
  year={2014},
  publisher={APS}
}

@article{chubb2021statistical,
  title={Statistical mechanical models for quantum codes with correlated noise},
  author={Chubb, Christopher T and Flammia, Steven T},
  journal={Annales de l’Institut Henri Poincar{\'e} D},
  volume={8},
  number={2},
  pages={269--321},
  year={2021}
}

@article{song2022optimal,
  title={Optimal thresholds for fracton codes and random spin models with subsystem symmetry},
  author={Song, Hao and Sch{\"o}nmeier-Kromer, Janik and Liu, Ke and Viyuela, Oscar and Pollet, Lode and Martin-Delgado, Miguel Angel},
  journal={Physical Review Letters},
  volume={129},
  number={23},
  pages={230502},
  year={2022},
  publisher={APS}
}

@article{wang2003confinement,
  title={Confinement-Higgs transition in a disordered gauge theory and the accuracy threshold for quantum memory},
  author={Wang, Chenyang and Harrington, Jim and Preskill, John},
  journal={Annals of Physics},
  volume={303},
  number={1},
  pages={31--58},
  year={2003},
  publisher={Elsevier}
}

@article{katzgraber2009error,
  title={Error threshold for color codes and random three-body Ising models},
  author={Katzgraber, Helmut G and Bomb{\'\i}n, H{\'e}ctor and Martin-Delgado, Martin A},
  journal={Physical review letters},
  volume={103},
  number={9},
  pages={090501},
  year={2009},
  publisher={APS}
}

@article{bombin2012strong,
  title={Strong resilience of topological codes to depolarization},
  author={Bombin, H{\'e}ctor and Andrist, Ruben S and Ohzeki, Masayuki and Katzgraber, Helmut G and Martin-Delgado, Miguel A},
  journal={Physical Review X},
  volume={2},
  number={2},
  pages={021004},
  year={2012},
  publisher={APS}
}

@article{kubica2018three,
  title={Three-dimensional color code thresholds via statistical-mechanical mapping},
  author={Kubica, Aleksander and Beverland, Michael E and Brand{\~a}o, Fernando and Preskill, John and Svore, Krysta M},
  journal={Physical review letters},
  volume={120},
  number={18},
  pages={180501},
  year={2018},
  publisher={APS}
}

@article{Darmawan_2017,
   title={Tensor-Network Simulations of the Surface Code under Realistic Noise},
   volume={119},
   ISSN={1079-7114},
   url={http://dx.doi.org/10.1103/PhysRevLett.119.040502},
   DOI={10.1103/physrevlett.119.040502},
   number={4},
   journal={Physical Review Letters},
   publisher={American Physical Society (APS)},
   author={Darmawan, Andrew S. and Poulin, David},
   year={2017},
   month=jul }

@article{Darmawan_2018,
   title={Linear-time general decoding algorithm for the surface code},
   volume={97},
   ISSN={2470-0053},
   url={http://dx.doi.org/10.1103/PhysRevE.97.051302},
   DOI={10.1103/physreve.97.051302},
   number={5},
   journal={Physical Review E},
   publisher={American Physical Society (APS)},
   author={Darmawan, Andrew S. and Poulin, David},
   year={2018},
   month=may }

@article{bombin2010topological,
  title={Topological subsystem codes},
  author={Bomb{\'\i}n, H{\'e}ctor},
  journal={Physical Review A—Atomic, Molecular, and Optical Physics},
  volume={81},
  number={3},
  pages={032301},
  year={2010},
  publisher={APS}
}

@article{tang2025phases,
  title={Phases of Floquet code under local decoherence},
  author={Tang, Yuchen and Bao, Yimu},
  journal={Physical Review A},
  volume={112},
  number={6},
  pages={062437},
  year={2025},
  publisher={APS}
}

@article{lee2023quantum,
  title={Quantum criticality under decoherence or weak measurement},
  author={Lee, Jong Yeon and Jian, Chao-Ming and Xu, Cenke},
  journal={PRX quantum},
  volume={4},
  number={3},
  pages={030317},
  year={2023},
  publisher={APS}
}

@article{ludwig2016topological,
  title={Topological phases: classification of topological insulators and superconductors of non-interacting fermions, and beyond},
  author={Ludwig, Andreas WW},
  journal={Physica Scripta},
  volume={168},
  number={1},
  pages={014001},
  year={2016},
  publisher={IOP Publishing}
}

@article{ryu2010topological,
  title={Topological insulators and superconductors: tenfold way and dimensional hierarchy},
  author={Ryu, Shinsei and Schnyder, Andreas P and Furusaki, Akira and Ludwig, Andreas WW},
  journal={New Journal of Physics},
  volume={12},
  number={6},
  pages={065010},
  year={2010}
}

@article{zirnbauer1996riemannian,
  title={Riemannian symmetric superspaces and their origin in random-matrix theory},
  author={Zirnbauer, Martin R},
  journal={Journal of Mathematical Physics},
  volume={37},
  number={10},
  pages={4986--5018},
  year={1996},
  publisher={American Institute of Physics}
}

@article{lessa2025strong,
  title={Strong-to-weak spontaneous symmetry breaking in mixed quantum states},
  author={Lessa, Leonardo A and Ma, Ruochen and Zhang, Jian-Hao and Bi, Zhen and Cheng, Meng and Wang, Chong},
  journal={PRX Quantum},
  volume={6},
  number={1},
  pages={010344},
  year={2025},
  publisher={APS}
}

@article{wegner1971duality,
  title={Duality in generalized Ising models and phase transitions without local order parameters},
  author={Wegner, Franz J},
  journal={Journal of Mathematical Physics},
  volume={12},
  number={10},
  pages={2259--2272},
  year={1971},
  publisher={American Institute of Physics}
}

@article{dennis_surface_code,
   title={Topological quantum memory},
   volume={43},
   ISSN={1089-7658},
   url={http://dx.doi.org/10.1063/1.1499754},
   DOI={10.1063/1.1499754},
   number={9},
   journal={Journal of Mathematical Physics},
   publisher={AIP Publishing},
   author={Dennis, Eric and Kitaev, Alexei and Landahl, Andrew and Preskill, John},
   year={2002},
   month=Sept, pages={4452–4505}
}

@misc{yan_bao_coherent_error_nlsm,
      title={Non-linear Sigma Model for the Surface Code with Coherent Errors}, 
      author={Stephen W. Yan and Yimu Bao and Sagar Vijay},
      year={2026},
      eprint={2603.25665},
      archivePrefix={arXiv},
      primaryClass={cond-mat.stat-mech},
      url={https://arxiv.org/abs/2603.25665}, 
}

@misc{yang_coherent_hexagonal,
      title={Decoding coherent errors in toric codes on honeycomb and square lattices: duality to Majorana monitored dynamics and symmetry classes}, 
      author={Zhou Yang and Andreas W. W. Ludwig and Chao-Ming Jian},
      year={2026},
      eprint={2604.08650},
      archivePrefix={arXiv},
      primaryClass={cond-mat.stat-mech},
      url={https://arxiv.org/abs/2604.08650}, 
}

@article{behrends_general_x,
      title={The surface code beyond Pauli channels: Logical noise coherence, information-theoretic measures, and errorfield-double phenomenology}, 
      author={Jan Behrends and Benjamin Béri},
      year={2025},
      eprint={2412.21055},
      archivePrefix={arXiv},
      primaryClass={quant-ph},
      journal={arXiv preprint arXiv:2412.21055},
      url={https://arxiv.org/abs/2412.21055} 
}

@misc{bejan_projected_ensemble,
      title={Projected logical ensembles in surface codes via the random-matrix theory of quantum dots}, 
      author={Mircea Bejan and Jan Behrends and Max McGinley and Benjamin Béri},
      year={2026},
      eprint={2606.17140},
      archivePrefix={arXiv},
      primaryClass={quant-ph},
      url={https://arxiv.org/abs/2606.17140}, 
}

@article{merz_chalker,
  title = {Two-dimensional random-bond Ising model, free fermions, and the network model},
  author = {Merz, F. and Chalker, J. T.},
  journal = {Phys. Rev. B},
  volume = {65},
  issue = {5},
  pages = {054425},
  numpages = {18},
  year = {2002},
  month = {Jan},
  publisher = {American Physical Society},
  doi = {10.1103/PhysRevB.65.054425},
  url = {https://link.aps.org/doi/10.1103/PhysRevB.65.054425}
}

@article{fava2023nonlinear,
  title={Nonlinear sigma models for monitored dynamics of free fermions},
  author={Fava, Michele and Piroli, Lorenzo and Swann, Tobias and Bernard, Denis and Nahum, Adam},
  journal={Physical Review X},
  volume={13},
  number={4},
  pages={041045},
  year={2023},
  publisher={APS}
}

@article{shanker_rbim,
  title = {Exact critical behavior of a random bond two-dimensional Ising model},
  author = {Shankar, R.},
  journal = {Phys. Rev. Lett.},
  volume = {58},
  issue = {23},
  pages = {2466--2469},
  numpages = {0},
  year = {1987},
  month = {Jun},
  publisher = {American Physical Society},
  doi = {10.1103/PhysRevLett.58.2466},
  url = {https://link.aps.org/doi/10.1103/PhysRevLett.58.2466}
}

@article{zirnbauer_continuum_CC,
   title={Toward a theory of the integer quantum Hall transition: Continuum limit of the Chalker–Coddington model},
   volume={38},
   ISSN={1089-7658},
   url={http://dx.doi.org/10.1063/1.531921},
   DOI={10.1063/1.531921},
   number={4},
   journal={Journal of Mathematical Physics},
   publisher={AIP Publishing},
   author={Zirnbauer, Martin R.},
   year={1997},
   month=Apr, pages={2007–2036} }

@article{verbaarschot_critique,
    author = "Verbaarschot, J. J. M. and Zirnbauer, M. R.",
    title = "{Critique of the Replica Trick}",
    doi = "10.1088/0305-4470/18/7/018",
    journal = "J. Phys. A",
    volume = "18",
    pages = "1093--1109",
    year = "1985"
}

@misc{zirnbauer_another_critique,
      title={Another critique of the replica trick}, 
      author={Martin R. Zirnbauer},
      year={1999},
      eprint={cond-mat/9903338},
      archivePrefix={arXiv},
      primaryClass={cond-mat.mes-hall},
      url={https://arxiv.org/abs/cond-mat/9903338}, 
}

@article{wegner_four_loop_beta,
title = {Four-loop-order $\beta$-function of nonlinear $\sigma$-models in symmetric spaces},
journal = {Nuclear Physics B},
volume = {316},
number = {3},
pages = {663-678},
year = {1989},
issn = {0550-3213},
doi = {https://doi.org/10.1016/0550-3213(89)90063-1},
url = {https://www.sciencedirect.com/science/article/pii/0550321389900631},
author = {Franz Wegner}
}

@article{hikami_three_loop_beta,
title = {Three-loop $\beta$-functions of non-linear $\sigma$ models on symmetric spaces},
journal = {Physics Letters B},
volume = {98},
number = {3},
pages = {208-210},
year = {1981},
issn = {0370-2693},
doi = {https://doi.org/10.1016/0370-2693(81)90989-8},
url = {https://www.sciencedirect.com/science/article/pii/0370269381909898},
author = {S. Hikami}
}

@article{hikami_conductance_normalization,
  title = {Anderson localization in a nonlinear-$\ensuremath{\sigma}$-model representation},
  author = {Hikami, Shinobu},
  journal = {Phys. Rev. B},
  volume = {24},
  issue = {5},
  pages = {2671--2679},
  numpages = {0},
  year = {1981},
  month = {Sep},
  publisher = {American Physical Society},
  doi = {10.1103/PhysRevB.24.2671},
  url = {https://link.aps.org/doi/10.1103/PhysRevB.24.2671}
}

@article{venn2022coherent,
  title = {Coherent-Error Threshold for Surface Codes from Majorana Delocalization},
  author = {Venn, Florian and Behrends, Jan and B\'eri, Benjamin},
  journal = {Phys. Rev. Lett.},
  volume = {131},
  issue = {6},
  pages = {060603},
  numpages = {6},
  year = {2023},
  month = {Aug},
  publisher = {American Physical Society},
  doi = {10.1103/PhysRevLett.131.060603},
  url = {https://link.aps.org/doi/10.1103/PhysRevLett.131.060603}
}

@article{fulga_class_diii_helical_superconductor,
   title={Thermal metal-insulator transition in a helical topological superconductor},
   volume={86},
   ISSN={1550-235X},
   url={http://dx.doi.org/10.1103/PhysRevB.86.054505},
   DOI={10.1103/physrevb.86.054505},
   number={5},
   journal={Physical Review B},
   publisher={American Physical Society (APS)},
   author={Fulga, I. C. and Akhmerov, A. R. and Tworzydło, J. and Béri, B. and Beenakker, C. W. J.},
   year={2012},
   month=Aug 
}

@misc{lerner_non_perturbative,
      title={Nonlinear Sigma Model for Disordered Media: Replica Trick for Non-Perturbative Results and Interactions}, 
      author={Igor V. Lerner},
      year={2002},
      eprint={cond-mat/0205555},
      archivePrefix={arXiv},
      primaryClass={cond-mat.dis-nn},
      url={https://arxiv.org/abs/cond-mat/0205555}, 
}

@article{andrews_partial_1,
  author  = {Andrews, George E.},
  title   = {Ramanujan's ``Lost'' Notebook. I. Partial {$\theta$}-functions},
  journal = {Advances in Mathematics},
  volume  = {41},
  year    = {1981},
  pages   = {137--172},
  doi     = {10.1016/0001-8708(81)90013-X}
}

@article{andrews_partial_9,
  author  = {Andrews, George E.},
  title   = {Ramanujan's ``Lost'' Notebook IX: The Partial Theta Function as an Entire Function},
  journal = {Advances in Mathematics},
  volume  = {191},
  year    = {2005},
  pages   = {408--422},
  doi     = {10.1016/j.aim.2004.03.017}
}

@book{andrews_lostnotebook_book_1,
  author    = {Andrews, George E. and Berndt, Bruce C.},
  title     = {Ramanujan's Lost Notebook: Part I},
  publisher = {Springer},
  address   = {New York},
  year      = {2005},
  doi       = {10.1007/0-387-28124-X},
  isbn      = {978-0-387-25529-3},
  pages     = {437}
}

@book{krantz_handbook,
  author    = {Krantz, Steven G.},
  title     = {Handbook of Complex Variables},
  publisher = {Birkh{\"a}user},
  address   = {Boston, MA},
  year      = {1999},
  isbn      = {978-0-8176-4011-8},
  doi       = {10.1007/978-1-4612-1588-2},
  pages     = {xxiv+290}
}

@misc{erfc_asymptotics,
    key = "{DLMF}",
    title = "{NIST Digital Library of Mathematical Functions}",
    url = "https://dlmf.nist.gov/7.12",
    note = "F.~W.~J. Olver, A.~B. {Olde Daalhuis}, D.~W. Lozier, B.~I. Schneider,
            R.~F. Boisvert, C.~W. Clark, B.~R. Miller, B.~V. Saunders,
            H.~S. Cohl, and M.~A. McClain, eds."
}

@article{fan2024diagnostics,
  title={Diagnostics of mixed-state topological order and breakdown of quantum memory},
  author={Fan, Ruihua and Bao, Yimu and Altman, Ehud and Vishwanath, Ashvin},
  journal={PRX Quantum},
  volume={5},
  number={2},
  pages={020343},
  year={2024},
  publisher={APS}
}

@article{dixon_orbifold_cft,
    author = "Dixon, Lance J. and Friedan, Daniel and Martinec, Emil J. and Shenker, Stephen H.",
    title = "{The Conformal Field Theory of Orbifolds}",
    reportNumber = "EFI-86-42-CHICAGO",
    doi = "10.1016/0550-3213(87)90676-6",
    journal = "Nucl. Phys. B",
    volume = "282",
    pages = "13--73",
    year = "1987"
}

@book{peskin_schroeder,
    author = "Peskin, Michael E. and Schroeder, Daniel V.",
    title = "{An Introduction to quantum field theory}",
    doi = "10.1201/9780429503559",
    isbn = "978-0-201-50397-5, 978-0-429-50355-9, 978-0-429-49417-8",
    publisher = "Addison-Wesley",
    address = "Reading, USA",
    year = "1995"
}

@article{brezin_callan_symanzik,
   author = "Brezin, E. and Le Guillou, J. C. and Zinn-Justin, Jean",
   title = "{Wilson's theory of critical phenomena and callan-symanzik equations in 4-epsilon dimensions}",
   doi = "10.1103/PhysRevD.9.1121",
   journal = "Phys. Rev. D",
   volume = "8",
   pages = "434--440",
   year = "1973",
   note = "[Addendum: Phys.Rev.D 9, 1121--1124 (1974), Erratum: Phys.Rev.D 10, 2046--2046 (1974)]"
}

@article{calabrese_norm_replica,
   title={Off-diagonal correlations in one-dimensional anyonic models: a replica approach},
   volume={2009},
   ISSN={1742-5468},
   url={http://dx.doi.org/10.1088/1742-5468/2009/03/P03002},
   DOI={10.1088/1742-5468/2009/03/p03002},
   number={03},
   journal={Journal of Statistical Mechanics: Theory and Experiment},
   publisher={IOP Publishing},
   author={Calabrese, Pasquale and Santachiara, Raoul},
   year={2009},
   month=Mar, pages={P03002} 
}

@article{kurchan_norm_replica,
doi = {10.1088/0305-4470/24/21/011},
url = {https://doi.org/10.1088/0305-4470/24/21/011},
year = {1991},
month = {nov},
publisher = {},
volume = {24},
number = {21},
pages = {4969},
author = {J Kurchan},
title = {Replica trick to calculate means of absolute values: applications to stochastic equations},
journal = {Journal of Physics A: Mathematical and General}
}

@article{sang_stability,
   title={Stability of Mixed-State Quantum Phases via Finite Markov Length},
   volume={134},
   ISSN={1079-7114},
   url={http://dx.doi.org/10.1103/PhysRevLett.134.070403},
   DOI={10.1103/physrevlett.134.070403},
   number={7},
   journal={Physical Review Letters},
   publisher={American Physical Society (APS)},
   author={Sang, Shengqi and Hsieh, Timothy H.},
   year={2025},
   month=Feb 
}

@article{information_critical_phases,
    author = "Vijay, Akash and Lee, Jong Yeon",
    title = "{Information Critical Phases under Decoherence}",
    eprint = "2512.22121",
    archivePrefix = "arXiv",
    primaryClass = "quant-ph",
    month = "12",
    year = "2025"
}

@article{tobias2026optimal,
  title={Optimal Decoding with the Worm},
  author={Tobias, Zac and Breuckmann, Nikolas P and Placke, Benedikt},
  journal={arXiv preprint arXiv:2603.05428},
  year={2026}
}

@article{chernoff1952measure,
  title={A measure of asymptotic efficiency for tests of a hypothesis based on the sum of observations},
  author={Chernoff, Herman},
  journal={The Annals of Mathematical Statistics},
  pages={493--507},
  year={1952},
  publisher={JSTOR}
}

@article{fuchs1996distinguishability,
  title={Distinguishability and accessible information in quantum theory},
  author={Fuchs, Christopher A},
  journal={arXiv preprint quant-ph/9601020},
  year={1996}
}

@article{temkin2025charge,
  title={Charge-Informed Quantum Error Correction},
  author={Temkin, Vlad and Weinstein, Zack and Fan, Ruihua and Podolsky, Daniel and Altman, Ehud},
  journal={arXiv preprint arXiv:2512.22119},
  year={2025}
}

@article{sala2026climbing,
  title={Climbing the Fusion Tree: Center Ordering and Decoding Hierarchy in Non-Abelian Mixed States},
  author={Sala, Pablo and Temkin, Vlad and Xu, Cenke and Podolsky, Daniel and Altman, Ehud},
  journal={arXiv preprint arXiv:2610.06692},
  year={2026}
}

@article{sala2024spontaneous,
  title={Spontaneous strong symmetry breaking in open systems: Purification perspective},
  author={Sala, Pablo and Gopalakrishnan, Sarang and Oshikawa, Masaki and You, Yizhi},
  journal={Physical Review B},
  volume={110},
  number={15},
  pages={155150},
  year={2024},
  publisher={APS}
}

@article{bhattacharyya1946measure,
  title={On a measure of divergence between two multinomial populations},
  author={Bhattacharyya, Anil},
  journal={Sankhy{\=a}: The Indian Journal of Statistics (1933-1960)},
  volume={7},
  number={4},
  pages={401--406},
  year={1946},
  publisher={JSTOR}
}
\end{document}